\documentclass[preprint,aps,showpacs,preprintnumbers]{revtex4-2}

\usepackage{epsfig,rotating}
\usepackage{multirow}
\usepackage{amsbsy}
\usepackage{amsfonts}
\usepackage{stmaryrd} 
\preprint{\today}

\usepackage{latexsym}
\usepackage{gensymb}
\usepackage{textcomp}
\usepackage{MnSymbol}
\usepackage{subfig}
\usepackage{dashrule}
\usepackage[usenames, dvipsnames]{color}
\usepackage{floatrow}
\usepackage{float}
\usepackage[font=small,format=plain,justification=justified,singlelinecheck=off]{caption}
\usepackage{multirow}
\usepackage{epstopdf}
\usepackage{pifont}
\usepackage{rotating}
\usepackage{tikz}
\usepackage{color}
\usepackage{lineno}
\usepackage{array}
\usepackage{blindtext}
\usepackage{tabularx}
\usepackage{graphicx}
\usepackage[geometry]{ifsym}
\usepackage{amsmath,amsthm} 

\begin{document}

\title{Nuclear spin relaxation in aqueous paramagnetic ion solutions}

\author{David A. Faux}
\affiliation{Department of Physics, University of Surrey, Guildford, GU2 7XH, UK.}
\email{d.faux@surrey.ac.uk}

\author{\"{O}rs Ist\'{o}k}
\affiliation{Department of Physics, University of Surrey, Guildford, GU2 7XH, UK.}

\author{Arifah A. Rahaman}
\affiliation{Department of Physics, University of Surrey, Guildford, GU2 7XH, UK.}

\author{Peter J. McDonald}
\affiliation{Department of Physics, University of Surrey, Guildford, GU2 7XH, UK.}

\author{Eoin McKiernan}
\affiliation{School of Chemistry, University College Dublin, Belfield, Dublin 4, Ireland.
}%

\author{Dermot F. Brougham}
\affiliation{School of Chemistry, University College Dublin, Belfield, Dublin 4, Ireland.
}%

\date{\today}

\begin{abstract}
A Brownian shell model describing the random rotational motion of a spherical shell of uniform particle density is presented and validated by molecular dynamics simulations.  The model is applied to proton spin rotation in aqueous paramagnetic ion complexes to yield an expression for the Larmor-frequency-dependent nuclear magnetic resonance spin-lattice relaxation rate $T_1^{-1}(\omega)$ describing the dipolar coupling of the nuclear spin of the proton with the electronic spin of the ion.  The Brownian shell model provides a significant enhancement to existing particle-particle dipolar models without added complexity allowing fits to experimental $T_1^{-1}(\omega)$ dispersion curves without arbitrary scaling parameters.  The model is successfully applied to measurements of $T_1^{-1}(\omega)$ from aqueous manganese(II), iron(III) and copper(II) systems where the scalar coupling contribution is known to be small.  Appropriate combinations of Brownian shell and translational diffusion models, representing the inner and outer sphere relaxation contributions respectively, are shown to provide excellent fits.  For the first time, quantitative fits are obtained to the full dispersion curve of each aquoion with just five fit parameters with the distance and time parameters each taking a physically justifiable numerical value.

\end{abstract}

\maketitle

\newpage

\section{Introduction} \label{sec:Intro}

A model is presented to describe the rotational dynamics of water in the vicinity of paramagnetic ion complexes in aqueous solution. The model is verified by molecular dynamics (MD) simulations and then used to calculate nuclear magnetic resonance (NMR) spin-lattice relaxation rates due to the coupling of the unpaired electrons of the central paramagnetic ion of spin quantum number $\mathbf{S}$ with the surrounding   nuclear spins $\mathbf{I}$ of the water.   Fundamentally, the rotational and translation dynamics of water in the vicinity of paramagnetic ion complexes determines the NMR spin-lattice and spin-spin relaxation rates,  $R_1= T_1^{-1}$ and $R_2=T_2^{-1}$ respectively. Measurements of $R_1$ and $R_2$ can be exploited for a diverse range of applications in varied material systems \cite{jaouen2006bioorganometallics,aime2005advances, bertini2005advances,geraldes2009classification,pan2011manganese,bodart2020quantification, baroni2009relaxometric,shapiro2011structure,Faux.2017a,Faux.2017b,korb2011}.

For instance, ten metal elements in the human body are essential for life and eight of these are paramagnetic \cite{zoroddu2019essential}.  
Paramagnetic ion complexes are naturally present in the active centers of biomolecules \cite{jaouen2006bioorganometallics} such as haemoglobin.  Magnetic resonance imaging relies upon image contrast between healthy and diseased tissues which may be enhanced by targeted contrast agents incorporating high-spin paramagnetic ions \cite{aime2005advances, bertini2005advances,geraldes2009classification,pan2011manganese}. Relaxation rates measured from liquids in the food industry, such as vinegars and wines, contain many species of paramagnetic ion providing a ``fingerprint” that can potentially identify the product origin for anti-fraud applications \cite{bodart2020quantification, baroni2009relaxometric}.  Furthermore, aqueous paramagnetic ions contribute to NMR relaxation rates measured from porous systems such as cementitious materials, polymer systems, catalysts, silica pastes and glasses, rocks and clays, to name but a few \cite{shapiro2011structure,Faux.2017a,Faux.2017b,korb2011}.   

The novel Brownian shell model presented in this article yields simple expressions for the field-dependent relaxation rates $R_1$ and $R_2$ for paramagnetic ion systems dominated by nuclear dipolar interactions.  Paramagnetic ions are hydrated by water molecules with size and steric considerations limiting the immediate neighbors to normally six water molecules in the first hydration layer known as the ``inner sphere".   The hydrate complex is then surrounded by anions and more water, referred to as ``outer sphere” water, that diffuses and can exchange with the inner sphere. The coupling of the unpaired electrons of the central paramagnetic ion of spin $\mathbf{S}$  with nuclear spins $\mathbf{I}$ of neighboring water hydrogen protons results in relaxation rates that provide insight into water dynamics.  The novel Brownian shell model presented in this article yields simple expressions for the field-dependent NMR spin-lattice and spin-spin relaxation rates $R_1$ and $R_2$ respectively for paramagnetic ion systems dominated by nuclear dipolar interactions. 

Nuclear magnetic resonance was first used to study nuclear spin relaxation in aqueous paramagnetic ion solutions in 1948 by Bloembergen, Purcell and Pound \cite{Bloembergen.1948} measuring $R_1$ for aqueous iron(III) ions, with later   studies exploring a range of aquoions \cite{bloembergen1957proton,morgan1959proton,bloembergen1961proton}. 
The analysis of magnetic resonance involving paramagnetic ions started with consideration of the related problem of a \emph{single} paramagnetic ion with a nuclear spin coupled to its own unpaired electrons \cite{bleaney1953paramagnetic,abragam1955overhauser,bloembergen1961proton,Abragam}.
The tensor coupling between the nuclear and electronic spins leads to hyperfine splitting of the atomic electronic energy levels with a magnitude normally dominated by the $\ell = 0$, or $s$-orbital, contribution.  The $s$-orbital coupling  provides a scalar term $\hbar A \,\mathbf{I}.\mathbf{S}$ where $A$ is a coupling constant proportional to the $s$-orbital wavefunction overlap at the site of the atomic nucleus.  

The coupling of a paramagnetic ion of electronic spin quantum number $\mathbf{S}$ with nuclear spins $\mathbf{I}$ on {\em different} atoms is also described by Abragam \cite{Abragam}.  The tensor $\mathbf{I}$--$\mathbf{S}$ coupling now manifests as {\em two} dominant contributions: the scalar coupling of the electronic spin orbital wavefunction at the site of a nuclear spin $\mathbf{I}$ of a \emph{different} atom, added to the dipolar coupling of $\mathbf{I}$ with the magnetisation density of the ion electronic spin $\mathbf{S}$ determined by classical magnetostatics.  The latter is referred to as the dipolar contribution.

Most NMR relaxometry experiments are undertaken at a fixed magnetic field typically in the $^1$H Larmor frequency range 2—20\,MHz.  Fast field-cycling NMR (FFC NMR) spectrometry is a specialised NMR technique that measures $R_1(\omega)$ as a function of magnetic field strength typically across a frequency range $f=\omega/2\pi$  from 0.01 to 20\,MHz.  The dispersion curves are exquisitely sensitive to the {\em dynamics} of spins over timescales from picoseconds to microseconds and display features that can be associated with specific relaxation processes.

The interpretation of the $T^{-1}_1(\omega)$ dispersion curve from a FFC NMR experiment for an aqueous paramagnetic system poses a difficult challenge.  It is not possible {\em a priori} to establish whether the dispersion curve is dominated by scalar coupling, dipolar coupling, or whether both contribute.  First, consider the inner sphere water.  The scalar coupling arises due to the overlap of the $s$-orbital wave functions of the electronic spins of the ion at the site of each of the $^1$H nuclear spins of the inner sphere water, often referred to as the ``contact" term. Relaxation due to the $\mathbf{I}$--$\mathbf{S}$ {\em dipolar} coupling, by contrast, is associated with the rotation of the shell of inner sphere $^1$H nuclear spins, or due to the exchange of water between the inner and outer spheres, or a combination of both dynamical processes. For the outer sphere water, the scalar coupling is not considered important due to the distance between the $\mathbf{I}$ and $\mathbf{S}$ spins \cite{Abragam}.  Relaxation due to the $\mathbf{I}$--$\mathbf{S}$ dipolar coupling with outer sphere protons is expected to contain contributions from both the change in angle and length of the $\mathbf{I}$--$\mathbf{S}$ spin vector due to rotational and relative translational motion respectively.

Solomon, Bloembergen, Morgan and others \cite{solomon1955physical,solomon1956nuclear,laukien1956pulse,morgan1959proton,bloembergen1961proton} adopted a simplified  model of the dipolar contribution by considering only rotational motions. These authors fixed the ion-$^1$H distance $d$ and treated the rotational dipolar correlation function as a single exponential function.  The SBM model produces a good match to inflection features of experimental $T^{-1}_1(\omega)$ dispersion curves and has since been the first port of call for the interpretation of experimental dispersions from liquids that contain paramagnetic ions.  The book by Bertini, Luchinat and Parigi \cite{bertini2001solution} illustrates the application of the SBM model to a range of aquoions.  

However, the SBM model is known to be flawed and Pell {\em et al} provide an excellent summary of some of the shortcomings \cite{pell2019paramagnetic}. The SBM model applied to paramagnetic aquoions originally associated an inflection feature normally seen in dispersion curves at 1—10\,MHz with the inner sphere water.  The associated dynamical time constant lies typically in the range 20—40\,ps \cite{bertini2016nmr} and was attributed to the reorientation of the inner sphere water.  However, these translational and rotational motions are  too short, typically a few picoseconds (\cite{faux2021a} and sources therein).  The inflection feature is now associated with the outer sphere water where the rotational time constant $\tau_{\rm r}$ of the second coordination shell lies in the range 20—40\,ps confirmed by molecular dynamics simulations in Sec.\,\ref{SectionResultsMn}.

Further evidence of the limitations of the SBM model emerge when quantitative fits are attempted using both the ion-$^1$H distance $d$ and $\tau_{\rm r}$.   
Fits revealed ion-$^1$H distances $d$ in the range 1.8--1.9\,\AA\ \cite{hernandez1991proton,powell199117o} which ``...is much too short to be reasonable'' \cite{hernandez1991proton}.  We demonstrate that the discrepancies are because the SBM model is a particle-particle model predicting a $d^{-6}$ interaction dependence  rather than a particle-shell model which leads to a $d^{-4}$ dependence.  Finally, and least significantly, the single-exponential approximation for the dynamical correlation function does not properly account for the angular boundary conditions of the rotational motion.

In this article we address these shortcomings by presenting a Brownian shell model that accommodates both the angular boundary conditions and yields the correct $d^{-4}$ distance dependence for $T^{-1}_1(\omega)$.   The Brownian shell model describes the rotational changes in the $\mathbf{I}$--$\mathbf{S}$ vector of the inner sphere water and captures the rotational (non-translational) dynamics of the outer sphere water.  The contribution to $T^{-1}_1(\omega)$ due to {\it translational} dynamics is described by an established model due to Hwang and Freed \cite{hwang1975dynamic}.    We conduct fresh experiments on de-oxygenated manganese(II), iron(III) and copper(II) chloride solutions and conduct fits 
to yield for the first time physically meaningful distance and dynamical time constants in each case.

Nonetheless, it remains a challenge to separate the scalar and dipolar contributions to dispersion features associated with the {\em inner} sphere water.  For example, the Mn(II) chloride system has a strong inflection feature at about 0.1\,MHz associated with the $\mathbf{I}$--$\mathbf{S}$ coupling with the inner sphere protons of the hexahydrate.   Both the SBM and scalar expressions for $T^{-1}_1(\omega)$ have a similar functional form \cite{la1973proton} and so a quality-of-fit criterion is ineffective at distinguishing between models.  The physical reasonableness of the fit parameters may provide clues, but the scalar contribution  contains a coupling constant that cannot be independently verified and a time constant whose origin is unclear.   As stated, the SBM model is a particle-particle model and yields an incorrect distance dependence. 

One frequently cited argument for dismissing the dipolar coupling as a contributor to the low-frequency dispersion feature in Mn(II) hexahydrate systems is that the rotational time constant must be of order microseconds to place the 0.1\,MHz inflection feature at the correct  frequency.  A microsecond rotational time constant is slow when compared to rotational time constants for outer-sphere water in the range 20--40\,ps \cite{pfeifer1962protonenrelaxation,koenig1984relaxation} or when deduced from models describing the complex as a uniform sphere in a viscous medium \cite{bloembergen1961proton,Abragam}.  In this article we show that the dipolar contribution cannot be dismissed on this basis.   We model the rotation of the Mn(II) chloride aqueous complex using both a classical collision model and molecular dynamics (MD) simulations.  Both suggest inner sphere rotational time constants of order microseconds.

Temperature-dependent measurements of $T^{-1}_1(\omega)$ provide evidence as to the relative contributions of the scalar and dipolar coupling.  Dipolar relaxation associated with dynamical processes should present an Arrhenius temperature dependence resulting in a decrease in the low frequency relaxation rate with temperature, whereas the scalar contribution is expected to increase as the temperature is raised \cite{ hernandez1991proton}. The temperature dependence of $T^{-1}_1(\omega)$ for the hexahydrates of both Mn(III) and Mn(II) were studied by Hernandez and Bryant \cite{ hernandez1991proton}.  The dispersions of the two ions show vastly different temperature dependencies.  The Mn(III) dispersion suggests contributions from both scalar and dipolar coupling. The Mn$^{3+}$--$^1$H distance is sufficiently short for the electron spin wave function overlap at the site of the $^1$H spin to be significant.  By contrast, for Mn(II) hexahydrate, a temperature dependence consistent with an activation law was found implying ``… that the electron relaxation rate makes a much smaller contribution if any to the effective correlation time for the electron-nuclear coupling” \cite{ hernandez1991proton}.  

On the basis that the experimental evidence allows the scalar contribution to the $T_1^{-1}(\omega)$ dispersion to be neglected for Mn(II) hexahydrate, we fit the Brownian shell model to the inner sphere feature of aqueous Mn(II) chloride.  Applying the same Brownian shell model to the outer sphere provides an exquisite match to the full dispersion curve of Mn(II) aquoions with just five physically-meaningful parameters yielding length- and time-scales in agreement with experiment and supported by fresh MD simulations.  For Fe(III), the absence of an inner sphere feature allows the dispersion to be fit using outer-sphere-only models. By contrast, for Cu(II), the water is loosely bound as evidenced by the very short inner--outer sphere exchange rate and is considered to be the dominant relaxation process for the inner sphere relaxation rate contribution.  This assumption is justified by the excellent agreement of the fit exchange time constant to experiment presented in Sec.\,\ref{SectionResults}.  The results provide fresh insight into water dynamics in and around aqueous metal ions in general and supply equations describing the frequency-dependent relaxation rate for aqueous Mn(II), Fe(III) and Cu(II) ions in particular.  

The article is structured as follows.  The outline theory for the Brownian shell model and summary of alternative models are presented in Sec.\,\ref{Sec:NMRmodels} with more detail contained in the Supplementary Material where the  underpinning assumptions are tested and validated through MD simulations.  Fast field cycling NMR measurements of $R_1(\omega)$ from aqueous Mn(II), Fe(III) and Cu(II) chlorides are described in Sec.\,\ref{SectionExperiments} and interpreted in Sec.\,\ref{SectionResults}.  Finally, the key outcomes are summarised in Sec.\,\ref{Section:conclusions}.

\section{The Brownian Shell and Alternative Models} \label{Sec:NMRmodels}

 The Brownian shell model describes the random rotational motion of a fixed length vector and provides significant enhancements to the SBM rotational model.  Both the SBM and Brownian shell models are described together with two diffusion models characterising the {\em translational} dynamics of water in the vicinity of paramagnetic ions.  First, however, we present general theoretical background explaining how diffusion models are converted to predictions of NMR relaxation rates.
 
\subsection{General NMR background}

A contribution to the spin-lattice (longitudinal)  relaxation rate $T_1^{-1}(\omega) = R_1(\omega)$ arises due to fluctuations in the dipolar interaction between the electronic spin ($S$) of a paramagnetic ion and a nuclear spin ($I$) of a water proton.   On the assumption that the electronic spin is located at the center of the paramagnetic ion, and within the fast-diffusion approximation \cite{redfield1955nuclear}, the relaxation rate may be written  \cite{Abragam,Faux.2017b}
\begin{eqnarray}
T_1^{-1}(\omega) =  &\:\:R_1(\omega)=&  \frac{1}{3} \beta_{IS} \left[  7 J(\omega_S) + 3 J (\omega)\right] \label{T1sigma}
\end{eqnarray}
where $\beta_{IS}= \left( \mu_0 / 4\pi \right)^2 \gamma_I^2  \gamma_S^2 \: \hbar^2 S(S+1)$, and $\gamma_S$ ($\gamma_I$) is the gyromagnetic ratio for the electronic (nuclear) spin.  The Larmor angular frequency of the electronic spin in the applied static field is $\omega_S \!= \! 658.21 \omega$ and $\omega = 2 \pi f$.  The spectral density functions $J(\omega)$ are obtained from the Fourier transformation of the dipolar correlation function $G(t)$ using the expression
\begin{eqnarray}
J(\omega) & = & 2 \int^\infty_{0} \!\!  G(t)\,\cos \omega t \,dt \:. \label{Jeqn}
\end{eqnarray}

\begin{figure}[tbh!]
	
	\begin{center}
		\includegraphics[width=1.00\textwidth, trim={2cm 12cm 2cm 0.7cm}]{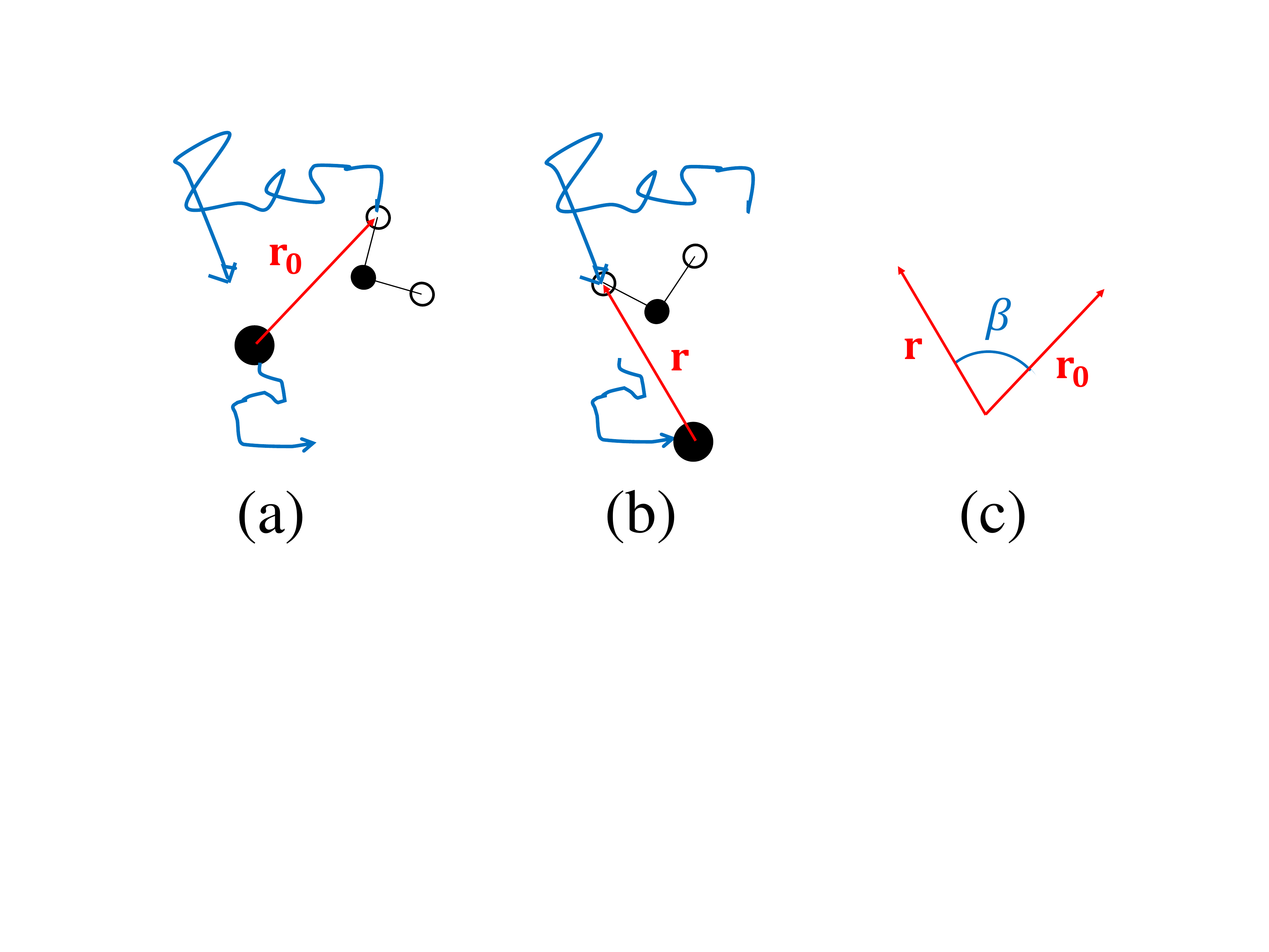} \\[-3mm]
		\caption{In (a), at $t=0$, a vector $\bf{r}_0$ extends from the paramagnetic ion to a hydrogen spin of a water molecule. In (b), time $t$ later, both the ion and the water molecule have moved and $\bf{r}$ links the ion to the same hydrogen spin. (c) shows the angle $\beta$ between the two spin-pair vectors.} 
		\label{Fig1inner_schematic}
	\end{center}
\end{figure}

Nuclear spin relaxation arises due to the relative motion of pairs of spins.  The spin-lattice relaxation rate $R_1(\omega)$ is obtained from the Fourier transform of the time-dependent dipolar correlation function $G(t)$. $G(t)$ captures all the relevant dynamical information describing how pairs of spins move relative to each other.
The key to developing a model describing the frequency-dependence of  NMR relaxation rates is therefore the determination of the dipolar correlation function $G(t)$.  
In three-dimensional (3D) systems, such as unconfined fluids, $G(t)$ may be determined using  the expression \cite{Messiah.1965,Sholl1974nuclear}
\begin{eqnarray}
G(t)  & =  &   \left< \frac{ P_2( \cos \beta )  }{r_0^3 r^3} \right> \label{Gst3Daveave}
\end{eqnarray}
where $r_0$ and $r$  are the magnitudes of vectors connecting the center of a paramagnetic ion to a proton spin at time $t\!=\!0$ and the same spin pair at time $t$ respectively. The angle between the vectors ${\bf r}$ and ${\bf r_0}$ is $\beta$ as illustrated in Fig.\ref{Fig1inner_schematic}.   The function  $P_2(x) = \frac{1}{2}(3 \cos^2 x - 1)$ is a Legendre polynomial.  The angular bracket of Eq.\,(\ref{Gst3Daveave}) indicates the average over an ensemble of spin pairs.

The water in aqueous ion systems can be identified with one of two  environments: the inner sphere of the ion, or the outer sphere.  The motion of the inner sphere water is restricted by its interaction with the ion whereas the outer sphere water has dynamical characteristics similar to that of bulk water. Water exchanges between the inner and outer spheres.  A description of key terms used in this article is presented in Table \ref{Table:terminology}.

The dipolar correlation function $G(t)$ is required for the $S\!-\!I$ interactions for $^1$H spins in each environment. A fraction $x$ of all $^1$H spins occupies the inner sphere and the remaining fraction $1\!-\!x$ occupies the outer sphere.  The spin-lattice relaxation rates associated with the inner and outer spheres are  $R_{{\rm1 i}}$ and $R_{{\rm 1o }}$ respectively.  Provided that the rate of exchange of $^1$H spins between the two environments is faster than the experimental relaxation rate,   $R_1(\omega)$ is the average of the relaxation rate for the inner and outer spheres weighted by the fraction of spins  in each environment \cite{Brownstein.1979}, thus
\begin{eqnarray}
R_1(\omega) & = & x R_{{\rm 1 i}}(\omega) + (1-x) R_{{\rm 1 o}}(\omega) \label{eqn:RRR} \:.
\end{eqnarray}

\linespread{0.95}
\begin{table}[ht]
	\caption{Key terms and a selection of dipolar models are summarised. }
	
	\begin{tabular}{p{0.25\linewidth}  p{0.75\linewidth}}\toprule \\[-2mm]
		{\bf TERM}	&  {\bf DESCRIPTION}  \\[1mm] \hline \\[-2mm]
		
		{\bf Inner sphere} & The inner sphere  (or first coordination sphere) refers to water contained in the first hydration shell of the ion. The inner sphere often contains six water molecules. \\[2mm] 
		
		{\bf Outer sphere} & The outer sphere refers to all water not contained in the inner sphere. \\[2mm] 
		
		{\bf SBM model} & The SBM model is due to Soloman, Bloembergen and Morgan   \cite{solomon1955physical,solomon1956nuclear,laukien1956pulse,
			morgan1959proton,bloembergen1961proton} and assumes Brownian rotation of spin pair vectors. \\[2mm] 
		
		{\bf Shell model} & In a shell model, a volume of water $^1$H spins are assumed to lie on a representative spherical  shell of fixed radius with the ion at the center. The spin pair vector connecting the ion and a $^1$H spin in the shell can change angle, but not length.   \\[2mm] 
		
		{\bf Inner shell model} &
		The inner shell model places all the $^1$H spins of the inner sphere water at a fixed distance $a$ from the ion as illustrated in Fig.\,\ref{Fig2inner_outer}(a). \\[2mm] 
		
		{\bf Outer shell model} &
		The outer shell model places all $^1$H spins of the outer sphere water at a representative effective distance $d$ as illustrated in Fig.\,\ref{Fig2inner_outer}(b) and described in section \ref{SubSecOuter}. {\it All} $^1$H spins not in the inner sphere are set to a distance $d$ from the ion.  The outer shell model therefore considers relaxation associated with rotational motion only.  \\[2mm] 		
		
		{\bf Continuum model} & A continuum model allows a uniform density of $^1$H spins to occupy a region of space.    The continuum model presented by Abragam in 1961 \cite{Abragam} is illustrated in Fig.\,\ref{Fig2inner_outer}(c). The spin pair vector connecting the ion and a $^1$H spin is assumed to change in length, but not angle. \\[2mm] 
		
		{\bf Hwang-Freed model} & The Hwang--Freed continuum model \cite{hwang1975dynamic}, obtained independently by Ayant {\em et al} \cite{ayant1975calcul}, imposes a boundary condition that prevents particles from moving closer together than $d_{\rm HF}$ as illustrated in Fig.\,\ref{Fig2inner_outer}(d). This model provides an improved physical description of the diffusion process and the considerable advantage of an analytical expression for the NMR spectral density functions.\\[1mm]
					\hline \hline 		
	\end{tabular} 
	\label{Table:terminology}
\end{table}

\begin{figure}[tbh!]
	
	\begin{center}
		\includegraphics[width=1.00\textwidth, trim={2cm 2cm 2cm 1cm}]{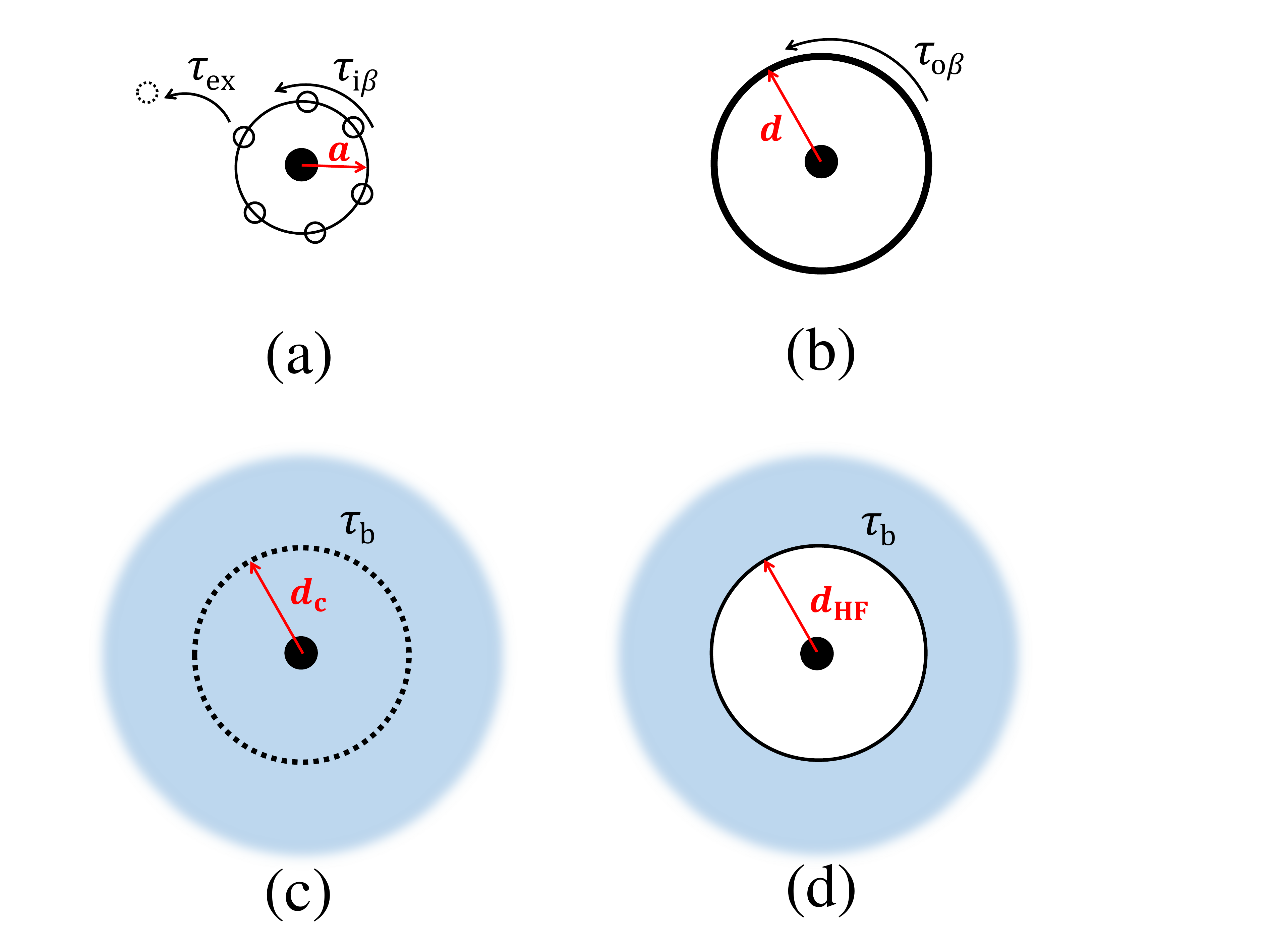}
		\caption{The figure presents four inner or outer sphere models.  Model (a) represents the inner shell model and shows a paramagnetic ion (\FilledCircle) surrounded by six representative $^1$H spins ({\SmallCircle}) constrained to a shell of radius $a$. The rotational time constant is $\tau_{\rm i \beta}$ and $\tau_{\rm ex}$ represents the inner-outer sphere water exchange lifetime.  Model (b) is the outer shell model with all outer sphere $^1$H spins placed at a fixed distance $d$.  The rotation of the shell is characterised by $\tau_{\rm o \beta}$ (see text). Model (c) is a continuum model describing the ion-$^1$H relative translational motion assuming a uniform $^1$H spin density allowing water to diffuse through the inner sphere space.  Characterising parameters are a distance $d_c$ and the bulk translational diffusion time constant $\tau_{\rm b}$.  Model (d) represents the Hwang-Freed \cite{hwang1975dynamic,ayant1975calcul} continuum model describing the ion-$^1$H relative translational motion with a reflective boundary condition at $r=d_{\rm HF}$.  }
		\label{Fig2inner_outer}
	\end{center}
\end{figure}

The objective is to identify a combination of  inner  and outer sphere models that captures the dynamics of $S\!-\!I$ spin pairs in each environment yielding expressions for the dipolar contribution to  $R_1(\omega)$ over the full frequency range of a FFC NMR experiment.  Candidate models are illustrated in Fig.\,\ref{Fig2inner_outer}.  

\subsection{The SBM particle--particle model} \label{inner-SBM}

The Soloman, Bloembergen and Morgan   \cite{solomon1955physical,solomon1956nuclear,laukien1956pulse,
	morgan1959proton,bloembergen1961proton} SBM model is a particle--particle model which can be applied to both the inner and outer sphere water.  The SBM model applied to the inner sphere considers the rotational decorrelation of the vector connecting the electronic point dipole at the ion center to an inner sphere $^1$H  spin.  The assumption of Brownian rotational motion (and ignoring angular boundary conditions) leads to a dipolar correlation function $G_{\rm i}(t) \!\propto \!\exp(-t/\tau_{\rm i})$ where $\tau_{\rm i}$ is the relaxation timescale.  The spectral density function may be expressed as
\begin{eqnarray}
J(\omega) & = & \frac{A}{a^6} \left( \frac{2 \tau_{\rm i}}{1 + \omega^2 \tau_{\rm i}^2} \right) \label{eqn:JwSBM}
\end{eqnarray}
where $A$ is a constant proportional to the volume density of paramagnetic ions.   The term $a^{-6}$ in the denominator of Eq.\,(\ref{eqn:JwSBM}) signals a particle-particle model.  The distance $a$ is normally considered the distance of nearest approach of the water $^1$H spins to the ion.

According to the SBM model, the time constant $\tau_{\rm i}$ may be identified with several processes via \cite{helm2006relaxivity}
\begin{eqnarray}
\frac{1}{\tau_{\rm i}}& = & \frac{1}{\tau_r} + \frac{1}{\tau_{\rm ex}} \label{eqn:SBMtimes}
\end{eqnarray}
where $\tau_r$ is a spin-pair-vector rotational time constant and the exchange time constant $\tau_{\rm ex}$ is the characteristic residence time of water molecules in the inner sphere. If there are $N_0$ proton spins in the inner sphere at $t\!=\!0$, there are $N_0 \exp(-t/\tau_{\rm ex})$ spins remaining at a later time $t$. Departing spins move into the outer sphere and are rapidly transported away. 

\begin{table}[ht]
	\caption{The dynamical time constants relevant to the SBM, shell and continuum models are presented and described. }
	
	\begin{tabular}{p{0.10\linewidth}  p{0.75\linewidth}}\toprule \\[-2mm]
		{\bf TIME}	 & {\bf DESCRIPTION}  \\[1mm] \hline \\[-2mm]
		
		$\tau_{\rm i}$ & A time constant that captures all dynamical processes associated with the inner sphere, for example through Eq.\,(\ref{eqn:SBMtimes}). \\[2mm] 
		
		$\tau_{\rm ex}$ & The exchange lifetime of a water molecule between the inner and outer sphere. If there are $N_0$ spins in the inner sphere at $t\!=\!0$,  $N_0 \exp(-t/\tau_{\rm ex})$ of those remain at later time $t$. 
		\\[2mm] 		
		
		$\tau_{\rm i \beta}$ & Used in shell models, this is the rotational time constant associated with the inner shell.  The time $2 \tau_{\rm i \beta}$ is the representative time for the ion complex comprising the inner sphere water to rotate through  one radian. \\[2mm] 
		
		$\tau_{\rm o \beta}$ & Used in shell models, this is the rotational time constant for the outer shell.  The time $2\tau_{\rm o \beta}$ is the representative time for the outer shell to rotate through one radian. In the context of $^1$H relaxation, it reflects the  $r^{-4}$ weighted average time for all the second coordination sphere ion--$^1$H rotors to move through one radian.  \\[2mm] 
		
		$\tau_{\rm b}$ &  The translation diffusion time constant of the bulk water is defined in terms of the 3D self-diffusion coefficient $D$  as $\tau_{\rm b} \!=\! \delta^2/6D$. $\delta$ is set to 0.27\,nm, the distance between one water oxygen and its nearest neighbor, so that a water molecule takes time $\tau_{\rm b}$ to move a distance 0.27\,nm in 3D. $\tau_{\rm b} \! \approx \! 5.5$\,ps for pure water at room temperature \cite{krynicki1978pressure}.   \\[2mm] 
		
		$\tau_{\rm r}$ & The rotational time constant used for the SBM model. 
		\\[1mm]	
		\hline \hline 
	\end{tabular} 
	\label{Table:timeConstants}
\end{table}

Equation (\ref{eqn:JwSBM}) may be substituted into Eq.\,(\ref{T1sigma}) to obtain a frequency-dependent relaxation rate that provides a good fit to many inflection features of FFC NMR dispersion curves obtained from aqueous paramagnetic systems indicating that the important dynamical processes are captured by the SBM model.  Its simplicity is attractive and the SBM model is often used to separately fit to both the inner and outer sphere contributions in systems where there are two separate inflection features at different Larmor frequencies. 

The SBM model is valid when dipolar interactions dominate and in the fast-diffusion limit \cite{redfield1955nuclear} whereby the proton spins have explored all dynamical and physical environments during time $T_1$.  The shortcomings of the SBM model are well known (see discussion by Kowalewski \cite{kowalewski1985theory} and summary by Pell {\em et al} \cite{pell2019paramagnetic}).  In particular, the description of the rotational correlation function describing the $S\!-\!I$  pair vector as a single exponential function is limited and the model does not provide the correct distance dependence for a  shell of $^1$H spin density, preventing a quantitative match to experiment without an arbitrary scaling factor.  

Due to more recent FFC NMR data, it is now timely to revisit the problem and to provide an enhanced dipolar model which, when fit to FFC NMR nuclear spin relaxation dispersion curves, yields relevant length- and time-scales consistent with independent experimental data obtained from aqueous paramagnetic ion solutions.  
Four alternative inner  and outer sphere models presented in Fig.\,\ref{Fig2inner_outer} reflect the most accessible and most common literature models for systems operating within the fast-diffusion limit.  The Brownian shell model is a shell model of the form illustrated in Fig.\,\ref{Fig2inner_outer}(a) and (b).  

Terminology and time constants associated with the models discussed below are summarised in Tables\,\ref{Table:terminology} and \ref{Table:timeConstants}.

\bigskip

\subsection{The Brownian shell model applied to inner sphere water} 
\label{inner-Brownian}

Steric hindrance ensures typically six water molecules at 2-3\,\AA\ from an aqueous paramagnetic ion center.  The inner sphere water remains in place for times from hundreds of picoseconds to hundreds of years \cite{helm2005inorganic} depending on the ion. 
The Brownian shell model is a dipolar model applied to both inner sphere and outer sphere water.  It assumes, as the SBM model does,  that the relaxation rate arises due to fluctuations of the spin-pair vector $\bf{r}$ connecting the electronic spin of the paramagnetic ion to the $^1$H spins of the water molecules set at a fixed distance $|{\bf r}_0|\!=\! |{\bf r}|\! =\! a$ as illustrated in Fig.\,\ref{Fig2inner_outer}(a). 

The dipolar correlation function for the inner shell, $G_{\rm i}(t)$, is determined from the time-dependent probability density function $P(\beta,t)$ where $\beta$ is the angle between $\bf{r}$ and $\bf{r}_0$ as defined in Fig.\,\ref{Fig1inner_schematic}.  $P(\beta,t)$ is derived in  Supplementary Material Note 1 \cite{SM1} through simplification of the analysis of a L\'{e}vy rotor \cite{faux2021a}. L\'{e}vy statistics is applied to systems where extreme angular changes occur more often than described by a Brownian model.  The extent to which the rotational diffusion of $\beta$ departs from Brownian motion is captured by the L\'{e}vy parameter $\alpha$.  Brownian rotational diffusion corresponds to $\alpha\!=\!2$ and L\'{e}vy behavior is signalled by $\alpha \!<\!2$. 
The \emph{outer} shell rotational dynamics is shown in Supplementary Material Note 2 \cite{SM21} 
to be Brownian. We later demonstrate through MD simulation that the \emph{inner} shell rotational dynamics is more complex. Nonetheless, the description of the inner shell rotation as  Brownian is assumed leading to 
\begin{eqnarray}
P(\beta,t) = N(t)  \left[ 1 +2 \sum_{p=1}^\infty e^{-p^2t/\tau} \cos p\beta  \right] \label{S9:Pbeta}
\end{eqnarray}
where the normalisation constant $N(t)$ may be found in Supplementary Material Note 1 \cite{SM1} and $\tau$ is the relevant rotational time constant.
The assumption of Brownian dynamics has the significant advantage of allowing the expansion of  $G(t)$ as an exponential series \cite{SM1} and providing a final expression for the relaxation rate dispersion as
\begin{eqnarray}
G_{\rm i}(t) & = & \frac{32 \pi \delta N_M}{15 a^4} \left[ e^{-t/\tau_{{\rm i}\beta}} + \tfrac{2}{3} e^{-2t/\tau_{{\rm i}\beta}} +\left(\tfrac{2}{3} \right)^2 e^{-3t/\tau_{{\rm i}\beta}} ... \right]   \label{eqn:GtXpan1}
\end{eqnarray}
where $N_M$ is the number of paramagnetic ions per unit volume, $\delta\!=\!0.27$ is a standard nanoscale distance.  The rotational inner shell time constant $\tau_{{\rm i}\beta}$ in Eq.\,(\ref{eqn:GtXpan1}) is equivalent to $\tau/4$ in the appendix so that $2\tau_{{\rm i}\beta}$ is the time taken for a rotation through one radian.
Implementation of the boundary conditions $0\!\leq \! \beta\! \leq \pi$ leads to the exponential series expression of Eq.\,(\ref{eqn:GtXpan1}). The dipolar correlation function $G(t)$ is proportional to $a^{-4}$ rather than $a^{-6}$ for particle-particle models due to the averaging of the spin-spin interaction over the shell area.

It is straightforward to show that the third and subsequent terms of the expansion make a negligible contribution to $G_{\rm i}(t)$. We therefore retain the first two terms of the expansion and multiply by a factor $\exp(-t/\tau_{\rm ex})$ to account for the exchange of water molecules between the inner and outer spheres to obtain
\begin{eqnarray}
G_{\rm i}(t) & = & \frac{32 \pi \delta N_M}{15 a^4} \left[ e^{-t/\tau_{\rm i1}} + \tfrac{2}{3} e^{-t/\tau_{\rm i2}} \right]   \label{eqn:GtXpan2}
\end{eqnarray}
where
\begin{eqnarray}
\frac{1}{\tau_{\rm i1}}  =  \frac{1}{\tau_{{\rm i}\beta}} + \frac{1}{\tau_{\rm ex}}   & \mbox{~~~~~~~~} & 
\frac{1}{\tau_{\rm i2}}  =  \frac{2}{\tau_{{\rm i}\beta}} + \frac{1}{\tau_{\rm ex}}  \label{eqn:tauoi} \: .
\end{eqnarray}
The Fourier transform of $G_{\rm i}(t)$ yields the spectral density function
\begin{eqnarray}
J_{\rm i}(\omega) & = & \frac{64 \pi \delta N_M}{15 a^4} \left[ \frac{\tau_{\rm i1}}{1 \!+\! \omega^2 \tau_{\rm i1}^2}   +     \frac{\frac{2}{3}\tau_{\rm i2}}{1 \!+\! \omega^2 \tau_{\rm i2}^2}  \right].   \label{eqn:Jw-inner}
\end{eqnarray}

Equation\,(\ref{eqn:Jw-inner}) is the spectral density function for the Brownian rotation of fixed-length spin-pair vectors with the time constant given by Eqs.\,(\ref{eqn:tauoi}). The inner shell time constants $\tau_{\rm i1}$ and $\tau_{\rm i2}$ are dominated by the shorter of $\tau_{\rm i\beta}$ or $\tau_{\rm ex}$ and so it is useful to generate reduced forms of Eq.\,(\ref{eqn:Jw-inner}) in the limits $\tau_{{\rm i}\beta} \gg \tau_{\rm ex}$ and $\tau_{{\rm i}\beta} \ll \tau_{\rm ex}$.  These are
\begin{eqnarray}
J_{{\rm i}\beta}(\omega) = \frac{64 \pi \delta N_M \tau_{{\rm i}\beta}}{45 a^4} \left[ \frac{3}{1 \!+\! \omega^2 \tau_{{\rm i}\beta}^2}   +     \frac{4}{4 \!+\! \omega^2 \tau_{{\rm i}\beta}^2}  \right]  & \mbox{~~~~~~~if~~~}& \tau_{{\rm i}\beta} \ll \tau_{\rm ex} \label{eqn:Jw-inner-beta} \\[4mm]
J_{\rm ex}(\omega)  =  \frac{64 \pi \delta N_M \tau_{\rm ex}}{9 a^4} \left[ \frac{1}{1 \!+\! \omega^2 \tau_{\rm ex}^2}  \right] &\mbox{~~~~~~~if~~~} &\tau_{{\rm i}\beta} \gg \tau_{\rm ex}. \label{eqn:Jw-inner-d} 
\end{eqnarray}
The spectral density functions given by Eq\,(\ref{eqn:Jw-inner}) and  simplifications given by Eqs.\,(\ref{eqn:Jw-inner-beta}) and (\ref{eqn:Jw-inner-d}) may be substituted into Eq.\,(\ref{T1sigma}) to obtain the frequency-dependent relaxation rate contribution for the Brownian inner shell model with parameters $a$, $\tau_{{\rm i}\beta}$ and/or $\tau_{\rm ex}$.

\subsection{The Brownian shell model applied to the outer sphere water}

The outer sphere refers to all water not in the inner sphere which, for 1\,mM equivalent ion concentration corresponds to about 55,000 water molecules per cation.  The water molecules are diffusing through the sample with a translational diffusion time constant $\tau_b$, and relaxation arises due to changes in the length and orientation of vectors connecting an ion to the $^1$H spins on the water molecules.  The shell model illustrated in Fig.\,\ref{Fig2inner_outer}(b) simplifies the complex relative spin dynamics by placing all outer sphere $^1$H spins in a shell at a fixed distance $d$ from the ion.  The distance $d$ represents an average, or effective, distance recognising the $r^{-4}$ dependence of the $S$--$I$ spin dipolar interaction which defines a spherical surface within which relaxation is effective.  The Brownian shell model applied to the outer sphere water therefore acts as a model to describe the rotational dynamics only.  The Brownian shell model is derived in Supplementary Material Note 1 \cite{SM1}.   Demonstration that the rotation dynamics of the outer shell water is Brownian may be found in Supplementary Material Note 2.1 \cite{SM21}. 

The spectral density function $J_{{\rm o}\beta}(\omega)$ for the outer shell can be adapted from the inner shell expression given by Eq.\,(\ref{eqn:Jw-inner-beta}) to obtain
\begin{eqnarray}
J_{{\rm o}\beta}(\omega) & = & \frac{64 \pi \delta N_M \tau_{{\rm o}\beta}}{45 d^4} \left[ \frac{3}{1 \!+\! \omega^2 \tau_{{\rm o}\beta}^2}   +     \frac{4}{4 \!+\! \omega^2 \tau_{{\rm o}\beta}^2}  \right] \label{eqn:Jw-outer-beta}
\end{eqnarray}
where the rotational time constant for ion-$^1$H spin-pair vectors is $\tau_{{\rm o}\beta}$. The outer shell rotational time constant is typically 20--40\,ps so that even the shortest $\tau_{\rm ex}$ satisfies $\tau_{\rm ex} \! \gg \tau_{{\rm o}\beta}$ \cite{helm2005inorganic}, and exchange of spins from the outer shell to the inner shell can be safely neglected.  \\

\noindent
\subsection{The continuous diffusion dipolar model} \label{outer-continuous} 

The continuous diffusion model illustrated in Fig.\,\ref{Fig2inner_outer}(c) has been widely used to describe  intermolecular relaxation in liquids. The word ``continuous" reflects the assumption that spins can diffuse unhindered through 3D space, and can therefore diffuse through each other.  
The dipolar correlation function may be expressed in several ways \cite{torrey1953nuclear,Abragam,hwang1975dynamic,Faux.1986,Faux.2017b}
depending on the application. A suitable expression for the time-dependent dipolar correlation function is \cite{Faux.2017b}
\begin{eqnarray}
G_{\rm o}^{\rm (c)}(t)  & = & \frac{4 \pi N_M}{d_c^3} \int_0^\infty \kappa^{-1} \,e^{-\delta^2 \kappa^2 \,t/{3d_c^2 \tau_{\rm b}}} \,J_{3/2}^2(\kappa)\: d\kappa  \mbox{~~~~}\label{Gff_final}
\end{eqnarray}
where $J_{3/2}(x)$ is a half-integer Bessel function and $\kappa$ is a Fourier variable. $\tau_{\rm b}$ is defined in Table \ref{Table:timeConstants}.  The superscript ``(c)" and subscript ``c" refer to continuous diffusion.  The derivation of Eq.\,(\ref{Gff_final}) requires a volume integral to be completed whereupon, to avoid a singularity, it is necessary to introduce a lower integral limit as distance $d_c$.  $d_c$ can be interpreted as an effective minimum spin-pair distance. The  $d_c^{-6}$  distance dependence of the spin--spin dipolar interaction is reduced to $d_c^{-3}$ due to the volume integral leading to Eq.\,(\ref{Gff_final}).  

Equation \,(\ref{Gff_final}) is associated with Torrey \cite{torrey1953nuclear} and Abragam \cite{Abragam} who described the relative motion of pairs of nuclear spins in a liquid.  
Equation\,(\ref{Gff_final}) has been widely used \cite{hwang1975dynamic,friedman1979epr,abernathy1997spin,Faux.2017b} not least as a correction term that accounts for the contribution to the dipolar relaxation rate associated with the relative translational diffusion of pairs of spins that cannot be accounted for by a numerical simulation \cite{Faux.1986,Faux.2015}.    Equation\,(\ref{Gff_final}) has to be computed numerically at each $t$, numerically Fourier transformed to find the spectral density function $J_{\rm o}^{\rm (c)}(\omega)$, then finally Eq.\,(\ref{T1sigma}) yields $R_1(\omega)$.  \\

\noindent
\subsection{The Hwang-Freed dipolar model} \label{outer-HF}

The Hwang-Freed model \cite{hwang1975dynamic} illustrated in Fig.\,\ref{Fig2inner_outer}(d) is a continuum diffusion with a hard-wall boundary condition added to prohibit spins from diffusing into a sphere of radius $d_{\rm HF}$ centered on the paramagnetic ion \cite{hwang1975dynamic}.  The result is an analytic expression for the spectral density function $J_{\rm o}^{\rm (HF)}(\omega)$ as \cite{hwang1975dynamic}
\begin{equation}
J_{\rm o}^{\rm (HF)}(\omega) =  \frac{12 \pi^2 N_M \tau_{\rm b}}{d_{\rm HF}^3}  \frac{2\sigma^2 + 5\sqrt{2}\sigma +8}{\sigma^6 + 4\sqrt{2}\sigma^5 + 16\sigma^4 + 27\sqrt{2}\sigma^3 + 81\sigma^2 + 81\sqrt{2}\sigma + 81} \label{eqn:Jw-HF}
\end{equation}
where $\sigma\!=\!\sqrt{\omega \tau_{\rm b}}$. Equation\,(\ref{eqn:Jw-HF}) may be substituted into Eq.\,(\ref{T1sigma}) to yield the relaxation rate which can be used to fit to experimental data with two fit parameters $d_{\rm HF}$ and translational diffusion time constant $\tau_{\rm b}$.  

The outer shell model and the continuum models for the outer sphere water constitute two extreme representations. The outer shell model places all the water, regardless of its distance from the ion, at a precise distance $d$ and therefore characterises the rotational dynamics but not the translational dynamics. By contrast, the continuum models captured by Eqs.\,(\ref{Gff_final}) and (\ref{eqn:Jw-HF}) are derived from a model that considers only the changes is the distance between pairs of spins and therefore captures  relative translational diffusion only with rotational motion playing no role. 

Combinations of the inner  and outer sphere models are now fit to experimental $R_1(\omega)$ results from aqueous Mn(II), Fe(III) and Cu(II) chloride where the use of dipolar models are justified.  We seek to establish the extent to which the dipolar models can quantitatively describe the dispersions in these systems.  Chloride salts are chosen for  consistency across ion type and because each chloride is soluble.  The models do not depend directly on the anion, but the anion is a constituent of ion complexes and therefore influences the rate of rotation of the inner sphere of spins through the time constant $\tau_{i\beta}$. The data collated by Helm and Merbach \cite{helm2005inorganic} suggest that the anion also has a modest impact on the water exchange lifetime $\tau_{\rm ex}$.

\section{Experiments} \label{SectionExperiments}

Fast-field-cycling NMR experiments were undertaken for samples of MnCl$_2$, FeCl$_3$ and CuCl$_2$ at different concentrations.  Initially, a 50$\pm$0.06\,mL of solution is prepared for each salt using deionised water at pH 7.
Both the salt and water are weighed separately. The solutions are diluted to obtain 10\,mL samples at the desired concentration before de-gassing to remove dissolved paramagnetic O$_2$ using the freeze-thaw method as follows.  A small volume of  approximately 0.1\,mL of solution is transferred to $\diameter$5$\times$220\,mm Wilmad economy borosilicate NMR 100\,MHz tubes. Each tube is connected to a vacuum pump with the
pressure limiter pre-set to 20\,kPa.  The sample is frozen in liquid N$_2$ for 18\,s and then evacuated before opening up the safety valve for 5-10\,s. After the system is closed and the ice turns to liquid, the process is repeated two more times. The glass tube is sealed with a butane fuelled torch, de-greased with ethanol, then placed and centered
in $\diameter$10\,mm tubes for the experiment.

A Stelar Spinmaster FFC-2000 fast field-cycling NMR relaxometer was used to acquire the frequency-dependent spin-lattice relaxation rate, $T_1^{-1} = R_1$, from solutions of MnCl$_2$ at 6.01, 2.00 and 0.50\,mM, for FeCl$_3$ at 2.5 and 1.5\,mM and CuCl$_2$ at 9.20, 3.57 and 1.50\,mM. Each sample was placed in the apparatus and left for approximately 10 minutes to equilibrate to the pre-calibrated air-flow temperature of 25\textdegree C. The equipment allows the user to measure $R_1(\omega)$ at low frequency ($< 0.328$ T, or 14\,MHz $^1$H Larmor frequency, $\nu_{\rm L} = \gamma B/2 \pi$) using pre-polarized (PP) pulse sequences at low frequency and non-polarized (NP) pulse sequences at high frequency ($>$\,14\,MHz). 

The PP sequence involves switching to a high polarization field, $B_p$, for a time $t_p\!>\!5 T_1(B_p)$, sufficient to maximise the initial $^1$H magnetization. In this case $B_p\!=\!0.587$\,T (equivalent to 25\,MHz) was used. The field is subsequently reduced to $B_r$ for a time $t_r$. The switching time $t_{\rm off} \! \ll \! T_1(B_r)$ of typically 3\,ms must be sufficiently short that no magnetization is lost during the switch and $B_r$  must subsequently be sufficiently stable that the spin system evolves during $t_r$ according to $T_1(B_r)$, i.e. it decreases to the equilibrium value at this field. This is followed by a switch to $B_a$, where $B_a \!>\! B_r$, and after the field stabilises a $\pi/2$ radiofrequency pulse, resonant at $\nu_a$ (16.3\,MHz in this case equal to $\gamma B_a/2 \pi$), is applied which results in a free-induction decay (FID) during $t_a$, the integral of which is a measure of the magnetization present at the end of $t_r$. Following a recycle delay, RD,  of several seconds at zero field, the process is repeated.  Typically eight scans are used to complete an excitation pulse phase cycle and improve the signal-to-noise ratio. To measure $R_1(B_r)$ the process is repeated for, in this case, 16 different values of $t_r$. The FID, conventionally, is exponentially decaying at a rate of $M(t_a) = M(B_r) \exp(-t_a/T_2^*)$ where $M(B_r)$  is the magnetisation recovered at $B_r$, $t_a$ is the post pulse time and  $1/T_2^*$, the effective $T_2$ relaxation rate.

During a NP sequence, any initial longitudinal $^1$H magnetization is `quenched' by preparing the spin system at zero field for several seconds (RD) and switching then to $B_r$ for duration $t_r$, during which time the magnetization increases to the equilibrium value at $B_r$. After switching to $B_a\!=\!0.383$\,T (equivalent to 16.3 MHz which in this case typically $<B_r$), an FID is acquired as before. Similarly to PP, the cycle is repeated for different $t_r$ values \cite{meledandri2012low}.
The FFC-NMR profile $R_1(\omega)$ was acquired at typically 20 logarithmically-spaced frequencies.

The acquisition is quite straightforward from an NMR perspective with a single $\pi/2$ read pulse used.  A single dummy pulse was applied and simple 4-phase cycle (with $\pi/2$ increments) is enabled. Hence 16 (typically) or 32 scans (repeats) are acquired to complete an integral number of phase cycles.

The individual FIDs are not determined by $T_2$, but rather by $T_2^*$ as the acquisition field is rather inhomogeneous. Hence the FID is not a single exponential. Critically, the magnet design is such that the field and field inhomogeneity, and hence the $T_2^*$ ``shape" of the FID, is recovered after each field cycle. Therefore the magnetisation is sampled by integrating the FID (excluding a dead time far shorter than $T_2^*$).  Sixteen recovery times are acquired ranging from circa 3\,ms to circa \,$5T_1$. 

This study is only presenting $T_1$ which is independent of magnet inhomogeneities.
The standard errors in the fitted $T_1$ values are low, typically $\approx$1.5\%, and so are smaller than the data mark size used in the figures. There is no evidence for any deviation from mono-exponentiality during the $^1$H magnetisation recovery to its equilibrium value at any measurement field (Larmor frequency); the residuals were found to be randomly distributed across the tau range sampled. This aspect was checked early in the study by evaluating with 32 tau values. 

The experimental $R_1(\omega)$ relaxation rate dispersion data for aqueous MnCl$_2$, FeCl$_3$ and CuCl$_2$ systems are processed as described below.

\section{Results} \label{SectionResults}

\subsection{Fitting preliminaries} \label{Section:fittingPrelims}


Fast-field cycling NMR measurements from aqueous solutions of Mn(II), Fe(III) and Cu(II) chlorides were completed at  $m$ different concentrations for each salt.  The experimental relaxation rates $ R_{1}(f_{ij}) $ at $j = 1 ... N_i$ frequencies $\omega_{ij} = 2 \pi f_{ij}$ at $i = 1 ... m$ molar concentrations (in units of mM) were converted to a 1\,mM master relaxation rate $R_{1}^{\rm (1)}(f_{ij})$, sometimes referred to as a relaxivity, using the equation
\begin{eqnarray}
R_{1}^{\rm (1)}(\omega_{ij}) &= &\frac{ R_{1}(\omega_{ij}) - R_{\rm 1,offset} }{c_i}  \label{eqn:XpConversion}
\end{eqnarray}
where $c_i$ is the concentration used for the scaling.  The $c_i$ are expected to match the concentration of the sample, but relaxation rates are not necessarily directly proportional to concentration at high paramagnetic ion concentrations presumably  due to interactions between hydrated ion complexes.

\begin{table}[ht]
	\caption{Sample ion concentrations and scaling parameters for each aquoion. The experimental ion concentrations are measured to $\pm 0.01$\,mM. Both the scaling concentration $c_i$ and offset rate $R_{\rm 1,offset}$ are used in Eq.\,(\ref{eqn:XpConversion}). The $^*$ indicates an estimated offset (see text).  }
	
	\begin{tabular}{rcccccccc}\toprule \\[-3mm]
		&  \multicolumn{3}{c}{Mn$^{2+}$} &  \multicolumn{2}{c}{Fe$^{3+}$} &  \multicolumn{3}{c}{Cu$^{2+}$}\\ \hline \\[-3mm]
		Experimental ion concentrations (mM) & ~~~6.01~ & ~2.0~ & ~0.5~ & ~~~2.5~ & ~1.5~~~ & ~~9.2~ & ~3.57~ & ~1.5~~ \\
		Scaling concentration $c_i$ (mM) & ~~~5.56~ & ~2.0~ & ~0.5~ & ~~~2.9~ & ~1.5~~~ & ~~9.2~ & ~3.57~ & ~1.5~~ \\
		Offset relaxation rate $R_{\rm 1,offset}$ ($s^{-1}$) &   \multicolumn{3}{c}{0.70} &  \multicolumn{2}{c}{0.4$^*$} &  \multicolumn{3}{c}{0.43}
		\\ \hline \hline
	\end{tabular} 
	\label{Table:samples}
\end{table}

Each dispersion curve includes an offset relaxation rate $R_{\rm 1,offset}$ to account for frequency-independent contributions {\it not} associated with dipolar $S\!-\!I$ interactions, such as proton-proton interactions, residual contributions due to impurities, un-removed dissolved dioxygen, and/or frequency-independent electron spin relaxation. 
$R_{\rm 1,offset}$ is found by minimising a quality-of-fit parameter
\begin{eqnarray}
Q & =  & \sum_{j=1}^{N_i} \: \sum_{i=1}^{m} \left[ R_{1}^{(1)}(\omega_{ij}) - \langle R_{1}(\omega_{ij}) \rangle_i \right]^2 \label{eqn:minimiseThis}
\end{eqnarray}
where the angular brackets indicate an average over concentrations $i=1...m$.

The values of $R_{\rm 1,offset}$ and $c_i$ used for each ion are listed in Table \ref{Table:samples}.  In the case of aqueous Cu(II) chloride, measurements at $m\!=\!3$ different concentrations are scaled using the actual ion concentration.  The $Q$ is minimised with a physically-reasonable offset of 0.43\,s$^{-1}$.  For Mn(II), it was necessary to minimise $Q$ by changing both the highest scaling concentration and $R_{\rm 1,offset}$ leading to a scaling concentration of 5.56\,mM compared to the experimental concentration of 6.01\,mM and an offset of 0.70\,s$^{-1}$.  Only two concentrations were available for Fe(III) and so a typical estimated offset of 0.4\,s$^{-1}$ was set.  $Q$ was then minimised with respect to the highest scaling concentration to yield 2.9\,mM compared to the experimental concentration of 2.5\,mM. 

Models used to fit to dispersion data for each aqueous ion system  include expressions for both inner and outer sphere water.   The equations for all models are presented in Sec.\,\ref{Sec:NMRmodels}.  The values of general physical quantities are presented in Table\,\ref{Table:general}. 

\begin{table}[ht]
	\caption{Numerical values of physical quantities used in calculations.}
	
	\begin{tabular}{rcccc}\toprule \\[-3mm]
		& &  ~~~Mn$^{2+}$~~~ & ~~~Fe$^{3+}$~~~ & ~~~Cu$^{2+}$~~~ \\ \hline \\[-3mm]
		Electronic spin quantum number  & $S$ & $ \frac{5}{2}$   & $ \frac{5}{2}$ & $ \frac{1}{2}$ \\[0.5mm]
		Number of inner sphere proton spins & ~~~~~$n$~~~~~~ & 12 & 12 & 5 \\[0.5mm]
		Constant in Eq.\,(\ref{T1sigma}) & ~~~~~$\beta_{IS}$~~~~~~  &\multicolumn{3}{c}{2.47$\times$10$^{11} \,S(S+1)$ m$^6$s$^{-2}$}  \\[0.5mm]
		Paramagnetic ion density (1\,mM) & ~~~~~$N_M$~~~~~~ & \multicolumn{3}{c}{6.02$\times$10$^{-4}$\,ions/nm$^{3}$} \\[0.5mm]
		Bulk water spin density & ~~~~~$N_H$~~~~~~ & \multicolumn{3}{c}{66.6\,spins/nm$^{3}$} 
		\\ \hline \hline
	\end{tabular} 
	\label{Table:general}
\end{table}

The quality of fit was assessed by the coefficient $Q$ given by
\begin{eqnarray}
Q & = & \sum_{j=1}^{N_i} \: \sum_{i=1}^{m} \left[ R_{1}^{(1)}(\omega_{ij}) - R_{1}^{\rm (model)}(\omega_{ij}) \right]^2  \label{eqn:quality}
\end{eqnarray}
where $R_{1}^{\rm (model)}(\omega_{ij})$ is the prediction of the chosen model at the $j^{\rm th}$ frequency $(\omega_{ij})$ for all $i = 1 ... m$ normalised datasets for each ion.

All combinations of inner and outer sphere models were explored during the fitting process.  The Brownian shell model only is applied to the inner sphere and justified by the sharp peaks in the experimental and MD radial density functions for the first coordination shell.  The characteristic dynamical time constant is either the rotational time constant $\tau_\beta$ or the exchange time constant $\tau_{\rm ex}$ (see Table\,\ref{Table:timeConstants}).  

The dynamics of the outer sphere is more complex as the ion-$^1$H spin-pair vector changes in both length and angle.  The Hwang-Freed model is used to describe the relative translational motion (change in vector length but not angle) and the Brownian shell model is used to describe the rotational motion (change in vector angle but not length).  The final choice of Hwang-Freed or Brownian shell model for the outer sphere water, or whether a combination of both is necessary, is decided solely by the quality-of-fit parameter $Q$ for each paramagnetic ion system.

The model combinations with the best fits are discussed in detail in the following sections and are presented in summary in  Table\,\ref{Table:modelCombinations}.  

\begin{table}[ht]
	\caption{Model combinations presented and discussed in Secs. \ref{SectionResultsMn} to \ref{SectionResultsCu}. }
	\begin{tabular}{ccccccc}\toprule 
	&	&  \multicolumn{2}{c}{~inner sphere~~~~~~~~} &~~~~~& \multicolumn{2}{c}{~~~~~~~~outer sphere~~~~~~~~}  \\[-0mm] \cline{3-7} \\[-4mm]
&		&  \multicolumn{2}{c}{shell~~~~~~~~} &~~~~~& shell & Hwang-Freed \\[-0mm] \cline{3-7} \\[-4mm]
& ~Label~		&  $a, \tau_{\rm ex}$ & $a, \tau_{\rm i\beta}$ &~~~~~& $d, \tau_{\rm o\beta}$ & $d_{\rm HF}, \tau_{\rm b}$ \\	\hline \\[-4mm]
		\multirow{3}{*}{Mn$^{2+}$}~~~& ~~Fit-Mn1~~ & \ding{51}  &~~~~~& & \ding{51}  &  \\[-0mm]
&	Fit-Mn2	&  & \ding{51}  &~~~~~& \ding{51} &  \\[-0mm]
& Fit-Mn3		&\ding{51}  &~~~~~& & & \ding{51} \\ \hline \\[-4mm]
		\multirow{2}{*}{Fe$^{3+}$}~~~	& FitFe1 &  &  &~~~~~& \ding{51}  & \ding{51}  \\[-0mm]
	& FitFe2	&  &\ding{51} &~~~~~& \ding{51} &  \\ \hline \\[-4mm]
		Cu$^{2+}$~~~ & 	&  \ding{51}  &~~~~~& & \ding{51}  &  \\[-5mm]	
		\\ \hline \hline
	\end{tabular} 
	\label{Table:modelCombinations}
\end{table}


The water molecules in the inner sphere of an aquoion may exchange with a water molecule in the outer sphere and move into the bulk.   The water exchange process is described by
\begin{center}
	[M(H$_2$O)$_6$]$^{2+}$ + H$_2$O$^*$ $\rightleftharpoons$  [M(H$_2$O)$_5$\,H$_2$O$^*$]$^{2+}$ + H$_2$O
\end{center}
for a metal ion M$^{2+}$. The exchange time constant $\tau_{\rm ex}$  was introduced through Eq.\,(\ref{eqn:SBMtimes}) where it can be seen that the inner sphere contribution is dominated by the shorter of the two time constants $\tau_{\rm ex}$ and $\tau_{\rm i \beta}$.  Helm and Merbach \cite{helm2005inorganic} provide an excellent review of exchange mechanisms between the inner and outer sphere and gather data on exchange rates $k$, measured primarily by $^{17}$O NMR, for a wide range of ions including the hexahydrates of Mn(II), Fe(III) and Cu(II) of interest here. The reciprocal of the exchange rate is described as the mean lifetime of a water molecule in the inner sphere \cite{helm2005inorganic} so that $\tau_{\rm ave}\! = \!k^{-1}$.  Note that $\tau_{\rm ex} = \ln(2)\sqrt{2}\,\tau_{\rm ave} = 0.98 \tau_{\rm ave}$ so that $\tau_{\rm ex} \approx \tau_{\rm ave}$.  Throughout, the equivalence of $\tau_{\rm ave}$ and $\tau_{\rm ex}$ is assumed.  

The experimental data for $\tau_{\rm ex}$ listed in Table \ref{Table:rates} guides the fitting process.  For [Fe(H$_2$O)$_6$]$^{3+}$ for instance, it is clear that $\tau_{\rm ex} \gg \tau_{\rm i \beta}$ and so $\tau_{\rm i \beta}$ dominates the dispersion.  For [Cu(H$_2$O)$_6$]$^{2+}$, $\tau_{\rm ex}$ is very short and dominates.  For [Mn(H$_2$O)$_6$]$^{2+}$ it is unclear which time constant dominates as discussed in Sec.\,\ref{SectionResultsMn}.  This information is duly reflected in the best model fits for $\tau_{\rm ex}$ in Table \ref{Table:modelCombinations} with fit outcomes for $\tau_{\rm ex}$ in Table \ref{Table:rates}. 

\begin{table}[ht]
	\caption{The lifetime of water molecules in the inner sphere of selected aquoions are presented \cite{helm2005inorganic}.}
	
	\begin{tabular}{ccccl}\toprule \\[-4mm]
		~~cation complex~~	&  method & ~~~~anion~~~~ &  ~~~~$\tau_{\rm ex} $ (s)~~~~ & source \\ \hline \\[-3mm]
		[Mn(H$_2$O)$_6$]$^{2+}$& $^{17}$O NMR &  (ClO$_4$)$^-$&  4.8$\times$10$^{-8}$  &  Ducommun {\it et al} \cite{ducommun1980high} \\[0mm]
		&  FFC NMR &  Cl$^-$&  1.5$\times$10$^{-6}$ & this work \\[0.5mm]
		[Fe(H$_2$O)$_6$]$^{3+}$ & $^{17}$O NMR & (OH)$^-$ &  6.3$\times$10$^{-3}$  & Swaddle {\it et al} \cite{swaddle1981high}; Grant {\it et al} \cite{grant1981kinetics} \\[0mm]
		& FFC NMR  & Cl$^-$ &  $>$ 1$\times$10$^{-4}$ & this work  \\[0.5mm]
		[Cu(H$_2$O)$_6$]$^{2+}$ & $^{17}$O NMR & (ClO$_4$)$^-$ &  2.3$\times$10$^{-10}$  & Powell {\it et al} \cite{powell199117o} \\[0mm]
		&  FFC NMR&  Cl$^-$&  2.4$\times$10$^{-10}$ & this work \\
		\hline \hline
	\end{tabular} 
	\label{Table:rates}
\end{table}

The NMRD curves for Mn(II), Fe(III) and Cu(II) are now analysed individually.

\subsection{Manganese} 
\label{SectionResultsMn} 

\begin{figure}[tbh!]
	\unitlength1cm
	\begin{center}
		\includegraphics[width=0.800\textwidth, trim={1cm 7cm 1cm 7cm}]{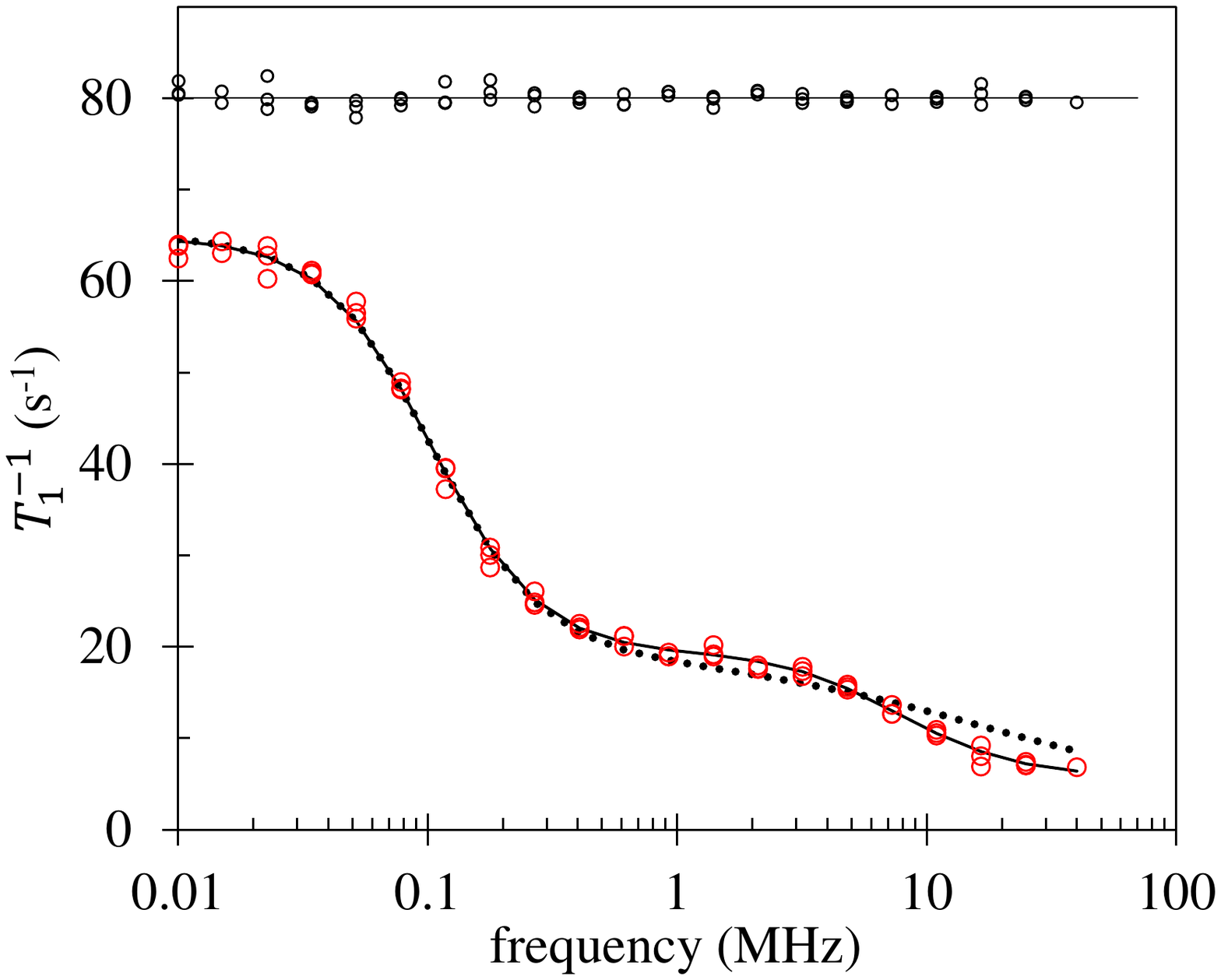}
		\caption{Experimental FFC NMR data for aqueous manganese chloride solution
			({\color{red}{\large $ \circ$}}) scaled to 1\,mM (see text) are presented.  
			The best fit Fit-Mn1 is shown by the solid line with residuals ({$\circ$}) drawn to scale and offset at 80\,s$^{-1}$. Fit-Mn3 is presented as a dotted line. }
		\label{Fig3_Mn}
	\end{center}
\end{figure}

The first frequency-dependent $T_1^{-1}(\omega)$ data for a dilute aqueous manganese ion system were presented by Morgan and Nolle in 1959 \cite{morgan1959proton}.  More recently, literature field-cycling measurements for Mn(II) for a range of anions are summarised and interpreted in Chapter 7 of the book by Bertini {\it et al} \cite{bertini2016nmr}.  
In the present study, FFC NMR relaxation rate measurements on aqueous MnCl$_2$ solutions at concentrations 0.5, 2.0, and 6.01\,mM are  transformed to a 1\,mM master curve as described in Sec.\,\ref{Section:fittingPrelims} and presented in Fig.\,\ref{Fig3_Mn}.

Mn(II) presents two inflection features at about 0.1 and 10\,MHz which are associated with water in the inner and outer spheres respectively \cite{bertini2016nmr}.  As discussed in the introduction, Hernandez and Bryant \cite{hernandez1991proton} provide compelling evidence that the 0.1\,MHz feature is associated with a dipolar relaxation process (rather than the electronic spin scalar contribution) and justifies trialling the Brownian shell model in the analysis that follows.

Two versions of the shell model describe the inner sphere's contribution to the relaxation rate. One version assumes $\tau_{{\rm i}\beta} \gg \tau_{\rm ex}$ so that the exchange time constant $\tau_{\rm ex}$ dominates the relaxation (Eq.\,\ref{eqn:Jw-inner-d}).  The inner sphere water exchanges with outer sphere water faster than the rate of rotation of the ion complex.   This version of the inner sphere shell model uses the fit parameter set $(a,  \tau_{\rm ex})$ and is referred to as shell($\tau_{\rm ex}$).  The second version of the inner sphere shell model takes $\tau_{{\rm i}\beta} \ll \tau_{\rm ex}$ (Eq.\,\ref{eqn:Jw-inner-beta}) and uses the fit parameter set $(a, \tau_{{\rm i}\beta})$.  Here, rotation of the ion complex is faster than the exchange lifetime and is abbreviated to shell($\tau_{{\rm i}\beta}$).  

Each inner sphere shell model was explored in tandem with one of the outer sphere models illustrated in Fig.\,\ref{Fig2inner_outer}(b)--(d).  The continuous diffusion model illustrated in Fig.\,\ref{Fig2inner_outer}(c), requires the numerical calculation of the dipolar correlation function $G(t)$ given by Eq.\,(\ref{Gff_final}) followed by a second numerical integration to execute the Fourier transform to finally obtain the spectral density function $J(\omega)$.  The continuous diffusion model is not discussed further because the Hwang-Freed model (Eq.\,\ref{eqn:Jw-HF}) in Fig.\,\ref{Fig2inner_outer}(d) not only provides an improved physical description of the bulk diffusion of fluid in the vicinity of an ion and the convenience of an analytic expression for $J(\omega)$, but in trial fits were also found to provide a superior fit to the experimental data. The Hwang-Freed outer sphere model uses the parameter set $(d_{\rm HF},  \tau_{\rm b})$, where $\tau_{\rm b}$ is the water translational diffusion time constant.  The second outer sphere model is the shell($\tau_{{\rm o}\beta}$) model with parameter set $(d, \tau_{{\rm o}\beta})$ (Eq.\,\ref{eqn:Jw-outer-beta}).

\begin{table}[ht]
	\caption{The Table presents the best fit parameters for the relaxation rate dispersion from aqueous  manganese chloride solution for three model combinations.}
	
	\begin{tabular}{rclll}\toprule \\[-4mm]
		& & {\bf Fit-Mn1} & {\bf Fit-Mn2} & {\bf Fit-Mn3} \\ \hline  \\[-4mm]
		Inner sphere model: & &shell($\tau_{\rm ex}$)~~~~~&  shell($\tau_{{\rm i}\beta}$)~~~~~& shell($\tau_{\rm ex}$) \\ \hline \\[-4mm]
		radius of inner shell &  $a$ & $0.272$\,nm  &  $0.263$\,nm  &  $0.270$\,nm   \\[-1mm]
		exchange time constant & ~~~$\tau_{\rm ex}$~~~~  &  1.5\,$\mu$s &  & 1.5\,$\mu$s  \\[-1mm]
		rotational time constant (inner)  & $\tau_{{\rm i}\beta}$  & & 1.8\,$\mu$s    &  \\[1mm] \hline \hline \\[-4mm]
		Outer sphere model: & & shell~~~~~~&  shell~~~~~~& Hwang-Freed  \\[0mm] \hline  \\[-4mm]
		radius of outer shell & $d$ &  $0.446$\,nm &  $0.435$\,nm & \\[-1mm]
		rotational time constant (outer) & $\tau_{{\rm o}\beta}$ & 37\,ps  & 32\,ps &      \\[-1mm]
		Hwang-Freed distance & $d_{\rm HF}$ &   & &  $0.408$\,nm  \\[-1mm]
		bulk diffusion time constant & $\tau_{\rm b}$  &  & & 27\,ps      \\ \hline \hline  \\[-4mm]
		inner sphere volume fraction &  $x$  & 3.67$\times 10^{-5}$ & 3.38$\times 10^{-5}$  &  3.47$\times 10^{-5}$   \\[-1mm]
		
		quality-of-fit parameter &   $Q$  &  39.5 &  52.5  & 134
		\\ \hline \hline
	\end{tabular} 
	\label{Table:Mn}
\end{table}

The optimum fit parameters for three of the four model combinations are presented as Fit-Mn1, Fit-Mn2 and Fit-Mn3 in Table \ref{Table:Mn}. The dispersion curves for Fit-Mn1 and Fit-Mn3 are displayed in Fig.\,\ref{Fig3_Mn} (the curves for Fit-Mn1 and Fit-Mn2 are very similar). The dispersion curves coincide at frequencies less than about 0.2\,MHz because both Fit-Mn1 and Fit-Mn3 use the same inner sphere model  which characterises the low-frequency region of the dispersion curve. By contrast, at frequencies above 1\,MHz, both shell($\tau_{{\rm o}\beta}$) model fits outperform the Hwang-Freed model.  This observation is consistent with Pfeifer \cite{pfeifer1962protonenrelaxation} who noted that translational diffusion processes contribute only 10-20\% to the $R_1$ relaxation rate for aqueous Mn(II) systems. 

\begin{figure}[tbh!]
	\unitlength1cm
	\begin{center}
		\includegraphics[width=0.70\textwidth, trim={2cm 1cm 3cm 1cm}]{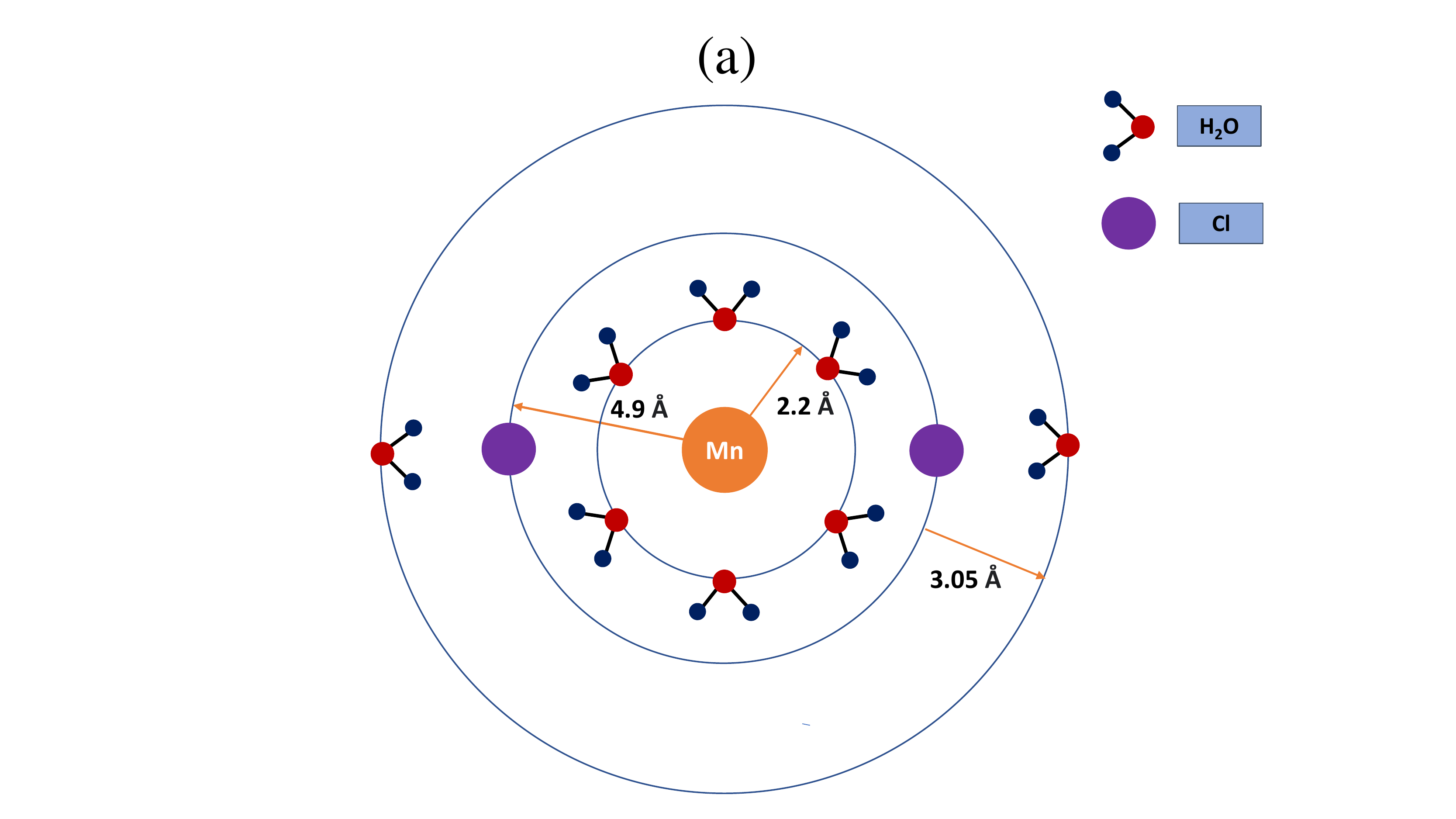}
		\includegraphics[width=0.70\textwidth, trim={2cm 3cm 2cm 0cm}]{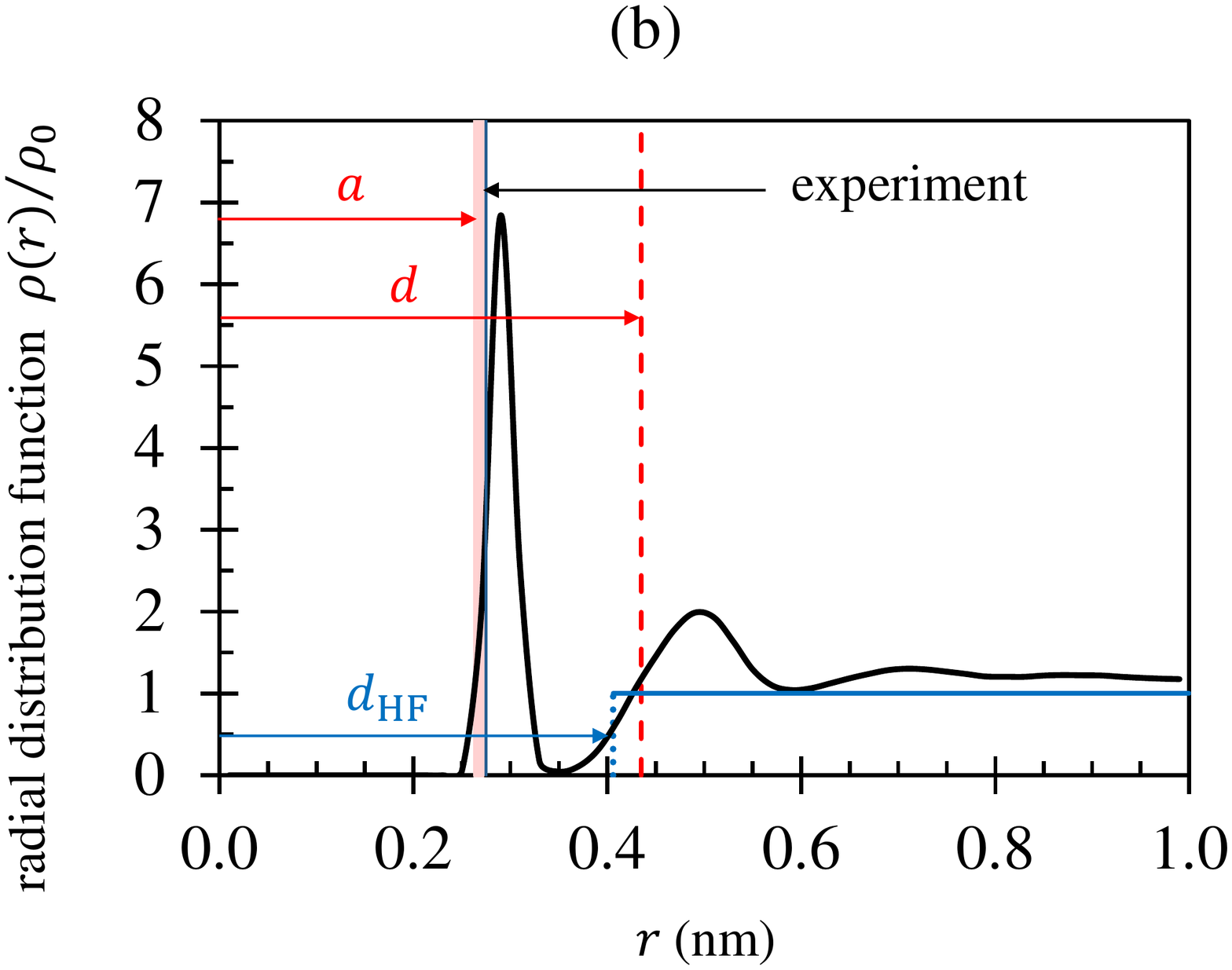}
		\caption{(a) a two-dimensional schematic representation of an aqueous Mn(II) chloride complex with distances obtained from radial distribution functions (RDFs) from MD simulation. (b) shows the Mn--H RDF.  The range of inner shell model radius $a$, outer shell model radius $d$, and the Hwang-Freed distance of nearest approach $d_{\rm HF}$ are shown as obtained from fits to the experimental relaxation rate dispersion curve. The experimental inner sphere is from Ref.\,\cite{smirnov2019coordination} }
		\label{Fig4_rdf}
	\end{center}
\end{figure}

The numerical values of the fit parameters presented in Table \ref{Table:Mn} are compared to  MD simulations of aqueous Mn(II) chloride and to prior work.  Figure\,\ref{Fig4_rdf}(a) presents a two-dimensional schematic representation of a Mn(II) chloride complex  with distances determined from radial density functions (RDFs)  obtained  from  MD simulations.  Fig.\,\ref{Fig4_rdf}(b) presents the  Mn--H RDF only.  Simulation details are provided in Supplementary Material Note 2.2 \cite{SM22}.  The Mn--O distance of 2.2\,\AA\ in Fig.\,\ref{Fig4_rdf}(a) is in accord with numerous X-ray measurements \cite{ohtaki1976x,magini1981hydration,caminiti1982comparative,smirnov2019coordination} as expected from MD simulations with optimised interatomic potentials.  The accuracy of the Mn--H RDF in Fig.\,\ref{Fig4_rdf}(b) is therefore assured.

The shell model and Hwang-Freed model provide alternative extreme representations of the dynamics of the outer sphere $^1$H spins.  The shell model assumes all spins are located at a distance $d$ from the paramagnetic ion and yields a rotational time constant, whereas the Hwang-Freed model assumes a uniform distribution of spins beyond a distance $d_{\rm HF}$ and provides a translational diffusion time constant.  The Hwang-Freed model yields a best fit bulk diffusion time constant  $\tau_{\rm b}\!=\!27$\,ps representing a diffusion coefficient five times smaller than pure water at room temperature \cite{krynicki1978pressure}.  The time constant reflects the slower motion of water in the second coordination shell and the $r^{-4}$ dependence of the relaxation rate which enhances the contribution of the water closest to the ion.  The best-fit Hwang-Freed distance $d_{\rm HF}$ is approximately the distance of nearest approach of the outer sphere water.  The quality-of-fit parameter for Fit-Mn3, however, indicates a significantly inferior fit compared to Fit-Mn1 and Fit-Mn2 so that the shell model provides the best description of the outer sphere water for manganese.  

Fits Fit-Mn1 and Fit-Mn2 yield the rotational time constant for the outer shell, $\tau_{{\rm o}\beta}$, of 37\,ps and 32\,ps respectively.  The time $\tau_{{\rm o}\beta}/2$ represents the time for a rotation through 1 radian \cite{faux2021a}.  A guide value for $\tau_{{\rm o}\beta}$ is found from the MD simulations of Mn(II) chloride yielding 32\,ps as described in the Supplementary Material \cite{SM22}.  These outcomes are in good agreement with the experimentl value of 38\,ps found by Pfeifer \cite{pfeifer1962protonenrelaxation} at room temperature for MnSO$_4$ and approximately 25\,ps fouand by Koenig and Brown \cite{koenig1984relaxation} at 35$\degree$C (anion unspecified). Outer sphere rotational time constants obtained from different aqueous ion systems presented in the literature normally lie in the range 20--40\,ps.  The independence of $\tau_{{\rm o}\beta}$ on ion type arises because the relaxation rate contribution from the outer sphere is most sensitive to the dynamics of the bulk water in the region immediately outside the inner sphere. The water dynamics in this region is similar for most ions.  The $r^{-4}$ distance-dependence of the relaxation rate leads to $d$  falling on the side of the second hydration peak closest to the origin for both fits Fit-Mn1 and Fit-Mn2.

We now describe the outcomes of modelling the {\em inner sphere} relaxation using the shell model introduced in Sec.\,\ref{Sec:NMRmodels}.  The MD simulations confirm that the Mn(II) aquoion forms a hexahydrate.  The ion complex illustrated in Fig.\,\ref{Fig4_rdf}(a) also shows two chloride ions with distances in accordance with RDF results.  

The values of the inner shell radius $a$ from fits Fit-Mn1 and Fit-Mn2 fall into a narrow band 0.263--0.272\,nm situated on the near side of the peak of the simulation Mn--H radial distribution function as illustrated in Fig.\,\ref{Fig4_rdf}(b).  Smirnov and Grechin \cite{smirnov2019coordination} used X-ray diffraction to determine the radial distribution functions for aqueous Mn(II) chloride solutions for a range of concentrations. These authors saw a sharpening of the radial distribution function as the salt solution became increasingly dilute presumably due to a reduction in ion-ion interactions. The lowest concentration studied is still substantially higher than 1\,mM for the FFC NMR measurements presented here. Nonetheless,  Table 1  of Smirnov and Grechin \cite{smirnov2019coordination} shows that the Mn--O distance is nearly independent of concentration at about 0.216\,nm.  Taking this value and assuming that the oxygen atom of the water molecule is pointing towards the manganese ion and the H atoms away from the ion, and using the O--H bond length of 0.096\,nm and a H-O-H angle of 104.5\degree, the H atoms are placed 0.275\,nm from the center of the manganese ion.  This value, also shown in Fig.\,\ref{Fig4_rdf}(b), is in excellent agreement with the fit outcomes in Table \ref{Table:Mn}. Moreover, the relaxation rate is proportional to $r^{-4}$ so that the model best fit value of $a$ should lie slightly to the left of the radial density function peak as observed. This result provides initial confidence in the shell model presented in Sec.\,\ref{Sec:NMRmodels} as a quantitative description of the inner sphere water dynamics.  

\begin{figure}[tbh!]
	\unitlength1cm
	\begin{center}
		\includegraphics[width=0.70\textwidth, trim={2cm 3cm 2cm 1cm}]{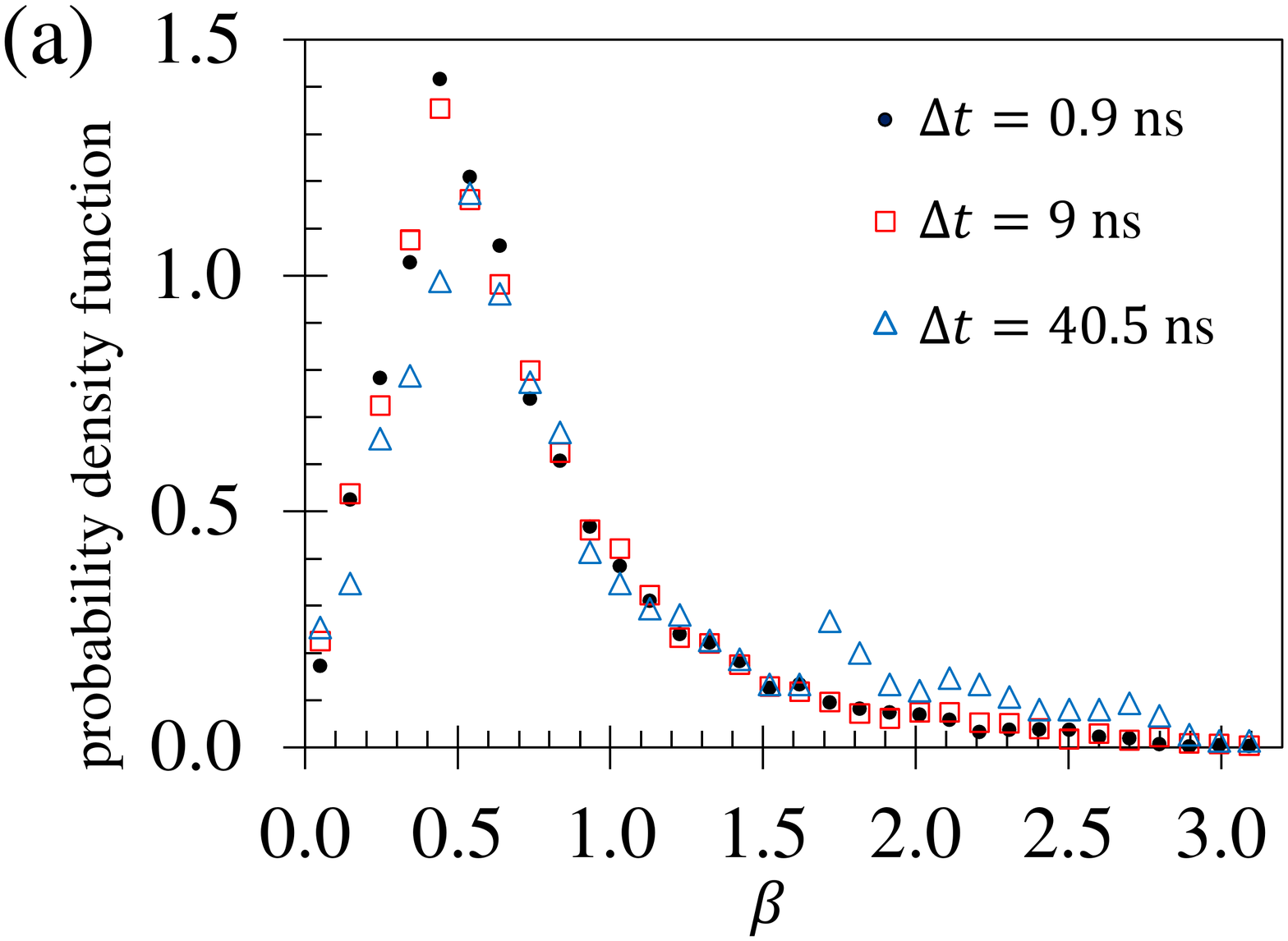}
		\includegraphics[width=0.70\textwidth, trim={2.6cm 3cm 1.6cm 1cm}]{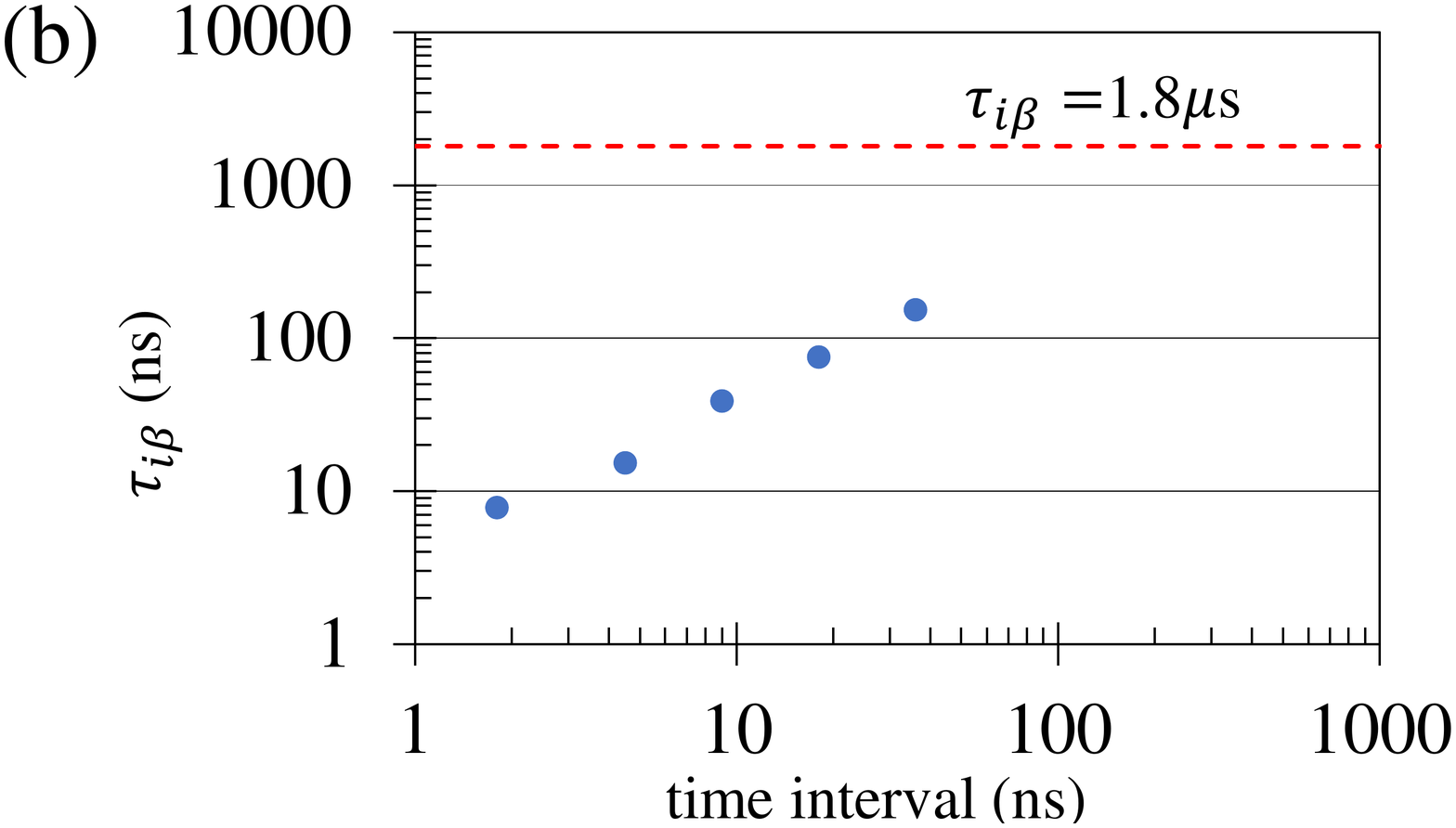}
		\caption{Results of MD simulations of aqueous Mn(II) chloride.  In (a), the probability density function $P(\beta, \Delta t) \sin \beta$ is presented as a function of angle of rotation $\beta$ for the inner sphere Mn--H vector at three different time intervals.  In (b), the inner sphere rotational time constant $\tau_{i\beta}$ is obtained by fitting Eq.\,(S9) in the Supplementary Material \cite{SM1} to the distributions in (a).}
		\label{Fig5_sphereRot}
	\end{center}
\end{figure}

The fit rotational time constant $\tau_{{\rm i}\beta} \approx 1.8\,\mu$s   is associated with the inner sphere $^1$H spins in the extended ion complex illustrated in Fig.\,\ref{Fig4_rdf}(a).  This is initially surprising because $1.8\,\mu$s is more than four orders of magnitude longer than typical outer sphere rotational time constants.  However, FFC NMR measurements from water-bearing porous solids also see an inflection feature in the same frequency range as Fig.\,\ref{Fig3_Mn}  due to the slow {\em translational} dynamics of $^1$H spins water at the solid surfaces with dynamical time constants in the range 0.1--10\,$\mu$s \cite{Korb.2011,faux2019advances}.  By contrast, the dynamics of the water in the second hydration layer above the solid surface is characterised by time constants typically in the range 10--20\,ps \cite{Faux.2015,faux2019advances}.  A rotational time constant of 1.8\,$\mu$s for $\tau_{{\rm i}\beta}$ for the Mn(II) aquoion should not, therefore, be surprising because the inner sphere water is a single layer of bound water surrounding and ion and adjacent to bulk-like water.

Nonetheless, we explore two alternative approaches to estimating $\tau_{{\rm i}\beta}$ for the aqueous Mn(II) chloride ion complex. The first estimate is obtained using molecular dynamics (MD) simulations with simulation details presented in Supplementary Material Note 2.2 \cite{SM22}. The results are presented in Fig.\,\ref{Fig5_sphereRot}.
Figure \ref{Fig5_sphereRot}(a) shows the probability density function describing the distribution of angular change of the Mn--H vector for protons in the inner sphere. The mean angular change over a time interval $\Delta t$ is presented.  Figure \ref{Fig5_sphereRot}(a) illustrates the complexity of the inner sphere dynamics. There is an initial rapid change in angle due to rotation and jostling of inner sphere protons apparent as a peak at about 0.5\,rad which is virtually unchanged for time intervals from 1\,ns to 40\,ns. The ion complex is not rotating on the timescale of the simulation as demonstrated in Fig.\,\ref{Fig5_sphereRot}(b) which presents $\tau_{{\rm i}\beta}$ as a function of time interval.  The $\tau_{{\rm i}\beta}$  are obtained from fits to the probability density function using Eq.\,(S9) of the Supplementary Material Note 1 \cite{SM1}. Figure\,\ref{Fig5_sphereRot}(b)  shows an almost linear relationship of $\tau_{{\rm i}\beta}$ as a function of time interval. If longer MD simulations could have been performed (so that  $\Delta t > \tau_{{\rm i}\beta}$),   $\tau_{{\rm i}\beta}$ would level off and become independent of $\Delta t$. 

The length of the MD simulations are  too short to estimate  $\tau_{{\rm i}\beta}$ for aqueous Mn(II) chloride.  A large simulation box size is necessary to prevent ion complexes interacting and statistical accuracy diminishes for increasing $\Delta t$.  The trade-off is that simulations are limited to timescales of about 50\,ns. Simulations of microsecond duration should see the probability density function in Fig.\,\ref{Fig5_sphereRot}(a) evolve to an approximate gaussian centered on $\pi/2$\,rad and yield a $\tau_{{\rm i}\beta}$ value that captures the temporal evolution of the system. Figure\,\ref{Fig5_sphereRot}(b) clearly shows that $\tau_{{\rm i}\beta}$ is at least 0.1\,$\mu$s and likely of order microseconds. We are not aware of experimental estimates of $\tau_{{\rm i}\beta}$ for aqueous Mn(II) chloride.  

Finally we note that the increased probability density at angles in excess of 1\,rad in Fig.\,\ref{Fig5_sphereRot}(a) is probably due to switching of positions between pairs of inner sphere water molecules providing a hint as to the primary mechanism for inner sphere rotation in ion complexes.  The slow rotation of the ion complex arises because inner sphere water is bound to the Mn(II) ion, as are the anions and water bound to the anions. The complex forms a shape rather like a paddle.  Not only must repeated collisions by fast moving outer sphere water molecules cause the ion complex to rotate but the loosely-bound associated water contained within the sphere must rotate with it.  Evidence confirming the large size of aqueous Mn(II) complexes and the presence of connected water was provided by Caminiti {\it et al} \cite{caminiti1982comparative} for Mn(II) sulfate.  X-ray diffraction studies identified radial structure functions extending to 15\,\AA\ with the [Mn(H$_2$O)$_4$(SO$_4$)$_2$]$^{-2}$ complex interacting with about 10 water molecules and each sulfate with about seven water molecules.  

The second estimate of $\tau_{{\rm i}\beta}$ is obtained by considering a sphere rotating due to bombardment of its surface by water molecules.  
The rotational model is presented in Supplementary Material Note 3 \cite{SM3}.  The ion complex in Fig.\,\ref{Fig4_rdf}(a) plus associated water forming the sphere will, on average, experience a net zero collision force.  Random rotations of the sphere arise due to fluctuations from the average collision force  in the same manner as conventional single particle self-diffusion in a liquid.  We show that this model yields a rotational time constant that scales to the fifth power of the radius of the sphere $R$ \cite{SM3}.  A direct estimate of $\tau_{{\rm i}\beta}$ is not possible, but taking the ratio with a ``standard'' rotor with known $\tau_{{\rm i}\beta}$ permits a crude estimate of the time constant for the ion complex.  Since $\tau_{{\rm i}\beta}$ for water is known, we have the expression \cite{SM3}
\begin{equation}
\frac{\tau_{{\rm i}\beta,c}}{\tau_{\beta,w}} = \frac{R_c^5}{R_w^5} \label{eqn:beta-ave-ratio}
\end{equation}
where the subscripts ``c" and ``w" refer to the complex and water respectively.   

Water is taken as the standard rotor with a sphere comprising a single water molecule. The radius of a single water molecule is the distance from the water center-of-mass to its H atom taken as 0.0938\,nm.  The
 rotational time constant for the intramolecular H--H vector has been shown to be 8\,ps \cite{faux2021a}. The rotational time constant for the O--H vector is expected to be similar and, indeed, from MD simulations described in the Supplementary Material Note 3 \cite{SM3}, is found to be 9\,ps. Taking the radius of the Mn(II) chloride complex from Fig.\,\ref{Fig4_rdf}(a) as 0.79\,nm leads to an estimate of  $\tau_{{\rm i}\beta,c}$  as 0.4\,$\mu$s. 
 
 There are dangers to this approach.  First, scaling arguments should be used with caution when scaling powers are high and, second, water is an atypical rotor. Water molecules are subject to very rapid hydrogen bond breaking and reforming leading to an angular change of about 1 radian every 4\,ps.  The rapid hydrogen bond switching in water produces a value $\tau_{\beta,w}$ much shorter than would otherwise be the case.  On that basis, and accepting the scaling argument, 
 the value of 0.4\,$\mu$s for $\tau_{{\rm i}\beta,c}$ can reasonably be considered a minimum value. 
 
 In summary, experience from porous material systems, the compelling evidence provided by the MD simulations, and the sphere rotation model all justify a microsecond rotational time constant for inner sphere water in ion hexahydrates.  The caveat is that the 0.1\,MHz inflection feature can be assigned to inner sphere rotation only if the inner-to-outer sphere exchange time constant is similar to, or longer than $\tau_{{\rm i}\beta}$.
 
We now consider the exchange time constant $\tau_{\rm ex}$. The best-fit value of $\tau_{\rm ex}$ is $1.5\,\mu$s obtained from both Fit-Mn1 and Fit-Mn3.  Again, there is no experimental result that provides a direct comparison.  Helm and Merbach list experimental results for $\tau_{\rm ex}$ in the range 8--450\,ns for different anions. The most relevant result is obtained from $^{17}$O NMR measurements of Ducommun {\em et al} \cite{ducommun1980high} who studied manganese hexahydrate perchlorate, Mn(ClO$_4$)$_2$.6H$_2$O, and found  $\tau_{\rm ex}$ to be 30 times shorter than the value presented here albeit with a high uncertainty \cite{ducommun1980high}.  If the best-fit value of 1.5\,$\mu$s found here is indeed a measure of $\tau_{\rm ex}$ for manganese chloride hexahydrate, {\em both} $\tau_{\rm ex}$ and $\tau_{{\rm i}\beta}$ would contribute to the inner sphere relaxation rate. 

Finally, the inner sphere volume fraction $x$ in Table \ref{Table:Mn} is approximately 0.35$\times$10$^{-5}$.  In principle, $x$ can be converted to an estimate of the number of water molecules in the inner sphere, but an estimate relies on assumptions about other parameters.  The issues and complexities are  discussed in detail in Supplementary Material \cite{SM4}.  

In summary, the inner shell model presented in Sec.\,\ref{inner-Brownian} yields an exquisite fit to the inner sphere dispersion feature of aqueous Mn(II) chloride. The full Mn(II) chloride dispersion curve is captured with just five fit parameters and each parameter is physically justifiable when compared to MD simulation results and/or independent measurements.

\subsection{Iron} \label{FittingFe} \label{SectionResultsFe} 

\begin{figure}[tbh!]
	\unitlength1cm
	\begin{center}
		\includegraphics[width=0.800\textwidth, trim={2cm 8.5cm 2cm 9cm}]{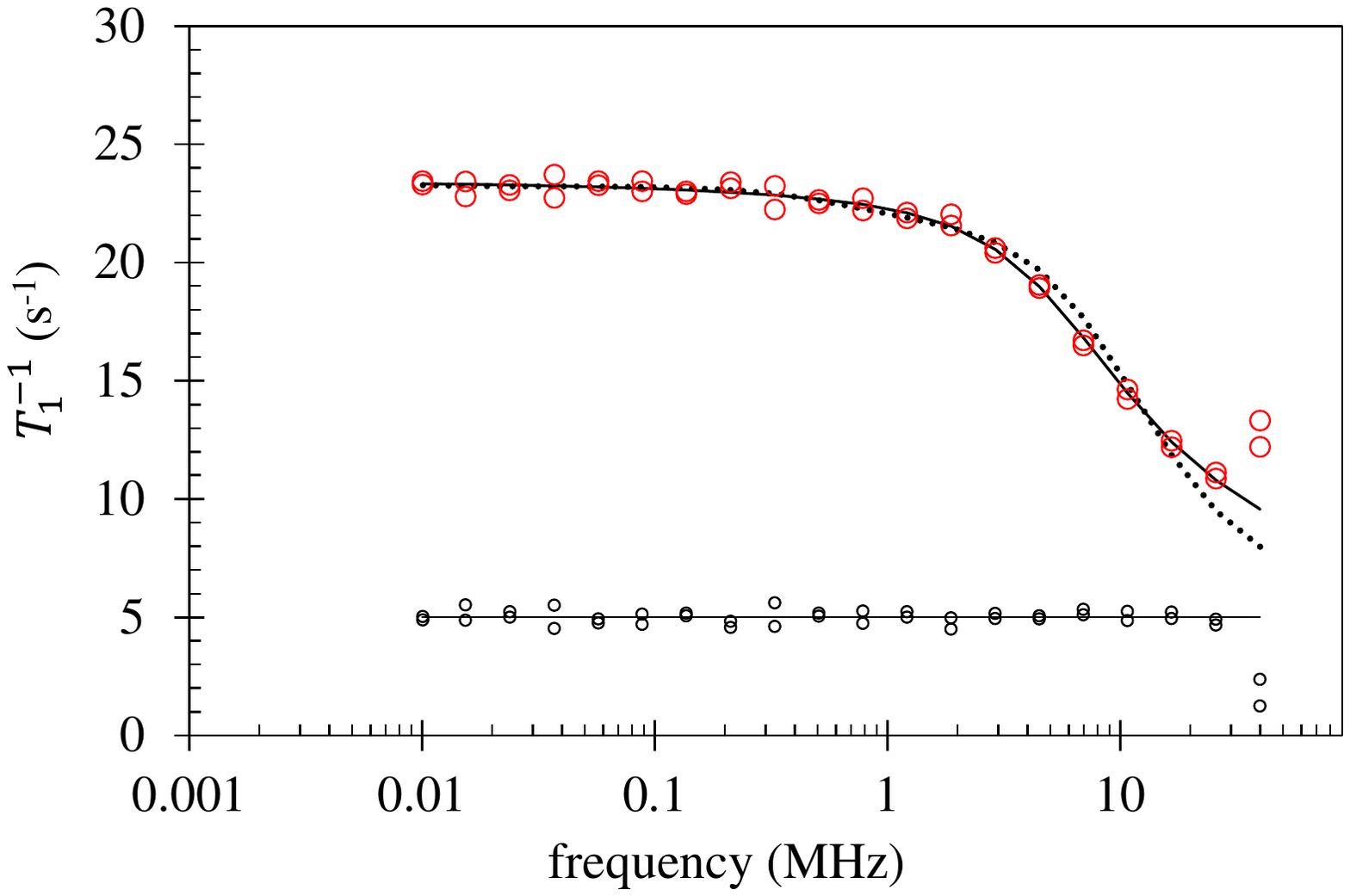}
		\caption{The experimental relaxation rate dispersion curve for aqueous Fe(III) chloride solution
			({\color{red}{\large $\circ$}}) scaled to 1\,mM is presented.  
			The  model fit combination Fit-Fe1 (see text) is shown by the solid line with  residuals ($\circ$) drawn to scale and offset at 5\,s$^{-1}$.  The dotted line is the fit Fit-Fe2. The origin of the two highest-frequency data points is discussed in the text.}
		\label{Fig6_Fe}
	\end{center}
\end{figure}

The primary motivation for the study of Fe(III) is that the ion appears in the solid component of many porous materials including rocks, cement material and silicas.  Fe(III) at the solid surfaces may desorb into the pore water, contributing to the measured NMR relaxation rate. 
The FFC NMR dispersion curves for an aqueous solution of Fe(III) chloride at 1.5\,mM and 2.5\,mM were converted to a master 1\,mM curve with an assumed offset of 0.4\,s$^{-1}$ (see Table \ref{Table:general}) and displayed in Fig.\,\ref{Fig6_Fe}.   Iron is next to manganese in the periodic table and both Fe(III) and Mn(II) have the same number of electrons. Nevertheless, the dispersion curves of aqueous  Fe(III) chloride and Mn(II) chloride differ considerably. 

The Fe(III) chloride dispersion reveals a single inflection at about 10\,MHz associated with the interaction of Fe(III) with fast-moving (outer sphere)  $^1$H spins.  The dispersion curve is featureless at lower frequencies.  However, there is a sharp peak at 40\,MHz. Bertini {\it et al} \cite{bertini1993water,bertini1993nuclear} suggest that the feature is consistent with electronic spin relaxation associated with the bonding of Fe(III) with hydroxide with a time constant of a few picoseconds.  The final two data points are excluded from the fits.

Table \ref{Table:rates} presents $^{17}$O NMR measurements of the exchange lifetime  $\tau_{\rm ex}$.  The lifetime of a water molecule in  the  [Fe(H$_2$O)$_6$]$^{3+}$ complex  is long, of order milliseconds.  Water exchange from the inner to the outer sphere on a timescale of milliseconds is too long to appear in the dispersion curve over the range of the experiment.  The absence of any inflection associated with inner sphere water suggests that  the rotational time constant $\tau_{{\rm i}\beta} \gtrsim 100\,\mu$s. The very slow rotation of the inner sphere protons is justified by the large complexes formed by Fe(III) as demonstrated by MD simulation \cite{zhang2015molecular}.  The complexes are substantially larger than those seen for Mn(II) chloride.  Indeed, Fe(III) compounds readily precipitate unless stabilising acid is added. Here, nitric acid was added to stabilise the complexes. Typical Fe(III) cluster sizes noted in Ref.\,\cite{zhang2015molecular} support $\tau_{{\rm i}\beta} \gtrsim 100\,\mu$s and would explain the absence of inner sphere features from the dispersion.

Two model combinations, labelled Fit-Fe1 and Fit-Fe2, are presented in Fig.\,\ref{Fig6_Fe} and Table \ref{Table:Fe}. Fit-Fe1 
assumes that inner sphere relaxation is not seen in the dispersion curve and Fit-Fe2 assumes that such a contribution exists.  In the case of Fit-Fe1, neither the shell($\tau_{{\rm o}\beta}$) nor Hwang-Freed model alone provides a satisfactory fit to the dispersion curve, but a combination of the two models provides an excellent fit evidenced by the small residuals. The parameter sets $(d, \tau_{{\rm o}\beta})$ and $(d_{\rm HF}, \tau_{\rm b})$ are presented as Fit-Fe1 in Table \ref{Table:Fe}. Both the shell$(\tau_{{\rm o}\beta})$ and Hwang-Freed models apply to the outer sphere so that $y$ is the symbol now used to  represent the fraction of spins contributing to the outer sphere shell model leaving the fraction $1-y$ of spins contributing to the Hwang-Freed model. There is no difference in the number of fit parameters between Fit-Fe1 and Fit-Fe2.  The outer shell rotational time constant $\tau_{{\rm o}\beta}$ obtained from Fit-Fe1 is 37\,ps consistent with 30\,ps reported by Bertini {\em et al} \cite{bertini1993nuclear} albeit through a different relaxation mechanism.  The outer shell radius $d$ in Fit-Fe1 is consistent with a second hydration peak and the Hwang-Freed distance of nearest approach is shorter at 0.343\,nm, again as expected.  The bulk diffusion time constant $\tau_{\rm b}\!=\!15$\,ps is exactly in the expected range 10--20\,ps.

\begin{table}[ht]
	\caption{The table presents the best fit parameters for the relaxation rate dispersion for two alternative models for aqueous  Fe(III) chloride solution. Parameter $y$ replaces $x$ in Fit-Fe1 to represent the fraction of water associated with the  the outer sphere shell model with fraction $1-y$ contributing to the Hwang-Freed continuum model. }	
	
	\begin{tabular}{rcll}\toprule \\[-4mm]
		& & {\bf Fit-Fe1} & {\bf Fit-Fe2} \\ \hline \\[-4mm]
		Inner sphere model: &~~~~~~~~~~~ &  ~~~~~~~~~~~~~~~~~~~~~& shell($\tau_{\rm i \beta}$) \\ \hline \\[-4mm]
		radius of inner shell &  $a$ &    &  $0.176$\,nm   \\[0mm]
		rotational time constant (inner)  & $\tau_{{\rm i}\beta}$  & & 0.12\,$\mu$s   \\[1mm] \hline \hline \\[-4mm]
		Outer sphere model: & & shell\,+\,HF &  shell~~~~~~  \\[0mm] \hline \\[-4mm]
		radius of outer shell & $d$ &  $0.406$\,nm &  $0.381$\,nm  \\[0mm]
		rotational time constant (outer) & $\tau_{{\rm o}\beta}$ & 37\,ps  & 21\,ps    \\[0mm]
		Hwang-Freed distance & $d_{\rm HF}$ & 0.343\,nm  &     \\[0mm]
		bulk diffusion time constant & $\tau_{\rm b}$  & 15\,ps  &        \\ \hline \hline \\[-3mm]
		inner sphere volume fraction &  $x$  &  &  3.64$\times 10^{-6}$   \\[0mm]
		shell model fraction &  $y$  &  0.464 &  \\[0mm]
		quality-of-fit parameter &   $Q$  &  2.66 &  9.4
		\\ \hline \hline
	\end{tabular} 
	\label{Table:Fe}
\end{table}

The Fit-Fe1 combines an outer shell model with an outer shell continuum model.  Recall that these models provide extreme representations of the spin dynamics. The outer shell model characterises the rotational dynamics but not the translational dynamics and the Hwang-Freed model captures relative translational diffusion only.  It is not unexpected to find that the optimum fit is a combination of the two.

Despite the success of Fit-Fe1, an alternative fit Fit-Fe2 is attempted which assumes that an inner sphere contribution exists. Here, it is assumed that  $\tau_{\rm ex}\! \gg \! \tau_{{\rm i}\beta}$ so that $\tau_{{\rm i}\beta}$ is the dominant time constant characterising the inner sphere contribution.  A trial shell($\tau_{{\rm i}\beta}$) + Hwang-Freed combination provided a poor quality of fit but a reasonable fit was found for the shell($\tau_{{\rm i}\beta}$) + shell($\tau_{{\rm o}\beta}$) model combination and the fit parameters are listed as Fit-Fe2.  However, the Fit-Fe2 fit parameters are not consistent with experiment.   Magini and Ragnai \cite{magini1979x} undertook X-ray diffraction studies of aqueous Fe(III) chlorides and found that the Fe-O distance for the  [Fe(H$_2$O)$_6$]$^{3+}$ complex was 0.22--0.23\,nm   \cite{magini1979x}.  Making the normal adjustment to determine the distance of the shell of $^1$H spins from the iron center suggests a Fe-H radial distribution function maximum at 0.28-0.29\,nm, in marked disagreement with the fit value 0.176\,nm listed as Fit-Fe2 in Table \ref{Table:Fe}.  Moreover, $\tau_{{\rm i}\beta}\!=\!0.12\,\mu$s for Fit-Fe2 suggests a much faster rotating Fe(III) complex compared to the Mn(II) aquoion in Fit-Mn2 of Table \ref{Table:Mn}, an outcome that seems unlikely in view of the larger sizes of Fe(III) complexes.  The values of the fit parameters emerging from Fit-Fe2 are not supported by experiment. The fit Fit-Fe1 is therefore the model combination of choice for aqueous Fe(III) chloride.  

Porous systems such as cementitious materials, rocks, clays and silicas contain small concentrations of Fe(III) in the solid that may desorb into the aqueous component when hydrated.  The fit parameters for Fit-Fe1 in Table \ref{Table:Fe} may be used as a standard 1\,mM dispersion curve that can be used to characterise the ``background" component from FFC NMR measurements from iron-containing porous material.

\subsection{Copper} \label{FittingCu}  \label{SectionResultsCu} 

\begin{figure}[tbh!]
	\unitlength1cm
	\begin{center}
		\includegraphics[width=0.8\textwidth, trim={0cm 6cm 0cm 7cm}]{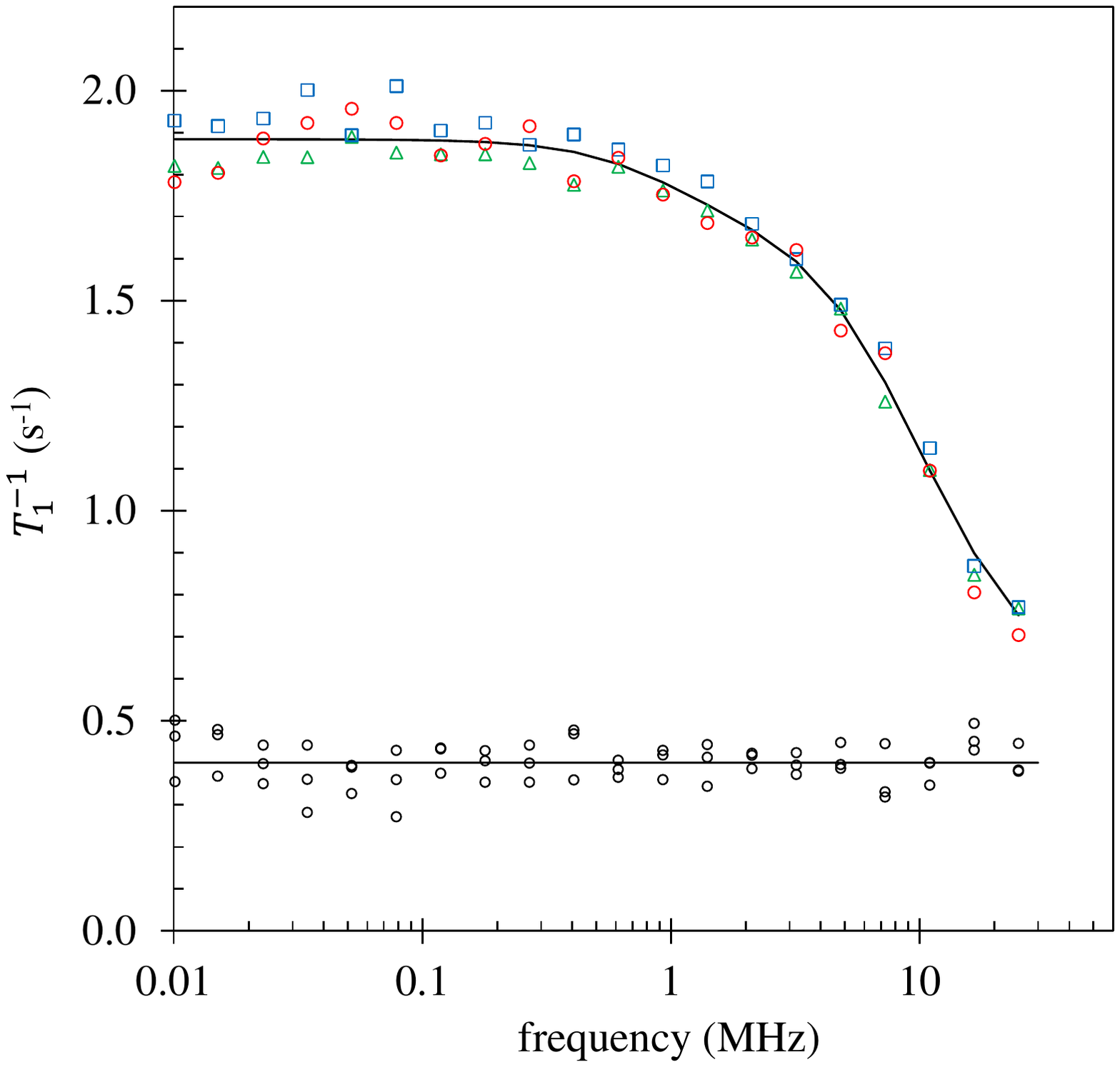} \\
		\caption{The experimental relaxation rate dispersion curve for aqueous Cu(II) chloride solution
			scaled to 1\,mM from 1.5\,mM ({\color{red}$\circ$}), 3.57\,mM ({\color{blue}$\square$}) and 9.2\,mM({\color{green}$\medtriangleup$}) is presented.  
			The best fit model combination is shown by the solid line with  residuals ({$\circ$}) drawn to scale and offset at 0.4\,s$^{-1}$. }
		\label{Fig7_Cu} 
	\end{center}
\end{figure}

Morgan and Nolle completed frequency-dependent measurements  of $T_1$ for dilute aqueous Cu(II) in 1959 \cite{morgan1959proton} and, in 1964, Hausser and Noack identified the single inflection feature as due to dipolar interactions \cite{hausser1964}.  Here, fast field-cycling NMR measurements on aqueous Cu(II) chloride solution were obtained at concentrations 9.2, 3.57 and 1.5\,mM. 
The 1\,mM master curve was obtained using Eq.\,(\ref{eqn:quality}) and is presented in Fig.\,\ref{Fig7_Cu}.  The relaxation is far weaker than for Mn(II) or Fe(III) because Cu(II) has an electronic spin $S=\frac{1}{2}$ compared to $\frac{5}{2}$ for both Mn(II) and Fe(III).  Only a single dispersion feature is evident.  Aqueous Cu(II) chloride is unusual because, at high chloride concentrations, the complexes link to form extended chains. However, at the chloride molarities used for the present experiments, the Cu(II) ions form a hexahydrate and application of the Brownian shell model is justified.  A summary of previous research and justification is provided in Supplementary Material Note 5 \cite{SM5}.

We are not aware of experimental water exchange lifetime $\tau_{\rm ex}$ estimates for Cu(II) chloride. Table \ref{Table:rates} reports, however, that $\tau_{\rm ex}$ in an aqueous Cu(II) chlorate complex is just 230\,ps \cite{powell199117o} suggesting that the inner shell water is very weakly bound, indeed one of the most weakly bound of all aquoions  \cite{helm2005inorganic}.   Experimental radial distribution functions \cite{magini1981hydration}  shows a number of broad peaks without evidence of a sharp first or second coordination shell. This supports the $^{17}$O NMR evidence for Cu(II) chlorate \cite{powell199117o} of weak bonding and suggests an unusually short $\tau_{\rm ex}$ also for Cu(II) chloride. On the basis of the experimental evidence, it can be assumed that $\tau_{{\rm i}\beta}\!\gg\!\tau_{\rm ex}$ so that the inner shell contribution to the relaxation rate is dominated by the exchange lifetime $\tau_{\rm ex}$.  The Hwang-Freed continuum model for the outer sphere water was tested but a satisfactory fit could not be obtained.  By contrast, the shell($\tau_{{\rm o}\beta}$)  model yielded an excellent fit.  Consequently, the results of the shell($\tau_{\rm ex}$) + shell($\tau_{{\rm o}\beta}$) combination only is presented in Table \ref{Table:Cu}.

\begin{table}[ht]
	\caption{The table presents the fit parameters for the single successful fit to the relaxation rate dispersion for aqueous Cu(II) chloride solution in column 1.  The second column indicates the range of fit parameter corresponding to a 5\% variation in the quality-of-fit parameter.}
	
	\begin{tabular}{rclll}\toprule \\[-3mm]
		Inner sphere model: &~~~~~~~~~~~ &   shell($\tau_{\rm ex}$)~~~~~~ &   shell($\tau_{\rm ex}$) \\ \hline \\[-4mm]
		radius of inner shell &  $a$ &  0.180\,nm &  0.162--0.252\,nm   \\[0mm]
		exchange time constant (inner)  & $\tau_{\rm ex}$  & 240\,ps & 230--370\,ps   \\[1mm] \hline \hline \\[-3mm]
		Outer sphere model: &  & shell & shell~~~~~~  \\[0mm] \hline  \\[-4mm]
		radius of outer shell & $d$ &  0.419\,nm &  0.413--0.419\,nm  \\[0mm]
		rotational time constant (outer) & $\tau_{{\rm o}\beta}$ & 27\,ps & 24--27\,ps     \\ \hline \hline  \\[-4mm]
		inner sphere volume fraction &  $x$  & 11.2$\times 10^{-4}$ & 6.2$\times 10^{-4}$--35.6$\times 10^{-4}$  \\[0mm]		
		quality-of-fit parameter &   $Q$  &  0.142 & 0.142--0.149
		\\ \hline \hline
	\end{tabular} 
	\label{Table:Cu}
\end{table} 
The best fit dispersion curve is presented in Fig.\,\ref{Fig7_Cu} using optimum fit parameters presented in the first column of Table \ref{Table:Cu}.
The agreement between $\tau_{\rm ex}= 240$\,ps from the fit and 230\,ps for $\tau_{\rm ex}$ as measured by Powell {\em et al} \cite{powell199117o} is excellent albeit the two experiments are with different anions.  
The rotational time constant for the outer shell at 27\,ps is shorter than either the Mn(II) or Fe(III) systems consistent with more loosely bound outer sphere water. 

In contrast to the Fe(III) and Mn(II) systems, good-quality fits ($Q$ within 5\% of best fit) can be obtained for a spread of values of $x$, $a$ and $\tau_{\rm ex}$.  The best-fit value of $a$ of 0.18\,nm is shorter than the spread of the first coordination shell observed experimentally of about  0.203--0.234\,$\AA$\,nm \cite{SM5} but the spread of inner shell radii presented in the second column of Table \ref{Table:Cu} spans the experimental range \cite{SM5}.   The value of $x$ is significantly larger than that seen for  Mn(II) or Fe(III), suggesting a broad Cu--$^1$H radial  distribution function consistent with experiment \cite{magini1981hydration}.

\section{Summary and conclusions} \label{Section:conclusions}

A Brownian shell model is presented that describes the Brownian rotational dynamics of ion-$^1$H spin-pair vectors of fixed length  and used to calculate the dipolar spin-lattice relaxation rate dispersion $R_1(\omega)$.  The Brownian shell model includes the angular boundary conditions, provides the correct distance dependence for $R_1(\omega)$, and yields a simple expression suitable to fit to experimental $R_1(\omega)$ dispersion curves.  

Model fits to FFC NMR relaxation rate dispersion curves over the frequency range 0.01--25\,MHz are executed for de-oxygenated aqueous Mn(II), Fe(III) and Cu(II) chloride solutions.  Combinations of shell models and/or the Hwang-Freed continuum model provide fits to FFC NMR $R_1(\omega)$ dispersion curves over the full experimental frequency range.  The optimum fit parameters are presented in Tables \ref{Table:Mn}, \ref{Table:Fe} and \ref{Table:Cu} for each aquoion.   The fit parameters are compared to the outputs of tailored MD simulations and to experimental data, where available, and found to take physically justifiable numerical values in each case.  

The inner-outer sphere water exchange lifetime $\tau_{\rm ex}$ is found to play a pivotal role in determining the shape of the relaxation rate dispersion curves and influences which combination of inner and outer shell models is appropriate in each case. For Fe(III) chloride, $\tau_{\rm ex}$ is too long to influence the dispersion in the frequency range of the experiment.  By contrast, for the Cu(II) system, $\tau_{\rm ex}$ is very short and what appears to be a single inflection feature is instead found to be a combination of two inflections  due to inner and outer shell water with similar dynamical time constants.  The Mn(II) dispersion is fit exquisitely to inner and outer sphere models for the low and high frequency inflection features respectively supporting previous experimental results indicating that the full Mn(II) dispersion is dominated by dipolar interactions.  The time constant characterising the rotation of the inner sphere spins is long, or order microseconds.  This result is supported by MD simulations which reveal an initial rapid small-angle rotation of ion-$^1$H vectors superposed on a very slow rotation of the Mn(II) chloride aquoion complex.

The Brownian shell model  constitutes a significant enhancement to the widely used dipolar SBM model without added complexity. Models based on dipolar interactions are however not suited to aqueous ion systems where the scalar interaction dominates.  Unfortunately, it is rare {\it a priori} to know which of the two interactions prevail.  If the dipolar models applied here yield fit parameters inconsistent with experiment, the evidence would point firmly to a dispersion curve with a scalar interaction origin. 

The expressions for the relaxation rate $R_1(\omega)$ with fit parameters for Mn(II), Fe(III) and Cu(II) aquoions may be used to estimate their concentrations in mixed liquids, to quantify the ``background" contribution from aqueous paramagnetic ions in porous material, to isolate paramagnetic ion contributions in biological material and to describe the frequency-dependent relaxation rate contributions in non-ionic systems, for instance due to dissolved paramagnetic dioxygen.  Finally, the elucidation of the complex dynamics of hydrated ion complexes through MD simulations clarifies the influence of metal ions on the water dynamics, particularly the slow rotation of ionic complexes and the significant slowing of water diffusion in nano-sized pores due to hydrated calcium in cementitious material.

\bigskip
{\bf Acknowledgment}

This work received funding from the European Union Horizon 2020 Research and Innovation Programme under the Marie Sk\l odowska-Curie Innovative Training Networks programme grant agreement No. 764691.  The authors thank Dr Annika Lohstroh for translating reference \cite{pfeifer1962protonenrelaxation} from German to English.

\bigskip
{\bf Conflict of interest}

The authors have no conflicts to disclose.

\bigskip
{\bf Data Availability Statement}

The data that support the findings of this study are available from the corresponding author upon reasonable request.

\newpage

\newpage
\bibliographystyle{apsrev4-1}
\bibliography{References_master_2022a}

\begin{thebibliography}{79}%
\makeatletter
\providecommand \@ifxundefined [1]{%
 \@ifx{#1\undefined}
}%
\providecommand \@ifnum [1]{%
 \ifnum #1\expandafter \@firstoftwo
 \else \expandafter \@secondoftwo
 \fi
}%
\providecommand \@ifx [1]{%
 \ifx #1\expandafter \@firstoftwo
 \else \expandafter \@secondoftwo
 \fi
}%
\providecommand \natexlab [1]{#1}%
\providecommand \enquote  [1]{``#1''}%
\providecommand \bibnamefont  [1]{#1}%
\providecommand \bibfnamefont [1]{#1}%
\providecommand \citenamefont [1]{#1}%
\providecommand \href@noop [0]{\@secondoftwo}%
\providecommand \href [0]{\begingroup \@sanitize@url \@href}%
\providecommand \@href[1]{\@@startlink{#1}\@@href}%
\providecommand \@@href[1]{\endgroup#1\@@endlink}%
\providecommand \@sanitize@url [0]{\catcode `\\12\catcode `\$12\catcode
  `\&12\catcode `\#12\catcode `\^12\catcode `\_12\catcode `\%12\relax}%
\providecommand \@@startlink[1]{}%
\providecommand \@@endlink[0]{}%
\providecommand \url  [0]{\begingroup\@sanitize@url \@url }%
\providecommand \@url [1]{\endgroup\@href {#1}{\urlprefix }}%
\providecommand \urlprefix  [0]{URL }%
\providecommand \Eprint [0]{\href }%
\providecommand \doibase [0]{http://dx.doi.org/}%
\providecommand \selectlanguage [0]{\@gobble}%
\providecommand \bibinfo  [0]{\@secondoftwo}%
\providecommand \bibfield  [0]{\@secondoftwo}%
\providecommand \translation [1]{[#1]}%
\providecommand \BibitemOpen [0]{}%
\providecommand \bibitemStop [0]{}%
\providecommand \bibitemNoStop [0]{.\EOS\space}%
\providecommand \EOS [0]{\spacefactor3000\relax}%
\providecommand \BibitemShut  [1]{\csname bibitem#1\endcsname}%
\let\auto@bib@innerbib\@empty
\bibitem [{\citenamefont {Jaouen}(2006)}]{jaouen2006bioorganometallics}%
  \BibitemOpen
  \bibfield  {author} {\bibinfo {author} {\bibfnamefont {G.}~\bibnamefont
  {Jaouen}},\ }\href@noop {} {\emph {\bibinfo {title} {Bioorganometallics:
  biomolecules, labeling, medicine}}}\ (\bibinfo  {publisher} {John Wiley \&
  Sons},\ \bibinfo {year} {2006})\BibitemShut {NoStop}%
\bibitem [{\citenamefont {Aime}\ \emph {et~al.}(2005)\citenamefont {Aime},
  \citenamefont {Botta},\ and\ \citenamefont {Terreno}}]{aime2005advances}%
  \BibitemOpen
  \bibfield  {author} {\bibinfo {author} {\bibfnamefont {I.}~\bibnamefont
  {Aime}}, \bibinfo {author} {\bibfnamefont {M.}~\bibnamefont {Botta}}, \ and\
  \bibinfo {author} {\bibfnamefont {E.}~\bibnamefont {Terreno}},\ }\href@noop
  {} {\emph {\bibinfo {title} {Advances in Inorganic Chemistry: Relaxometry of
  Water-Metal Ion Interactions}}},\ edited by\ \bibinfo {editor} {\bibfnamefont
  {I.}~\bibnamefont {Bertini}}\ and\ \bibinfo {editor} {\bibfnamefont
  {R.}~\bibnamefont {van Eldik}}\ (\bibinfo  {publisher} {Elsevier, San
  Diego},\ \bibinfo {year} {2005})\ pp.\ \bibinfo {pages}
  {173--237}\BibitemShut {NoStop}%
\bibitem [{\citenamefont {Bertini}\ \emph {et~al.}(2005)\citenamefont
  {Bertini}, \citenamefont {Luchinat},\ and\ \citenamefont
  {Parigi}}]{bertini2005advances}%
  \BibitemOpen
  \bibfield  {author} {\bibinfo {author} {\bibfnamefont {I.}~\bibnamefont
  {Bertini}}, \bibinfo {author} {\bibfnamefont {C.}~\bibnamefont {Luchinat}}, \
  and\ \bibinfo {author} {\bibfnamefont {G.}~\bibnamefont {Parigi}},\
  }\href@noop {} {\emph {\bibinfo {title} {Advances in Inorganic Chemistry:
  Relaxometry of Water-Metal Ion Interactions}}},\ edited by\ \bibinfo {editor}
  {\bibfnamefont {I.}~\bibnamefont {Bertini}}\ and\ \bibinfo {editor}
  {\bibfnamefont {R.}~\bibnamefont {van Eldik}}\ (\bibinfo  {publisher}
  {Elsevier, San Diego},\ \bibinfo {year} {2005})\ pp.\ \bibinfo {pages}
  {105--172}\BibitemShut {NoStop}%
\bibitem [{\citenamefont {Geraldes}\ and\ \citenamefont
  {Laurent}(2009)}]{geraldes2009classification}%
  \BibitemOpen
  \bibfield  {author} {\bibinfo {author} {\bibfnamefont {C.~F.}\ \bibnamefont
  {Geraldes}}\ and\ \bibinfo {author} {\bibfnamefont {S.}~\bibnamefont
  {Laurent}},\ }\href@noop {} {\bibfield  {journal} {\bibinfo  {journal}
  {Contrast Media \& Mol. I.}\ }\textbf {\bibinfo {volume} {4}},\ \bibinfo
  {pages} {1} (\bibinfo {year} {2009})}\BibitemShut {NoStop}%
\bibitem [{\citenamefont {Pan}\ \emph {et~al.}(2011)\citenamefont {Pan},
  \citenamefont {Schmieder}, \citenamefont {Wickline},\ and\ \citenamefont
  {Lanza}}]{pan2011manganese}%
  \BibitemOpen
  \bibfield  {author} {\bibinfo {author} {\bibfnamefont {D.}~\bibnamefont
  {Pan}}, \bibinfo {author} {\bibfnamefont {A.~H.}\ \bibnamefont {Schmieder}},
  \bibinfo {author} {\bibfnamefont {S.~A.}\ \bibnamefont {Wickline}}, \ and\
  \bibinfo {author} {\bibfnamefont {G.~M.}\ \bibnamefont {Lanza}},\ }\href@noop
  {} {\bibfield  {journal} {\bibinfo  {journal} {Tetrahedron}\ }\textbf
  {\bibinfo {volume} {67}},\ \bibinfo {pages} {8431} (\bibinfo {year}
  {2011})}\BibitemShut {NoStop}%
\bibitem [{\citenamefont {Bodart}\ \emph {et~al.}(2020)\citenamefont {Bodart},
  \citenamefont {Rachocki}, \citenamefont {Tritt-Goc}, \citenamefont
  {Michalke}, \citenamefont {Schmitt-Kopplin}, \citenamefont {Karbowiak},\ and\
  \citenamefont {Gougeon}}]{bodart2020quantification}%
  \BibitemOpen
  \bibfield  {author} {\bibinfo {author} {\bibfnamefont {P.~R.}\ \bibnamefont
  {Bodart}}, \bibinfo {author} {\bibfnamefont {A.}~\bibnamefont {Rachocki}},
  \bibinfo {author} {\bibfnamefont {J.}~\bibnamefont {Tritt-Goc}}, \bibinfo
  {author} {\bibfnamefont {B.}~\bibnamefont {Michalke}}, \bibinfo {author}
  {\bibfnamefont {P.}~\bibnamefont {Schmitt-Kopplin}}, \bibinfo {author}
  {\bibfnamefont {T.}~\bibnamefont {Karbowiak}}, \ and\ \bibinfo {author}
  {\bibfnamefont {R.~D.}\ \bibnamefont {Gougeon}},\ }\href@noop {} {\bibfield
  {journal} {\bibinfo  {journal} {Talanta}\ }\textbf {\bibinfo {volume}
  {209}},\ \bibinfo {pages} {120561} (\bibinfo {year} {2020})}\BibitemShut
  {NoStop}%
\bibitem [{\citenamefont {Baroni}\ \emph {et~al.}(2009)\citenamefont {Baroni},
  \citenamefont {Consonni}, \citenamefont {Ferrante},\ and\ \citenamefont
  {Aime}}]{baroni2009relaxometric}%
  \BibitemOpen
  \bibfield  {author} {\bibinfo {author} {\bibfnamefont {S.}~\bibnamefont
  {Baroni}}, \bibinfo {author} {\bibfnamefont {R.}~\bibnamefont {Consonni}},
  \bibinfo {author} {\bibfnamefont {G.}~\bibnamefont {Ferrante}}, \ and\
  \bibinfo {author} {\bibfnamefont {S.}~\bibnamefont {Aime}},\ }\href@noop {}
  {\bibfield  {journal} {\bibinfo  {journal} {J. Agric. Food Chem.}\ }\textbf
  {\bibinfo {volume} {57}},\ \bibinfo {pages} {3028} (\bibinfo {year}
  {2009})}\BibitemShut {NoStop}%
\bibitem [{\citenamefont {Shapiro}(2011)}]{shapiro2011structure}%
  \BibitemOpen
  \bibfield  {author} {\bibinfo {author} {\bibfnamefont {Y.~E.}\ \bibnamefont
  {Shapiro}},\ }\href@noop {} {\bibfield  {journal} {\bibinfo  {journal} {Prog.
  Polym. Sci.}\ }\textbf {\bibinfo {volume} {36}},\ \bibinfo {pages} {1184}
  (\bibinfo {year} {2011})}\BibitemShut {NoStop}%
\bibitem [{\citenamefont {Faux}\ \emph {et~al.}(2017)\citenamefont {Faux},
  \citenamefont {McDonald},\ and\ \citenamefont {Howlett}}]{Faux.2017a}%
  \BibitemOpen
  \bibfield  {author} {\bibinfo {author} {\bibfnamefont {D.~A.}\ \bibnamefont
  {Faux}}, \bibinfo {author} {\bibfnamefont {P.~J.}\ \bibnamefont {McDonald}},
  \ and\ \bibinfo {author} {\bibfnamefont {N.~C.}\ \bibnamefont {Howlett}},\
  }\href@noop {} {\bibfield  {journal} {\bibinfo  {journal} {Phys. Rev. E}\
  }\textbf {\bibinfo {volume} {95}},\ \bibinfo {pages} {033116} (\bibinfo
  {year} {2017})}\BibitemShut {NoStop}%
\bibitem [{\citenamefont {Faux}\ and\ \citenamefont
  {McDonald}(2017)}]{Faux.2017b}%
  \BibitemOpen
  \bibfield  {author} {\bibinfo {author} {\bibfnamefont {D.~A.}\ \bibnamefont
  {Faux}}\ and\ \bibinfo {author} {\bibfnamefont {P.~J.}\ \bibnamefont
  {McDonald}},\ }\href@noop {} {\bibfield  {journal} {\bibinfo  {journal}
  {Phys. Rev. E}\ }\textbf {\bibinfo {volume} {95}},\ \bibinfo {pages} {033117}
  (\bibinfo {year} {2017})}\BibitemShut {NoStop}%
\bibitem [{\citenamefont {Korb}(2011)}]{korb2011}%
  \BibitemOpen
  \bibfield  {author} {\bibinfo {author} {\bibfnamefont {J.-P.}\ \bibnamefont
  {Korb}},\ }\href@noop {} {\bibfield  {journal} {\bibinfo  {journal} {New J.
  Phys.}\ }\textbf {\bibinfo {volume} {13}},\ \bibinfo {pages} {035016}
  (\bibinfo {year} {2011})}\BibitemShut {NoStop}%
\bibitem [{\citenamefont {Zoroddu}\ \emph {et~al.}(2019)\citenamefont
  {Zoroddu}, \citenamefont {Aaseth}, \citenamefont {Crisponi}, \citenamefont
  {Medici}, \citenamefont {Peana},\ and\ \citenamefont
  {Nurchi}}]{zoroddu2019essential}%
  \BibitemOpen
  \bibfield  {author} {\bibinfo {author} {\bibfnamefont {M.~A.}\ \bibnamefont
  {Zoroddu}}, \bibinfo {author} {\bibfnamefont {J.}~\bibnamefont {Aaseth}},
  \bibinfo {author} {\bibfnamefont {G.}~\bibnamefont {Crisponi}}, \bibinfo
  {author} {\bibfnamefont {S.}~\bibnamefont {Medici}}, \bibinfo {author}
  {\bibfnamefont {M.}~\bibnamefont {Peana}}, \ and\ \bibinfo {author}
  {\bibfnamefont {V.~M.}\ \bibnamefont {Nurchi}},\ }\href@noop {} {\bibfield
  {journal} {\bibinfo  {journal} {J. Inorg. Biochem.}\ }\textbf {\bibinfo
  {volume} {195}},\ \bibinfo {pages} {120} (\bibinfo {year}
  {2019})}\BibitemShut {NoStop}%
\bibitem [{\citenamefont {Bloembergen}\ \emph {et~al.}(1948)\citenamefont
  {Bloembergen}, \citenamefont {Purcell},\ and\ \citenamefont
  {Pound}}]{Bloembergen.1948}%
  \BibitemOpen
  \bibfield  {author} {\bibinfo {author} {\bibfnamefont {N.}~\bibnamefont
  {Bloembergen}}, \bibinfo {author} {\bibfnamefont {E.~M.}\ \bibnamefont
  {Purcell}}, \ and\ \bibinfo {author} {\bibfnamefont {R.~V.}\ \bibnamefont
  {Pound}},\ }\href@noop {} {\bibfield  {journal} {\bibinfo  {journal} {Phys.
  Rev.}\ }\textbf {\bibinfo {volume} {73}},\ \bibinfo {pages} {679} (\bibinfo
  {year} {1948})}\BibitemShut {NoStop}%
\bibitem [{\citenamefont {Bloembergen}(1957)}]{bloembergen1957proton}%
  \BibitemOpen
  \bibfield  {author} {\bibinfo {author} {\bibfnamefont {N.}~\bibnamefont
  {Bloembergen}},\ }\href@noop {} {\bibfield  {journal} {\bibinfo  {journal}
  {J. Chem. Phys.}\ }\textbf {\bibinfo {volume} {27}},\ \bibinfo {pages} {572}
  (\bibinfo {year} {1957})}\BibitemShut {NoStop}%
\bibitem [{\citenamefont {Morgan}\ and\ \citenamefont
  {Nolle}(1959)}]{morgan1959proton}%
  \BibitemOpen
  \bibfield  {author} {\bibinfo {author} {\bibfnamefont {L.}~\bibnamefont
  {Morgan}}\ and\ \bibinfo {author} {\bibfnamefont {A.}~\bibnamefont {Nolle}},\
  }\href@noop {} {\bibfield  {journal} {\bibinfo  {journal} {J. Chem. Phys.}\
  }\textbf {\bibinfo {volume} {31}},\ \bibinfo {pages} {365} (\bibinfo {year}
  {1959})}\BibitemShut {NoStop}%
\bibitem [{\citenamefont {Bloembergen}\ and\ \citenamefont
  {Morgan}(1961)}]{bloembergen1961proton}%
  \BibitemOpen
  \bibfield  {author} {\bibinfo {author} {\bibfnamefont {N.}~\bibnamefont
  {Bloembergen}}\ and\ \bibinfo {author} {\bibfnamefont {L.}~\bibnamefont
  {Morgan}},\ }\href@noop {} {\bibfield  {journal} {\bibinfo  {journal} {J.
  Chem. Phys.}\ }\textbf {\bibinfo {volume} {34}},\ \bibinfo {pages} {842}
  (\bibinfo {year} {1961})}\BibitemShut {NoStop}%
\bibitem [{\citenamefont {Bleaney}\ and\ \citenamefont
  {Stevens}(1953)}]{bleaney1953paramagnetic}%
  \BibitemOpen
  \bibfield  {author} {\bibinfo {author} {\bibfnamefont {B.}~\bibnamefont
  {Bleaney}}\ and\ \bibinfo {author} {\bibfnamefont {K.}~\bibnamefont
  {Stevens}},\ }\href@noop {} {\bibfield  {journal} {\bibinfo  {journal} {Rep.
  Prog. Phys.}\ }\textbf {\bibinfo {volume} {16}},\ \bibinfo {pages} {108}
  (\bibinfo {year} {1953})}\BibitemShut {NoStop}%
\bibitem [{\citenamefont {Abragam}(1955)}]{abragam1955overhauser}%
  \BibitemOpen
  \bibfield  {author} {\bibinfo {author} {\bibfnamefont {A.}~\bibnamefont
  {Abragam}},\ }\href@noop {} {\bibfield  {journal} {\bibinfo  {journal} {Phys.
  Rev.}\ }\textbf {\bibinfo {volume} {98}},\ \bibinfo {pages} {1729} (\bibinfo
  {year} {1955})}\BibitemShut {NoStop}%
\bibitem [{\citenamefont {Abragam}(1961)}]{Abragam}%
  \BibitemOpen
  \bibfield  {author} {\bibinfo {author} {\bibfnamefont {A.}~\bibnamefont
  {Abragam}},\ }\href@noop {} {\emph {\bibinfo {title} {Principles of nuclear
  magnetism}}}\ (\bibinfo  {publisher} {Clarendon Press, Oxford},\ \bibinfo
  {year} {1961})\BibitemShut {NoStop}%
\bibitem [{\citenamefont {Solomon}(1955)}]{solomon1955physical}%
  \BibitemOpen
  \bibfield  {author} {\bibinfo {author} {\bibfnamefont {I.}~\bibnamefont
  {Solomon}},\ }\href@noop {} {\bibfield  {journal} {\bibinfo  {journal} {Phys.
  Rev.}\ }\textbf {\bibinfo {volume} {99}},\ \bibinfo {pages} {559} (\bibinfo
  {year} {1955})}\BibitemShut {NoStop}%
\bibitem [{\citenamefont {Solomon}\ and\ \citenamefont
  {Bloembergen}(1956)}]{solomon1956nuclear}%
  \BibitemOpen
  \bibfield  {author} {\bibinfo {author} {\bibfnamefont {I.}~\bibnamefont
  {Solomon}}\ and\ \bibinfo {author} {\bibfnamefont {N.}~\bibnamefont
  {Bloembergen}},\ }\href@noop {} {\bibfield  {journal} {\bibinfo  {journal}
  {J. Chem. Phys.}\ }\textbf {\bibinfo {volume} {25}},\ \bibinfo {pages} {261}
  (\bibinfo {year} {1956})}\BibitemShut {NoStop}%
\bibitem [{\citenamefont {Laukien}\ and\ \citenamefont
  {Schl\"{u}ter}(1956)}]{laukien1956pulse}%
  \BibitemOpen
  \bibfield  {author} {\bibinfo {author} {\bibfnamefont {G.}~\bibnamefont
  {Laukien}}\ and\ \bibinfo {author} {\bibfnamefont {J.}~\bibnamefont
  {Schl\"{u}ter}},\ }\href@noop {} {\bibfield  {journal} {\bibinfo  {journal}
  {Z. Phys.}\ }\textbf {\bibinfo {volume} {146}},\ \bibinfo {pages} {113}
  (\bibinfo {year} {1956})}\BibitemShut {NoStop}%
\bibitem [{\citenamefont {Bertini}\ \emph {et~al.}(2001)\citenamefont
  {Bertini}, \citenamefont {Luchinat},\ and\ \citenamefont
  {Parigi}}]{bertini2001solution}%
  \BibitemOpen
  \bibfield  {author} {\bibinfo {author} {\bibfnamefont {I.}~\bibnamefont
  {Bertini}}, \bibinfo {author} {\bibfnamefont {C.}~\bibnamefont {Luchinat}}, \
  and\ \bibinfo {author} {\bibfnamefont {G.}~\bibnamefont {Parigi}},\
  }\href@noop {} {\emph {\bibinfo {title} {Solution NMR of paramagnetic
  molecules: applications to metallobiomolecules and models}}}\ (\bibinfo
  {publisher} {Elsevier},\ \bibinfo {year} {2001})\BibitemShut {NoStop}%
\bibitem [{\citenamefont {Pell}\ \emph {et~al.}(2019)\citenamefont {Pell},
  \citenamefont {Pintacuda},\ and\ \citenamefont
  {Grey}}]{pell2019paramagnetic}%
  \BibitemOpen
  \bibfield  {author} {\bibinfo {author} {\bibfnamefont {A.~J.}\ \bibnamefont
  {Pell}}, \bibinfo {author} {\bibfnamefont {G.}~\bibnamefont {Pintacuda}}, \
  and\ \bibinfo {author} {\bibfnamefont {C.~P.}\ \bibnamefont {Grey}},\
  }\href@noop {} {\bibfield  {journal} {\bibinfo  {journal} {Progress in
  nuclear magnetic resonance spectroscopy}\ }\textbf {\bibinfo {volume}
  {111}},\ \bibinfo {pages} {1} (\bibinfo {year} {2019})}\BibitemShut {NoStop}%
\bibitem [{\citenamefont {Bertini}\ \emph {et~al.}(2016)\citenamefont
  {Bertini}, \citenamefont {Luchinat}, \citenamefont {Parigi},\ and\
  \citenamefont {Ravera}}]{bertini2016nmr}%
  \BibitemOpen
  \bibfield  {author} {\bibinfo {author} {\bibfnamefont {I.}~\bibnamefont
  {Bertini}}, \bibinfo {author} {\bibfnamefont {C.}~\bibnamefont {Luchinat}},
  \bibinfo {author} {\bibfnamefont {G.}~\bibnamefont {Parigi}}, \ and\ \bibinfo
  {author} {\bibfnamefont {E.}~\bibnamefont {Ravera}},\ }\href@noop {} {\emph
  {\bibinfo {title} {NMR of paramagnetic molecules: applications to
  metallobiomolecules and models}}}\ (\bibinfo  {publisher} {Elsevier},\
  \bibinfo {year} {2016})\ Chap.~\bibinfo {chapter} {7}\BibitemShut {NoStop}%
\bibitem [{\citenamefont {Faux}\ \emph {et~al.}(2021)\citenamefont {Faux},
  \citenamefont {Rahaman},\ and\ \citenamefont {McDonald}}]{faux2021a}%
  \BibitemOpen
  \bibfield  {author} {\bibinfo {author} {\bibfnamefont {D.~A.}\ \bibnamefont
  {Faux}}, \bibinfo {author} {\bibfnamefont {A.~A.}\ \bibnamefont {Rahaman}}, \
  and\ \bibinfo {author} {\bibfnamefont {P.~J.}\ \bibnamefont {McDonald}},\
  }\href@noop {} {\bibfield  {journal} {\bibinfo  {journal} {Phys. Rev. Lett.}\
  }\textbf {\bibinfo {volume} {127}},\ \bibinfo {pages} {256001} (\bibinfo
  {year} {2021})}\BibitemShut {NoStop}%
\bibitem [{\citenamefont {Hernandez}\ and\ \citenamefont
  {Bryant}(1991)}]{hernandez1991proton}%
  \BibitemOpen
  \bibfield  {author} {\bibinfo {author} {\bibfnamefont {G.}~\bibnamefont
  {Hernandez}}\ and\ \bibinfo {author} {\bibfnamefont {R.~G.}\ \bibnamefont
  {Bryant}},\ }\href@noop {} {\bibfield  {journal} {\bibinfo  {journal}
  {Bioconjug. Chem.}\ }\textbf {\bibinfo {volume} {2}},\ \bibinfo {pages} {394}
  (\bibinfo {year} {1991})}\BibitemShut {NoStop}%
\bibitem [{\citenamefont {Powell}\ \emph {et~al.}(1991)\citenamefont {Powell},
  \citenamefont {Helm},\ and\ \citenamefont {Merbach}}]{powell199117o}%
  \BibitemOpen
  \bibfield  {author} {\bibinfo {author} {\bibfnamefont {D.~H.}\ \bibnamefont
  {Powell}}, \bibinfo {author} {\bibfnamefont {L.}~\bibnamefont {Helm}}, \ and\
  \bibinfo {author} {\bibfnamefont {A.~E.}\ \bibnamefont {Merbach}},\
  }\href@noop {} {\bibfield  {journal} {\bibinfo  {journal} {J. Chem. Phys.}\
  }\textbf {\bibinfo {volume} {95}},\ \bibinfo {pages} {9258} (\bibinfo {year}
  {1991})}\BibitemShut {NoStop}%
\bibitem [{\citenamefont {Hwang}\ and\ \citenamefont
  {Freed}(1975)}]{hwang1975dynamic}%
  \BibitemOpen
  \bibfield  {author} {\bibinfo {author} {\bibfnamefont {L.-P.}\ \bibnamefont
  {Hwang}}\ and\ \bibinfo {author} {\bibfnamefont {J.~H.}\ \bibnamefont
  {Freed}},\ }\href@noop {} {\bibfield  {journal} {\bibinfo  {journal} {J.
  Chem. Phys.}\ }\textbf {\bibinfo {volume} {63}},\ \bibinfo {pages} {4017}
  (\bibinfo {year} {1975})}\BibitemShut {NoStop}%
\bibitem [{\citenamefont {La~Mar}\ and\ \citenamefont
  {Walker}(1973)}]{la1973proton}%
  \BibitemOpen
  \bibfield  {author} {\bibinfo {author} {\bibfnamefont {G.~N.}\ \bibnamefont
  {La~Mar}}\ and\ \bibinfo {author} {\bibfnamefont {F.~A.}\ \bibnamefont
  {Walker}},\ }\href@noop {} {\bibfield  {journal} {\bibinfo  {journal} {J. Am.
  Chem. Soc.}\ }\textbf {\bibinfo {volume} {95}},\ \bibinfo {pages} {6950}
  (\bibinfo {year} {1973})}\BibitemShut {NoStop}%
\bibitem [{\citenamefont {Pfeifer}(1962)}]{pfeifer1962protonenrelaxation}%
  \BibitemOpen
  \bibfield  {author} {\bibinfo {author} {\bibfnamefont {H.}~\bibnamefont
  {Pfeifer}},\ }\href@noop {} {\bibfield  {journal} {\bibinfo  {journal} {Z.
  Naturforsch. A}\ }\textbf {\bibinfo {volume} {17}},\ \bibinfo {pages} {279}
  (\bibinfo {year} {1962})}\BibitemShut {NoStop}%
\bibitem [{\citenamefont {Koenig}\ and\ \citenamefont
  {Brown~III}(1984)}]{koenig1984relaxation}%
  \BibitemOpen
  \bibfield  {author} {\bibinfo {author} {\bibfnamefont {S.~H.}\ \bibnamefont
  {Koenig}}\ and\ \bibinfo {author} {\bibfnamefont {R.~D.}\ \bibnamefont
  {Brown~III}},\ }\href@noop {} {\bibfield  {journal} {\bibinfo  {journal}
  {Magn. Reson. Med.}\ }\textbf {\bibinfo {volume} {1}},\ \bibinfo {pages}
  {478} (\bibinfo {year} {1984})}\BibitemShut {NoStop}%
\bibitem [{\citenamefont {Redfield}(1955)}]{redfield1955nuclear}%
  \BibitemOpen
  \bibfield  {author} {\bibinfo {author} {\bibfnamefont {A.~G.}\ \bibnamefont
  {Redfield}},\ }\href@noop {} {\bibfield  {journal} {\bibinfo  {journal}
  {Phys. Rev.}\ }\textbf {\bibinfo {volume} {98}},\ \bibinfo {pages} {1787}
  (\bibinfo {year} {1955})}\BibitemShut {NoStop}%
\bibitem [{\citenamefont {Messiah}(1965)}]{Messiah.1965}%
  \BibitemOpen
  \bibfield  {author} {\bibinfo {author} {\bibfnamefont {A.}~\bibnamefont
  {Messiah}},\ }\href@noop {} {\emph {\bibinfo {title} {Quantum Mechanics}}}\
  (\bibinfo  {publisher} {North Holland Press},\ \bibinfo {year}
  {1965})\BibitemShut {NoStop}%
\bibitem [{\citenamefont {Sholl}(1974)}]{Sholl1974nuclear}%
  \BibitemOpen
  \bibfield  {author} {\bibinfo {author} {\bibfnamefont {C.}~\bibnamefont
  {Sholl}},\ }\href@noop {} {\bibfield  {journal} {\bibinfo  {journal} {J.
  Phys. C: Solid State Phys.}\ }\textbf {\bibinfo {volume} {7}},\ \bibinfo
  {pages} {3378} (\bibinfo {year} {1974})}\BibitemShut {NoStop}%
\bibitem [{\citenamefont {Brownstein}\ and\ \citenamefont
  {Tarr}(1979)}]{Brownstein.1979}%
  \BibitemOpen
  \bibfield  {author} {\bibinfo {author} {\bibfnamefont {K.~R.}\ \bibnamefont
  {Brownstein}}\ and\ \bibinfo {author} {\bibfnamefont {C.~E.}\ \bibnamefont
  {Tarr}},\ }\href@noop {} {\bibfield  {journal} {\bibinfo  {journal} {Phys.
  Rev.A}\ }\textbf {\bibinfo {volume} {19}},\ \bibinfo {pages} {2446} (\bibinfo
  {year} {1979})}\BibitemShut {NoStop}%
\bibitem [{\citenamefont {Ayant}\ \emph {et~al.}(1975)\citenamefont {Ayant},
  \citenamefont {Belorizky}, \citenamefont {Aluzon},\ and\ \citenamefont
  {Gallice}}]{ayant1975calcul}%
  \BibitemOpen
  \bibfield  {author} {\bibinfo {author} {\bibfnamefont {Y.}~\bibnamefont
  {Ayant}}, \bibinfo {author} {\bibfnamefont {E.}~\bibnamefont {Belorizky}},
  \bibinfo {author} {\bibfnamefont {J.}~\bibnamefont {Aluzon}}, \ and\ \bibinfo
  {author} {\bibfnamefont {J.}~\bibnamefont {Gallice}},\ }\href@noop {}
  {\bibfield  {journal} {\bibinfo  {journal} {J. Phys.-Paris}\ }\textbf
  {\bibinfo {volume} {36}},\ \bibinfo {pages} {991} (\bibinfo {year}
  {1975})}\BibitemShut {NoStop}%
\bibitem [{\citenamefont {Helm}(2006)}]{helm2006relaxivity}%
  \BibitemOpen
  \bibfield  {author} {\bibinfo {author} {\bibfnamefont {L.}~\bibnamefont
  {Helm}},\ }\href@noop {} {\bibfield  {journal} {\bibinfo  {journal} {Prog.
  Nucl. Magn. Reson. Spectrosc.}\ }\textbf {\bibinfo {volume} {49}},\ \bibinfo
  {pages} {45} (\bibinfo {year} {2006})}\BibitemShut {NoStop}%
\bibitem [{\citenamefont {Krynicki}\ \emph {et~al.}(1978)\citenamefont
  {Krynicki}, \citenamefont {Green},\ and\ \citenamefont
  {Sawyer}}]{krynicki1978pressure}%
  \BibitemOpen
  \bibfield  {author} {\bibinfo {author} {\bibfnamefont {K.}~\bibnamefont
  {Krynicki}}, \bibinfo {author} {\bibfnamefont {C.~D.}\ \bibnamefont {Green}},
  \ and\ \bibinfo {author} {\bibfnamefont {D.~W.}\ \bibnamefont {Sawyer}},\
  }\href@noop {} {\bibfield  {journal} {\bibinfo  {journal} {Faraday Discuss.
  Chem. Soc.}\ }\textbf {\bibinfo {volume} {66}},\ \bibinfo {pages} {199}
  (\bibinfo {year} {1978})}\BibitemShut {NoStop}%
\bibitem [{\citenamefont {Kowalewski}\ \emph {et~al.}(1985)\citenamefont
  {Kowalewski}, \citenamefont {Nordenski\"{o}ld}, \citenamefont {Benetis},\
  and\ \citenamefont {Westlund}}]{kowalewski1985theory}%
  \BibitemOpen
  \bibfield  {author} {\bibinfo {author} {\bibfnamefont {J.}~\bibnamefont
  {Kowalewski}}, \bibinfo {author} {\bibfnamefont {L.}~\bibnamefont
  {Nordenski\"{o}ld}}, \bibinfo {author} {\bibfnamefont {N.}~\bibnamefont
  {Benetis}}, \ and\ \bibinfo {author} {\bibfnamefont {P.-O.}\ \bibnamefont
  {Westlund}},\ }\href@noop {} {\bibfield  {journal} {\bibinfo  {journal}
  {Prog. Nucl. Magn. Reson. Spectrosc.}\ }\textbf {\bibinfo {volume} {17}},\
  \bibinfo {pages} {141} (\bibinfo {year} {1985})}\BibitemShut {NoStop}%
\bibitem [{\citenamefont {Helm}\ and\ \citenamefont
  {Merbach}(2005)}]{helm2005inorganic}%
  \BibitemOpen
  \bibfield  {author} {\bibinfo {author} {\bibfnamefont {L.}~\bibnamefont
  {Helm}}\ and\ \bibinfo {author} {\bibfnamefont {A.~E.}\ \bibnamefont
  {Merbach}},\ }\href@noop {} {\bibfield  {journal} {\bibinfo  {journal} {Chem.
  Rev.}\ }\textbf {\bibinfo {volume} {105}},\ \bibinfo {pages} {1923} (\bibinfo
  {year} {2005})}\BibitemShut {NoStop}%
\bibitem [{SM1()}]{SM1}%
  \BibitemOpen
  \href@noop {} {}\bibinfo {note} {See Supplementary Material Note 1 for the
  derivation of the probability density function for a Brownian rotor, which
  includes Refs.
  \cite{Messiah.1965,Sholl1974nuclear,Faux.2017a,faux2021a}}\BibitemShut
  {NoStop}%
\bibitem [{SM2({\natexlab{a}})}]{SM21}%
  \BibitemOpen
  \href@noop {} {}\bibinfo {note} {See the details of the
  molecular dynamics simulations in Supplementary Material Note 2 and
  subsection Note 2.1, which includes
  Refs.\cite{plimpton1995fast,jewett2020moltemplate,Berendsen.1981,Berendsen.1987,mark2001structure,nose1984molecular,joung2008determination,joung2009molecular}}\BibitemShut
  {NoStop}%
\bibitem [{\citenamefont {Torrey}(1953)}]{torrey1953nuclear}%
  \BibitemOpen
  \bibfield  {author} {\bibinfo {author} {\bibfnamefont {H.~C.}\ \bibnamefont
  {Torrey}},\ }\href@noop {} {\bibfield  {journal} {\bibinfo  {journal} {Phys.
  Rev.}\ }\textbf {\bibinfo {volume} {92}},\ \bibinfo {pages} {962} (\bibinfo
  {year} {1953})}\BibitemShut {NoStop}%
\bibitem [{\citenamefont {Faux}\ \emph {et~al.}(1986)\citenamefont {Faux},
  \citenamefont {Ross},\ and\ \citenamefont {Sholl}}]{Faux.1986}%
  \BibitemOpen
  \bibfield  {author} {\bibinfo {author} {\bibfnamefont {D.~A.}\ \bibnamefont
  {Faux}}, \bibinfo {author} {\bibfnamefont {D.~K.}\ \bibnamefont {Ross}}, \
  and\ \bibinfo {author} {\bibfnamefont {C.~A.}\ \bibnamefont {Sholl}},\
  }\href@noop {} {\bibfield  {journal} {\bibinfo  {journal} {J. Phys. C: Solid
  State Phys.}\ }\textbf {\bibinfo {volume} {19}},\ \bibinfo {pages} {4115}
  (\bibinfo {year} {1986})}\BibitemShut {NoStop}%
\bibitem [{\citenamefont {Friedman}\ \emph {et~al.}(1979)\citenamefont
  {Friedman}, \citenamefont {Holz},\ and\ \citenamefont
  {Hertz}}]{friedman1979epr}%
  \BibitemOpen
  \bibfield  {author} {\bibinfo {author} {\bibfnamefont {H.}~\bibnamefont
  {Friedman}}, \bibinfo {author} {\bibfnamefont {M.}~\bibnamefont {Holz}}, \
  and\ \bibinfo {author} {\bibfnamefont {H.}~\bibnamefont {Hertz}},\
  }\href@noop {} {\bibfield  {journal} {\bibinfo  {journal} {J. Chem. Phys.}\
  }\textbf {\bibinfo {volume} {70}},\ \bibinfo {pages} {3369} (\bibinfo {year}
  {1979})}\BibitemShut {NoStop}%
\bibitem [{\citenamefont {Abernathy}\ and\ \citenamefont
  {Sharp}(1997)}]{abernathy1997spin}%
  \BibitemOpen
  \bibfield  {author} {\bibinfo {author} {\bibfnamefont {S.~M.}\ \bibnamefont
  {Abernathy}}\ and\ \bibinfo {author} {\bibfnamefont {R.~R.}\ \bibnamefont
  {Sharp}},\ }\href@noop {} {\bibfield  {journal} {\bibinfo  {journal} {J.
  Chem. Phys.}\ }\textbf {\bibinfo {volume} {106}},\ \bibinfo {pages} {9032}
  (\bibinfo {year} {1997})}\BibitemShut {NoStop}%
\bibitem [{\citenamefont {Faux}\ \emph {et~al.}(2015)\citenamefont {Faux},
  \citenamefont {Cachia}, \citenamefont {McDonald}, \citenamefont {Bhatt},
  \citenamefont {Howlett},\ and\ \citenamefont {Churakov}}]{Faux.2015}%
  \BibitemOpen
  \bibfield  {author} {\bibinfo {author} {\bibfnamefont {D.~A.}\ \bibnamefont
  {Faux}}, \bibinfo {author} {\bibfnamefont {S.-H.~P.}\ \bibnamefont {Cachia}},
  \bibinfo {author} {\bibfnamefont {P.~J.}\ \bibnamefont {McDonald}}, \bibinfo
  {author} {\bibfnamefont {J.~S.}\ \bibnamefont {Bhatt}}, \bibinfo {author}
  {\bibfnamefont {N.~C.}\ \bibnamefont {Howlett}}, \ and\ \bibinfo {author}
  {\bibfnamefont {S.~V.}\ \bibnamefont {Churakov}},\ }\href@noop {} {\bibfield
  {journal} {\bibinfo  {journal} {Phys. Rev. E}\ }\textbf {\bibinfo {volume}
  {91}},\ \bibinfo {pages} {032311} (\bibinfo {year} {2015})}\BibitemShut
  {NoStop}%
\bibitem [{\citenamefont {Meledandri}\ and\ \citenamefont
  {Brougham}(2012)}]{meledandri2012low}%
  \BibitemOpen
  \bibfield  {author} {\bibinfo {author} {\bibfnamefont {C.~J.}\ \bibnamefont
  {Meledandri}}\ and\ \bibinfo {author} {\bibfnamefont {D.~F.}\ \bibnamefont
  {Brougham}},\ }\href@noop {} {\bibfield  {journal} {\bibinfo  {journal}
  {Anal. Methods}\ }\textbf {\bibinfo {volume} {4}},\ \bibinfo {pages} {331}
  (\bibinfo {year} {2012})}\BibitemShut {NoStop}%
\bibitem [{\citenamefont {Ducommun}\ \emph {et~al.}(1980)\citenamefont
  {Ducommun}, \citenamefont {Newman},\ and\ \citenamefont
  {Merbach}}]{ducommun1980high}%
  \BibitemOpen
  \bibfield  {author} {\bibinfo {author} {\bibfnamefont {Y.}~\bibnamefont
  {Ducommun}}, \bibinfo {author} {\bibfnamefont {K.~E.}\ \bibnamefont
  {Newman}}, \ and\ \bibinfo {author} {\bibfnamefont {A.~E.}\ \bibnamefont
  {Merbach}},\ }\href@noop {} {\bibfield  {journal} {\bibinfo  {journal}
  {Inorg. Chem.}\ }\textbf {\bibinfo {volume} {19}},\ \bibinfo {pages} {3696}
  (\bibinfo {year} {1980})}\BibitemShut {NoStop}%
\bibitem [{\citenamefont {Swaddle}\ and\ \citenamefont
  {Merbach}(1981)}]{swaddle1981high}%
  \BibitemOpen
  \bibfield  {author} {\bibinfo {author} {\bibfnamefont {T.~W.}\ \bibnamefont
  {Swaddle}}\ and\ \bibinfo {author} {\bibfnamefont {A.~E.}\ \bibnamefont
  {Merbach}},\ }\href@noop {} {\bibfield  {journal} {\bibinfo  {journal}
  {Inorg. Chem.}\ }\textbf {\bibinfo {volume} {20}},\ \bibinfo {pages} {4212}
  (\bibinfo {year} {1981})}\BibitemShut {NoStop}%
\bibitem [{\citenamefont {Grant}\ and\ \citenamefont
  {Jordan}(1981)}]{grant1981kinetics}%
  \BibitemOpen
  \bibfield  {author} {\bibinfo {author} {\bibfnamefont {M.}~\bibnamefont
  {Grant}}\ and\ \bibinfo {author} {\bibfnamefont {R.}~\bibnamefont {Jordan}},\
  }\href@noop {} {\bibfield  {journal} {\bibinfo  {journal} {Inorg. Chem.}\
  }\textbf {\bibinfo {volume} {20}},\ \bibinfo {pages} {55} (\bibinfo {year}
  {1981})}\BibitemShut {NoStop}%
\bibitem [{SM2({\natexlab{b}})}]{SM22}%
  \BibitemOpen
  \href@noop {} {}\bibinfo {note} {See the details of the
  molecular dynamics simulations in Supplementary Material Note 2 and
  subsection Note 2.2, which includes
  Refs.\cite{plimpton1995fast,jewett2020moltemplate,Berendsen.1981,Berendsen.1987,mark2001structure,nose1984molecular}}\BibitemShut
  {NoStop}%
\bibitem [{\citenamefont {Ohtaki}\ \emph {et~al.}(1976)\citenamefont {Ohtaki},
  \citenamefont {Yamaguchi},\ and\ \citenamefont {Maeda}}]{ohtaki1976x}%
  \BibitemOpen
  \bibfield  {author} {\bibinfo {author} {\bibfnamefont {H.}~\bibnamefont
  {Ohtaki}}, \bibinfo {author} {\bibfnamefont {T.}~\bibnamefont {Yamaguchi}}, \
  and\ \bibinfo {author} {\bibfnamefont {M.}~\bibnamefont {Maeda}},\
  }\href@noop {} {\bibfield  {journal} {\bibinfo  {journal} {Bulletin of the
  Chemical Society of Japan}\ }\textbf {\bibinfo {volume} {49}},\ \bibinfo
  {pages} {701} (\bibinfo {year} {1976})}\BibitemShut {NoStop}%
\bibitem [{\citenamefont {Magini}(1981)}]{magini1981hydration}%
  \BibitemOpen
  \bibfield  {author} {\bibinfo {author} {\bibfnamefont {M.}~\bibnamefont
  {Magini}},\ }\href@noop {} {\bibfield  {journal} {\bibinfo  {journal} {J.
  Chem. Phys.}\ }\textbf {\bibinfo {volume} {74}},\ \bibinfo {pages} {2523}
  (\bibinfo {year} {1981})}\BibitemShut {NoStop}%
\bibitem [{\citenamefont {Caminiti}\ \emph {et~al.}(1982)\citenamefont
  {Caminiti}, \citenamefont {Marongiu},\ and\ \citenamefont
  {Paschina}}]{caminiti1982comparative}%
  \BibitemOpen
  \bibfield  {author} {\bibinfo {author} {\bibfnamefont {R.}~\bibnamefont
  {Caminiti}}, \bibinfo {author} {\bibfnamefont {G.}~\bibnamefont {Marongiu}},
  \ and\ \bibinfo {author} {\bibfnamefont {G.}~\bibnamefont {Paschina}},\
  }\href@noop {} {\bibfield  {journal} {\bibinfo  {journal} {Z. Naturforsch.
  A}\ }\textbf {\bibinfo {volume} {37}},\ \bibinfo {pages} {581} (\bibinfo
  {year} {1982})}\BibitemShut {NoStop}%
\bibitem [{\citenamefont {Smirnov}\ and\ \citenamefont
  {Grechin}(2019)}]{smirnov2019coordination}%
  \BibitemOpen
  \bibfield  {author} {\bibinfo {author} {\bibfnamefont {P.}~\bibnamefont
  {Smirnov}}\ and\ \bibinfo {author} {\bibfnamefont {O.}~\bibnamefont
  {Grechin}},\ }\href@noop {} {\bibfield  {journal} {\bibinfo  {journal} {Russ.
  J. Phys. Chem. A}\ }\textbf {\bibinfo {volume} {93}},\ \bibinfo {pages}
  {2213} (\bibinfo {year} {2019})}\BibitemShut {NoStop}%
\bibitem [{\citenamefont {Faux}\ \emph {et~al.}(2019)\citenamefont {Faux},
  \citenamefont {Kogon}, \citenamefont {Bortolotti},\ and\ \citenamefont
  {McDonald}}]{faux2019advances}%
  \BibitemOpen
  \bibfield  {author} {\bibinfo {author} {\bibfnamefont {D.}~\bibnamefont
  {Faux}}, \bibinfo {author} {\bibfnamefont {R.}~\bibnamefont {Kogon}},
  \bibinfo {author} {\bibfnamefont {V.}~\bibnamefont {Bortolotti}}, \ and\
  \bibinfo {author} {\bibfnamefont {P.}~\bibnamefont {McDonald}},\ }\href@noop
  {} {\bibfield  {journal} {\bibinfo  {journal} {Molecules}\ }\textbf {\bibinfo
  {volume} {24}},\ \bibinfo {pages} {3688} (\bibinfo {year}
  {2019})}\BibitemShut {NoStop}%
\bibitem [{SM3()}]{SM3}%
  \BibitemOpen
  \href@noop {} {}\bibinfo {note} {See Supplementary Material Note 3 for a
  model that describes the rotation of a sphere due to collisions of water
  molecules, which includes Ref. \cite{faux2021a}}\BibitemShut {NoStop}%
\bibitem [{SM4()}]{SM4}%
  \BibitemOpen
  \href@noop {} {}\bibinfo {note} {See Supplementary Material Note 4 for a
  discussion of the interpretation of the inner sphere volume fraction $x$ in
  the context of Mn-Fit1 results.}\BibitemShut {Stop}%
\bibitem [{\citenamefont {Bertini}\ \emph
  {et~al.}(1993{\natexlab{a}})\citenamefont {Bertini}, \citenamefont {Capozzi},
  \citenamefont {Luchinat}, \citenamefont {Nicastro},\ and\ \citenamefont
  {Xia}}]{bertini1993water}%
  \BibitemOpen
  \bibfield  {author} {\bibinfo {author} {\bibfnamefont {I.}~\bibnamefont
  {Bertini}}, \bibinfo {author} {\bibfnamefont {F.}~\bibnamefont {Capozzi}},
  \bibinfo {author} {\bibfnamefont {C.}~\bibnamefont {Luchinat}}, \bibinfo
  {author} {\bibfnamefont {G.}~\bibnamefont {Nicastro}}, \ and\ \bibinfo
  {author} {\bibfnamefont {Z.}~\bibnamefont {Xia}},\ }\href@noop {} {\bibfield
  {journal} {\bibinfo  {journal} {J. Phys. Chem.}\ }\textbf {\bibinfo {volume}
  {97}},\ \bibinfo {pages} {6351} (\bibinfo {year}
  {1993}{\natexlab{a}})}\BibitemShut {NoStop}%
\bibitem [{\citenamefont {Bertini}\ \emph
  {et~al.}(1993{\natexlab{b}})\citenamefont {Bertini}, \citenamefont {Capozzi},
  \citenamefont {Luchinat},\ and\ \citenamefont {Xia}}]{bertini1993nuclear}%
  \BibitemOpen
  \bibfield  {author} {\bibinfo {author} {\bibfnamefont {I.}~\bibnamefont
  {Bertini}}, \bibinfo {author} {\bibfnamefont {F.}~\bibnamefont {Capozzi}},
  \bibinfo {author} {\bibfnamefont {C.}~\bibnamefont {Luchinat}}, \ and\
  \bibinfo {author} {\bibfnamefont {Z.}~\bibnamefont {Xia}},\ }\href@noop {}
  {\bibfield  {journal} {\bibinfo  {journal} {J. Phys. Chem.}\ }\textbf
  {\bibinfo {volume} {97}},\ \bibinfo {pages} {1134} (\bibinfo {year}
  {1993}{\natexlab{b}})}\BibitemShut {NoStop}%
\bibitem [{\citenamefont {Zhang}\ \emph {et~al.}(2015)\citenamefont {Zhang},
  \citenamefont {Waychunas},\ and\ \citenamefont
  {Banfield}}]{zhang2015molecular}%
  \BibitemOpen
  \bibfield  {author} {\bibinfo {author} {\bibfnamefont {H.}~\bibnamefont
  {Zhang}}, \bibinfo {author} {\bibfnamefont {G.~A.}\ \bibnamefont
  {Waychunas}}, \ and\ \bibinfo {author} {\bibfnamefont {J.~F.}\ \bibnamefont
  {Banfield}},\ }\href@noop {} {\bibfield  {journal} {\bibinfo  {journal} {J.
  Phys. Chem. B}\ }\textbf {\bibinfo {volume} {119}},\ \bibinfo {pages} {10630}
  (\bibinfo {year} {2015})}\BibitemShut {NoStop}%
\bibitem [{\citenamefont {Magini}\ and\ \citenamefont
  {Radnai}(1979)}]{magini1979x}%
  \BibitemOpen
  \bibfield  {author} {\bibinfo {author} {\bibfnamefont {M.}~\bibnamefont
  {Magini}}\ and\ \bibinfo {author} {\bibfnamefont {T.}~\bibnamefont
  {Radnai}},\ }\href@noop {} {\bibfield  {journal} {\bibinfo  {journal} {J.
  Chem. Phys.}\ }\textbf {\bibinfo {volume} {71}},\ \bibinfo {pages} {4255}
  (\bibinfo {year} {1979})}\BibitemShut {NoStop}%
\bibitem [{\citenamefont {Hausser}\ and\ \citenamefont
  {Noack}(1964)}]{hausser1964}%
  \BibitemOpen
  \bibfield  {author} {\bibinfo {author} {\bibfnamefont {R.}~\bibnamefont
  {Hausser}}\ and\ \bibinfo {author} {\bibfnamefont {F.}~\bibnamefont
  {Noack}},\ }\href@noop {} {\bibfield  {journal} {\bibinfo  {journal} {Z.
  Phys.}\ }\textbf {\bibinfo {volume} {182}},\ \bibinfo {pages} {93} (\bibinfo
  {year} {1964})}\BibitemShut {NoStop}%
\bibitem [{SM5()}]{SM5}%
  \BibitemOpen
  \href@noop {} {}\bibinfo {note} {See Supplementary Material Note 5 for
  justification that Cu(II) chloride forms an inner sphere hexahydrate at the
  chloride concentrations used in the present work and an estimate of the range
  of Cu(II)--$^1$H distances seen in Cu(II) chloride. Note 5 includes Refs.
  \cite{bell1973solute,magini1981hydration,dangelo1997structural,khan1976stability,bjerrum1986weak,ramette1983copper}}\BibitemShut
  {NoStop}%
\bibitem [{\citenamefont {Plimpton}(1995)}]{plimpton1995fast}%
  \BibitemOpen
  \bibfield  {author} {\bibinfo {author} {\bibfnamefont {S.}~\bibnamefont
  {Plimpton}},\ }\href@noop {} {\bibfield  {journal} {\bibinfo  {journal} {J.
  Comput. Phys.}\ }\textbf {\bibinfo {volume} {117}},\ \bibinfo {pages} {1}
  (\bibinfo {year} {1995})}\BibitemShut {NoStop}%
\bibitem [{\citenamefont {Jewett}(2020)}]{jewett2020moltemplate}%
  \BibitemOpen
  \bibfield  {author} {\bibinfo {author} {\bibfnamefont {A.}~\bibnamefont
  {Jewett}},\ }\href@noop {} {\emph {\bibinfo {title} {Moltemplate manual}}}\
  (\bibinfo  {publisher} {University of California, Santa Barbara Shea Lab},\
  \bibinfo {year} {2020})\BibitemShut {NoStop}%
\bibitem [{\citenamefont {Berendsen}\ \emph {et~al.}(1981)\citenamefont
  {Berendsen}, \citenamefont {Postma}, \citenamefont {Van~Gunsteren},\ and\
  \citenamefont {Hermans}}]{Berendsen.1981}%
  \BibitemOpen
  \bibfield  {author} {\bibinfo {author} {\bibfnamefont {H.}~\bibnamefont
  {Berendsen}}, \bibinfo {author} {\bibfnamefont {J.}~\bibnamefont {Postma}},
  \bibinfo {author} {\bibfnamefont {W.}~\bibnamefont {Van~Gunsteren}}, \ and\
  \bibinfo {author} {\bibfnamefont {J.}~\bibnamefont {Hermans}},\ }\href@noop
  {} {\emph {\bibinfo {title} {Interaction models for water in relation to
  protein hydration}}}\ (\bibinfo  {publisher} {Springer},\ \bibinfo {year}
  {1981})\ p.\ \bibinfo {pages} {331}\BibitemShut {NoStop}%
\bibitem [{\citenamefont {Berendsen}\ \emph {et~al.}(1987)\citenamefont
  {Berendsen}, \citenamefont {Grigera},\ and\ \citenamefont
  {Straatsma}}]{Berendsen.1987}%
  \BibitemOpen
  \bibfield  {author} {\bibinfo {author} {\bibfnamefont {H.~J.~C.}\
  \bibnamefont {Berendsen}}, \bibinfo {author} {\bibfnamefont {J.~R.}\
  \bibnamefont {Grigera}}, \ and\ \bibinfo {author} {\bibfnamefont {T.~P.}\
  \bibnamefont {Straatsma}},\ }\href@noop {} {\bibfield  {journal} {\bibinfo
  {journal} {J.Phys. Chem.}\ }\textbf {\bibinfo {volume} {91}},\ \bibinfo
  {pages} {6269} (\bibinfo {year} {1987})}\BibitemShut {NoStop}%
\bibitem [{\citenamefont {Mark}\ and\ \citenamefont
  {Nilsson}(2001)}]{mark2001structure}%
  \BibitemOpen
  \bibfield  {author} {\bibinfo {author} {\bibfnamefont {P.}~\bibnamefont
  {Mark}}\ and\ \bibinfo {author} {\bibfnamefont {L.}~\bibnamefont {Nilsson}},\
  }\href@noop {} {\bibfield  {journal} {\bibinfo  {journal} {J. Phys. Chem. A}\
  }\textbf {\bibinfo {volume} {105}},\ \bibinfo {pages} {9954} (\bibinfo {year}
  {2001})}\BibitemShut {NoStop}%
\bibitem [{\citenamefont {Nos{\'e}}(1984)}]{nose1984molecular}%
  \BibitemOpen
  \bibfield  {author} {\bibinfo {author} {\bibfnamefont {S.}~\bibnamefont
  {Nos{\'e}}},\ }\href@noop {} {\bibfield  {journal} {\bibinfo  {journal} {Mol.
  Phys.}\ }\textbf {\bibinfo {volume} {52}},\ \bibinfo {pages} {255} (\bibinfo
  {year} {1984})}\BibitemShut {NoStop}%
\bibitem [{\citenamefont {Joung}\ and\ \citenamefont
  {Cheatham~III}(2008)}]{joung2008determination}%
  \BibitemOpen
  \bibfield  {author} {\bibinfo {author} {\bibfnamefont {I.~S.}\ \bibnamefont
  {Joung}}\ and\ \bibinfo {author} {\bibfnamefont {T.~E.}\ \bibnamefont
  {Cheatham~III}},\ }\href@noop {} {\bibfield  {journal} {\bibinfo  {journal}
  {J. Phys. Chem. B}\ }\textbf {\bibinfo {volume} {112}},\ \bibinfo {pages}
  {9020} (\bibinfo {year} {2008})}\BibitemShut {NoStop}%
\bibitem [{\citenamefont {Joung}\ and\ \citenamefont
  {Cheatham~III}(2009)}]{joung2009molecular}%
  \BibitemOpen
  \bibfield  {author} {\bibinfo {author} {\bibfnamefont {I.~S.}\ \bibnamefont
  {Joung}}\ and\ \bibinfo {author} {\bibfnamefont {T.~E.}\ \bibnamefont
  {Cheatham~III}},\ }\href@noop {} {\bibfield  {journal} {\bibinfo  {journal}
  {J. Phys. Chem. B}\ }\textbf {\bibinfo {volume} {113}},\ \bibinfo {pages}
  {13279} (\bibinfo {year} {2009})}\BibitemShut {NoStop}%
\bibitem [{\citenamefont {Bell}\ \emph {et~al.}(1973)\citenamefont {Bell},
  \citenamefont {Tyvoll},\ and\ \citenamefont {Wertz}}]{bell1973solute}%
  \BibitemOpen
  \bibfield  {author} {\bibinfo {author} {\bibfnamefont {J.~R.}\ \bibnamefont
  {Bell}}, \bibinfo {author} {\bibfnamefont {J.~L.}\ \bibnamefont {Tyvoll}}, \
  and\ \bibinfo {author} {\bibfnamefont {D.~L.}\ \bibnamefont {Wertz}},\
  }\href@noop {} {\bibfield  {journal} {\bibinfo  {journal} {J. Am. Chem.
  Soc.}\ }\textbf {\bibinfo {volume} {95}},\ \bibinfo {pages} {1456} (\bibinfo
  {year} {1973})}\BibitemShut {NoStop}%
\bibitem [{\citenamefont {D’angelo}\ \emph {et~al.}(1997)\citenamefont
  {D’angelo}, \citenamefont {Bottari}, \citenamefont {Festa}, \citenamefont
  {Nolting},\ and\ \citenamefont {Pavel}}]{dangelo1997structural}%
  \BibitemOpen
  \bibfield  {author} {\bibinfo {author} {\bibfnamefont {P.}~\bibnamefont
  {D’angelo}}, \bibinfo {author} {\bibfnamefont {E.}~\bibnamefont {Bottari}},
  \bibinfo {author} {\bibfnamefont {M.}~\bibnamefont {Festa}}, \bibinfo
  {author} {\bibfnamefont {H.-F.}\ \bibnamefont {Nolting}}, \ and\ \bibinfo
  {author} {\bibfnamefont {N.}~\bibnamefont {Pavel}},\ }\href@noop {}
  {\bibfield  {journal} {\bibinfo  {journal} {J. Chem. Phys.}\ }\textbf
  {\bibinfo {volume} {107}},\ \bibinfo {pages} {2807} (\bibinfo {year}
  {1997})}\BibitemShut {NoStop}%
\bibitem [{\citenamefont {Khan}\ and\ \citenamefont
  {Schwing-Weill}(1976)}]{khan1976stability}%
  \BibitemOpen
  \bibfield  {author} {\bibinfo {author} {\bibfnamefont {M.}~\bibnamefont
  {Khan}}\ and\ \bibinfo {author} {\bibfnamefont {M.}~\bibnamefont
  {Schwing-Weill}},\ }\href@noop {} {\bibfield  {journal} {\bibinfo  {journal}
  {Inorg. Chem.}\ }\textbf {\bibinfo {volume} {15}},\ \bibinfo {pages} {2202}
  (\bibinfo {year} {1976})}\BibitemShut {NoStop}%
\bibitem [{\citenamefont {Bjerrum}\ and\ \citenamefont
  {Skibsted}(1986)}]{bjerrum1986weak}%
  \BibitemOpen
  \bibfield  {author} {\bibinfo {author} {\bibfnamefont {J.}~\bibnamefont
  {Bjerrum}}\ and\ \bibinfo {author} {\bibfnamefont {L.}~\bibnamefont
  {Skibsted}},\ }\href@noop {} {\bibfield  {journal} {\bibinfo  {journal}
  {Inorg. Chem.}\ }\textbf {\bibinfo {volume} {25}},\ \bibinfo {pages} {2479}
  (\bibinfo {year} {1986})}\BibitemShut {NoStop}%
\bibitem [{\citenamefont {Ramette}\ and\ \citenamefont
  {Fan}(1983)}]{ramette1983copper}%
  \BibitemOpen
  \bibfield  {author} {\bibinfo {author} {\bibfnamefont {R.~W.}\ \bibnamefont
  {Ramette}}\ and\ \bibinfo {author} {\bibfnamefont {G.}~\bibnamefont {Fan}},\
  }\href@noop {} {\bibfield  {journal} {\bibinfo  {journal} {Inorg. Chem.}\
  }\textbf {\bibinfo {volume} {22}},\ \bibinfo {pages} {3323} (\bibinfo {year}
  {1983})}\BibitemShut {NoStop}%
\end{thebibliography}%


\begin{thebibliography}{19}%
\makeatletter
\providecommand \@ifxundefined [1]{%
 \@ifx{#1\undefined}
}%
\providecommand \@ifnum [1]{%
 \ifnum #1\expandafter \@firstoftwo
 \else \expandafter \@secondoftwo
 \fi
}%
\providecommand \@ifx [1]{%
 \ifx #1\expandafter \@firstoftwo
 \else \expandafter \@secondoftwo
 \fi
}%
\providecommand \natexlab [1]{#1}%
\providecommand \enquote  [1]{``#1''}%
\providecommand \bibnamefont  [1]{#1}%
\providecommand \bibfnamefont [1]{#1}%
\providecommand \citenamefont [1]{#1}%
\providecommand \href@noop [0]{\@secondoftwo}%
\providecommand \href [0]{\begingroup \@sanitize@url \@href}%
\providecommand \@href[1]{\@@startlink{#1}\@@href}%
\providecommand \@@href[1]{\endgroup#1\@@endlink}%
\providecommand \@sanitize@url [0]{\catcode `\\12\catcode `\$12\catcode
  `\&12\catcode `\#12\catcode `\^12\catcode `\_12\catcode `\%12\relax}%
\providecommand \@@startlink[1]{}%
\providecommand \@@endlink[0]{}%
\providecommand \url  [0]{\begingroup\@sanitize@url \@url }%
\providecommand \@url [1]{\endgroup\@href {#1}{\urlprefix }}%
\providecommand \urlprefix  [0]{URL }%
\providecommand \Eprint [0]{\href }%
\providecommand \doibase [0]{http://dx.doi.org/}%
\providecommand \selectlanguage [0]{\@gobble}%
\providecommand \bibinfo  [0]{\@secondoftwo}%
\providecommand \bibfield  [0]{\@secondoftwo}%
\providecommand \translation [1]{[#1]}%
\providecommand \BibitemOpen [0]{}%
\providecommand \bibitemStop [0]{}%
\providecommand \bibitemNoStop [0]{.\EOS\space}%
\providecommand \EOS [0]{\spacefactor3000\relax}%
\providecommand \BibitemShut  [1]{\csname bibitem#1\endcsname}%
\let\auto@bib@innerbib\@empty
\bibitem [{\citenamefont {Messiah}(1965)}]{Messiah.1965}%
  \BibitemOpen
  \bibfield  {author} {\bibinfo {author} {\bibfnamefont {A.}~\bibnamefont
  {Messiah}},\ }\href@noop {} {\emph {\bibinfo {title} {Quantum Mechanics}}}\
  (\bibinfo  {publisher} {North Holland Press},\ \bibinfo {year}
  {1965})\BibitemShut {NoStop}%
\bibitem [{\citenamefont {Sholl}(1974)}]{Sholl1974nuclear}%
  \BibitemOpen
  \bibfield  {author} {\bibinfo {author} {\bibfnamefont {C.}~\bibnamefont
  {Sholl}},\ }\href@noop {} {\bibfield  {journal} {\bibinfo  {journal} {J.
  Phys. C: Solid State Phys.}\ }\textbf {\bibinfo {volume} {7}},\ \bibinfo
  {pages} {3378} (\bibinfo {year} {1974})}\BibitemShut {NoStop}%
\bibitem [{\citenamefont {Faux}\ \emph {et~al.}(2017)\citenamefont {Faux},
  \citenamefont {McDonald},\ and\ \citenamefont {Howlett}}]{Faux.2017a}%
  \BibitemOpen
  \bibfield  {author} {\bibinfo {author} {\bibfnamefont {D.~A.}\ \bibnamefont
  {Faux}}, \bibinfo {author} {\bibfnamefont {P.~J.}\ \bibnamefont {McDonald}},
  \ and\ \bibinfo {author} {\bibfnamefont {N.~C.}\ \bibnamefont {Howlett}},\
  }\href@noop {} {\bibfield  {journal} {\bibinfo  {journal} {Phys. Rev. E}\
  }\textbf {\bibinfo {volume} {95}},\ \bibinfo {pages} {033116} (\bibinfo
  {year} {2017})}\BibitemShut {NoStop}%
\bibitem [{\citenamefont {Faux}\ \emph {et~al.}(2021)\citenamefont {Faux},
  \citenamefont {Rahaman},\ and\ \citenamefont {McDonald}}]{faux2021a}%
  \BibitemOpen
  \bibfield  {author} {\bibinfo {author} {\bibfnamefont {D.~A.}\ \bibnamefont
  {Faux}}, \bibinfo {author} {\bibfnamefont {A.~A.}\ \bibnamefont {Rahaman}}, \
  and\ \bibinfo {author} {\bibfnamefont {P.~J.}\ \bibnamefont {McDonald}},\
  }\href@noop {} {\bibfield  {journal} {\bibinfo  {journal} {Phys. Rev. Lett.}\
  }\textbf {\bibinfo {volume} {127}},\ \bibinfo {pages} {256001} (\bibinfo
  {year} {2021})}\BibitemShut {NoStop}%
\bibitem [{\citenamefont {Plimpton}(1995)}]{plimpton1995fast}%
  \BibitemOpen
  \bibfield  {author} {\bibinfo {author} {\bibfnamefont {S.}~\bibnamefont
  {Plimpton}},\ }\href@noop {} {\bibfield  {journal} {\bibinfo  {journal} {J.
  Comput. Phys.}\ }\textbf {\bibinfo {volume} {117}},\ \bibinfo {pages} {1}
  (\bibinfo {year} {1995})}\BibitemShut {NoStop}%
\bibitem [{\citenamefont {Jewett}(2020)}]{jewett2020moltemplate}%
  \BibitemOpen
  \bibfield  {author} {\bibinfo {author} {\bibfnamefont {A.}~\bibnamefont
  {Jewett}},\ }\href@noop {} {\emph {\bibinfo {title} {Moltemplate manual}}}\
  (\bibinfo  {publisher} {University of California, Santa Barbara Shea Lab},\
  \bibinfo {year} {2020})\BibitemShut {NoStop}%
\bibitem [{\citenamefont {Berendsen}\ \emph {et~al.}(1981)\citenamefont
  {Berendsen}, \citenamefont {Postma}, \citenamefont {Van~Gunsteren},\ and\
  \citenamefont {Hermans}}]{Berendsen.1981}%
  \BibitemOpen
  \bibfield  {author} {\bibinfo {author} {\bibfnamefont {H.}~\bibnamefont
  {Berendsen}}, \bibinfo {author} {\bibfnamefont {J.}~\bibnamefont {Postma}},
  \bibinfo {author} {\bibfnamefont {W.}~\bibnamefont {Van~Gunsteren}}, \ and\
  \bibinfo {author} {\bibfnamefont {J.}~\bibnamefont {Hermans}},\ }\href@noop
  {} {\emph {\bibinfo {title} {Interaction models for water in relation to
  protein hydration}}}\ (\bibinfo  {publisher} {Springer},\ \bibinfo {year}
  {1981})\ p.\ \bibinfo {pages} {331}\BibitemShut {NoStop}%
\bibitem [{\citenamefont {Berendsen}\ \emph {et~al.}(1987)\citenamefont
  {Berendsen}, \citenamefont {Grigera},\ and\ \citenamefont
  {Straatsma}}]{Berendsen.1987}%
  \BibitemOpen
  \bibfield  {author} {\bibinfo {author} {\bibfnamefont {H.~J.~C.}\
  \bibnamefont {Berendsen}}, \bibinfo {author} {\bibfnamefont {J.~R.}\
  \bibnamefont {Grigera}}, \ and\ \bibinfo {author} {\bibfnamefont {T.~P.}\
  \bibnamefont {Straatsma}},\ }\href@noop {} {\bibfield  {journal} {\bibinfo
  {journal} {J.Phys. Chem.}\ }\textbf {\bibinfo {volume} {91}},\ \bibinfo
  {pages} {6269} (\bibinfo {year} {1987})}\BibitemShut {NoStop}%
\bibitem [{\citenamefont {Mark}\ and\ \citenamefont
  {Nilsson}(2001)}]{mark2001structure}%
  \BibitemOpen
  \bibfield  {author} {\bibinfo {author} {\bibfnamefont {P.}~\bibnamefont
  {Mark}}\ and\ \bibinfo {author} {\bibfnamefont {L.}~\bibnamefont {Nilsson}},\
  }\href@noop {} {\bibfield  {journal} {\bibinfo  {journal} {J. Phys. Chem. A}\
  }\textbf {\bibinfo {volume} {105}},\ \bibinfo {pages} {9954} (\bibinfo {year}
  {2001})}\BibitemShut {NoStop}%
\bibitem [{\citenamefont {Nos{\'e}}(1984)}]{nose1984molecular}%
  \BibitemOpen
  \bibfield  {author} {\bibinfo {author} {\bibfnamefont {S.}~\bibnamefont
  {Nos{\'e}}},\ }\href@noop {} {\bibfield  {journal} {\bibinfo  {journal} {Mol.
  Phys.}\ }\textbf {\bibinfo {volume} {52}},\ \bibinfo {pages} {255} (\bibinfo
  {year} {1984})}\BibitemShut {NoStop}%
\bibitem [{\citenamefont {Joung}\ and\ \citenamefont
  {Cheatham~III}(2008)}]{joung2008determination}%
  \BibitemOpen
  \bibfield  {author} {\bibinfo {author} {\bibfnamefont {I.~S.}\ \bibnamefont
  {Joung}}\ and\ \bibinfo {author} {\bibfnamefont {T.~E.}\ \bibnamefont
  {Cheatham~III}},\ }\href@noop {} {\bibfield  {journal} {\bibinfo  {journal}
  {J. Phys. Chem. B}\ }\textbf {\bibinfo {volume} {112}},\ \bibinfo {pages}
  {9020} (\bibinfo {year} {2008})}\BibitemShut {NoStop}%
\bibitem [{\citenamefont {Joung}\ and\ \citenamefont
  {Cheatham~III}(2009)}]{joung2009molecular}%
  \BibitemOpen
  \bibfield  {author} {\bibinfo {author} {\bibfnamefont {I.~S.}\ \bibnamefont
  {Joung}}\ and\ \bibinfo {author} {\bibfnamefont {T.~E.}\ \bibnamefont
  {Cheatham~III}},\ }\href@noop {} {\bibfield  {journal} {\bibinfo  {journal}
  {J. Phys. Chem. B}\ }\textbf {\bibinfo {volume} {113}},\ \bibinfo {pages}
  {13279} (\bibinfo {year} {2009})}\BibitemShut {NoStop}%
\bibitem [{\citenamefont {Zhang}\ \emph {et~al.}(2021)\citenamefont {Zhang},
  \citenamefont {Tang}, \citenamefont {Ma}, \citenamefont {Liang},
  \citenamefont {Zeng},\ and\ \citenamefont {Hefter}}]{zhang2021spectroscopic}%
  \BibitemOpen
  \bibfield  {author} {\bibinfo {author} {\bibfnamefont {N.}~\bibnamefont
  {Zhang}}, \bibinfo {author} {\bibfnamefont {J.}~\bibnamefont {Tang}},
  \bibinfo {author} {\bibfnamefont {Y.}~\bibnamefont {Ma}}, \bibinfo {author}
  {\bibfnamefont {M.}~\bibnamefont {Liang}}, \bibinfo {author} {\bibfnamefont
  {D.}~\bibnamefont {Zeng}}, \ and\ \bibinfo {author} {\bibfnamefont
  {G.}~\bibnamefont {Hefter}},\ }\href@noop {} {\bibfield  {journal} {\bibinfo
  {journal} {Phys. Chem. Chem. Phys.}\ }\textbf {\bibinfo {volume} {23}},\
  \bibinfo {pages} {6807} (\bibinfo {year} {2021})}\BibitemShut {NoStop}%
\bibitem [{\citenamefont {Bell}\ \emph {et~al.}(1973)\citenamefont {Bell},
  \citenamefont {Tyvoll},\ and\ \citenamefont {Wertz}}]{bell1973solute}%
  \BibitemOpen
  \bibfield  {author} {\bibinfo {author} {\bibfnamefont {J.~R.}\ \bibnamefont
  {Bell}}, \bibinfo {author} {\bibfnamefont {J.~L.}\ \bibnamefont {Tyvoll}}, \
  and\ \bibinfo {author} {\bibfnamefont {D.~L.}\ \bibnamefont {Wertz}},\
  }\href@noop {} {\bibfield  {journal} {\bibinfo  {journal} {J. Am. Chem.
  Soc.}\ }\textbf {\bibinfo {volume} {95}},\ \bibinfo {pages} {1456} (\bibinfo
  {year} {1973})}\BibitemShut {NoStop}%
\bibitem [{\citenamefont {Magini}(1981)}]{magini1981hydration}%
  \BibitemOpen
  \bibfield  {author} {\bibinfo {author} {\bibfnamefont {M.}~\bibnamefont
  {Magini}},\ }\href@noop {} {\bibfield  {journal} {\bibinfo  {journal} {J.
  Chem. Phys.}\ }\textbf {\bibinfo {volume} {74}},\ \bibinfo {pages} {2523}
  (\bibinfo {year} {1981})}\BibitemShut {NoStop}%
\bibitem [{\citenamefont {D’angelo}\ \emph {et~al.}(1997)\citenamefont
  {D’angelo}, \citenamefont {Bottari}, \citenamefont {Festa}, \citenamefont
  {Nolting},\ and\ \citenamefont {Pavel}}]{dangelo1997structural}%
  \BibitemOpen
  \bibfield  {author} {\bibinfo {author} {\bibfnamefont {P.}~\bibnamefont
  {D’angelo}}, \bibinfo {author} {\bibfnamefont {E.}~\bibnamefont {Bottari}},
  \bibinfo {author} {\bibfnamefont {M.}~\bibnamefont {Festa}}, \bibinfo
  {author} {\bibfnamefont {H.-F.}\ \bibnamefont {Nolting}}, \ and\ \bibinfo
  {author} {\bibfnamefont {N.}~\bibnamefont {Pavel}},\ }\href@noop {}
  {\bibfield  {journal} {\bibinfo  {journal} {J. Chem. Phys.}\ }\textbf
  {\bibinfo {volume} {107}},\ \bibinfo {pages} {2807} (\bibinfo {year}
  {1997})}\BibitemShut {NoStop}%
\bibitem [{\citenamefont {Khan}\ and\ \citenamefont
  {Schwing-Weill}(1976)}]{khan1976stability}%
  \BibitemOpen
  \bibfield  {author} {\bibinfo {author} {\bibfnamefont {M.}~\bibnamefont
  {Khan}}\ and\ \bibinfo {author} {\bibfnamefont {M.}~\bibnamefont
  {Schwing-Weill}},\ }\href@noop {} {\bibfield  {journal} {\bibinfo  {journal}
  {Inorg. Chem.}\ }\textbf {\bibinfo {volume} {15}},\ \bibinfo {pages} {2202}
  (\bibinfo {year} {1976})}\BibitemShut {NoStop}%
\bibitem [{\citenamefont {Bjerrum}\ and\ \citenamefont
  {Skibsted}(1986)}]{bjerrum1986weak}%
  \BibitemOpen
  \bibfield  {author} {\bibinfo {author} {\bibfnamefont {J.}~\bibnamefont
  {Bjerrum}}\ and\ \bibinfo {author} {\bibfnamefont {L.}~\bibnamefont
  {Skibsted}},\ }\href@noop {} {\bibfield  {journal} {\bibinfo  {journal}
  {Inorg. Chem.}\ }\textbf {\bibinfo {volume} {25}},\ \bibinfo {pages} {2479}
  (\bibinfo {year} {1986})}\BibitemShut {NoStop}%
\bibitem [{\citenamefont {Ramette}\ and\ \citenamefont
  {Fan}(1983)}]{ramette1983copper}%
  \BibitemOpen
  \bibfield  {author} {\bibinfo {author} {\bibfnamefont {R.~W.}\ \bibnamefont
  {Ramette}}\ and\ \bibinfo {author} {\bibfnamefont {G.}~\bibnamefont {Fan}},\
  }\href@noop {} {\bibfield  {journal} {\bibinfo  {journal} {Inorg. Chem.}\
  }\textbf {\bibinfo {volume} {22}},\ \bibinfo {pages} {3323} (\bibinfo {year}
  {1983})}\BibitemShut {NoStop}%
\end{thebibliography}%

\end{document}



\begin{center}
\bf\LARGE{Nuclear spin relaxation in aqueous paramagnetic ion solutions}

{\bf\Large Supplementary material} 
\end{center}

\linenumbers
\renewcommand{\thefigure}{S\arabic{figure}}
\setcounter{figure}{0}

\section{Note 1: the Brownian shell model}

The objective is to calculate the spectral density function $J(\omega)$  for the interaction between an electronic spin ($S$) of a paramagnetic ion and a spherical shell containing a density of $^1$H spins ($I$) surrounding the ion at a radius $a$. All $^1$H spins are fixed at the distance $a$. 

The calculation starts with the general expression for the dipolar correlation function Eq.\,(\ref{S1:Gt_ave})\cite{Messiah.1965,Sholl1974nuclear}
\[
G(t) =  \left< \frac{ P_2( \cos \beta )  }{r_0^3 r^3} \right> \tag{S1} \label{S1:Gt_ave}
\]
where $P_2(x)=\tfrac{1}{2}(3x^2-1)$ is the second-rank Legendre polynomial and $\beta$ is the smallest angle between a specific $S$--$I$ spin-pair vector ${\bf r}_0$ at $t\!=\!0$ and ${\bf r}$ at a later time $t$.   The average is taken over the density of $^1$H spins contained in the shell. $G(t)$ captures all the relevant dynamical information describing how spin-pair vectors move in time. Retaining the generality of Eq.\,(\ref{S1:Gt_ave}) for the time being,  an equivalent expression yields
\[
G(t) =  \int_{\mathbb{R}^3} \!\!  \int_{\mathbb{R}^3_0} \frac{ P_2( \cos \beta )}{r_0^3 \: r^3}  P({\bf r},t  \cap {\bf r}_0)  \: d^3 {\bf r}_0 \, d^3 {\bf r} \tag{S2} \label{S2:Gt_integral}
\]
where $P({\bf r},t \cap \:{\bf r}_0)$ is the probability density function describing the probability distribution of pairs of spins separated by  ${\bf r}_0$ at $t=0$ {\em and} by ${\bf r}$ at time $t$. The subscript 0 on all quantities indicates the value at $t=0$. $P({\bf r},t \:\cap \:{\bf r}_0)$ may be expanded as
\[
P({\bf r},t \:\cap \, {\bf r}_0)  =  P({\bf r}_0) P({\bf r},t \: | \: {\bf r}_0) \tag{S3}  \label{S3:Pprod}
\]
where $P({\bf r},t \: | \: {\bf r}_0)$ is the time-dependent conditional probability density function describing a spin pair separated by ${\bf r}$ at time $t$ {\em given} that the same pair were separated by ${\bf r}_0$ at $t=0$. $P({\bf r}_0)$ is the {\it a priori} probability density describing the probability per unit volume of finding a spin pair, equivalent to the reciprocal of the ``volume per paramagnetic spin" equal to $N_M$, the number of paramagnetic ions per unit volume.

A shell model demands that the  $S\!-\!I$ distance is constant so that $|{\bf r}| \!= \!|{\bf r}_0|\! = \!a$, and  $P( r_0) \!= N_M$ at $r_0\!=\!a$ and zero otherwise.  Since $\beta=0$ at $t=0$, both $\mathbb{R}^3_0$ and $\mathbb{R}^3$ integrals can be executed, save for the integral involving $\beta$ yielding
\[
\!\!\!\!\!\!G(t)  \!=\!   \frac{4  \delta N_M }{a^4} \!\!\!\!\int_0^\pi \!\! P_2( \cos \! \beta)  P(\beta,t)  
\sin  \beta \, d \beta. \tag{S4}   \label{S4:G3App}
\] 
The distance $\delta$ is the effective thickness of the shell at radius $a$ and appears when volume integrals are undertaken over an areal distribution of spin, such as that leading to Eq.\,(\ref{S4:G3App}).  Normally, $\delta \! = \! 0.27$\,nm is assumed to equal the mean distance between water molecules or, equivalently, the thickness of a layer of water at a surface \cite{Faux.2017a}. This choice of $\delta$  is appropriate  for the outer-sphere water which is at the  bulk water density.  For the inner shell however, the value of $\delta$ is determined by considering a volume extending from $r = a \pm \delta/2$ containing $n$ proton spins.  The $\delta$ is chosen so that the inner-shell volume contains spins at the  normal bulk-water density of $N_H=66.6$\,spins/nm$^3$.  Elementary mathematics yields
\[
\delta = \left[ a^2 + \frac{n}{2 \pi a N_H} \right]^{1/2} - a. \tag{S5}   \label{S5:delta}
\]

An expression for $P(\beta,t)$ for an anomalous (L\'{e}vy) rotor has been derived as (see supplementary material of reference \cite{faux2021a}) 
\[
P_L(\beta,t) = N(t)  \left[ 1 +2 \sum_{p=1}^\infty e^{-p^\alpha t/\tau} \cos p\beta  \right]  \tag{S6}  \label{S6:PbetaLevy}
\]
where $\alpha$ is the L\'{e}vy coefficient. The time constant $\tau$ is an rotational diffusion time constant such that $\tau/2$ is the time taken for a rotor to move through an angle of one radian\cite{faux2021a}.   $N(t)$ is a time-dependent normalisation constant which ensures that
\[
\int_0^\pi P_L(\beta,t) \sin \beta \, d\beta = 1 \tag{S7} \label{S7:normalise}
\]
leading to 
\[
N(t)  = \frac{1}{2} \left[ 1 - 2 \!\! \sum_{p=2,4...}^\infty   \!\! \frac{e^{-p^\alpha t/\tau}}{(p^2-1)} \right]^{-1}. \tag{S8} \label{S8:LevyPnorm}
\]
Equation (\ref{S6:PbetaLevy}) was introduced by Faux {\it et al} \cite{faux2021a} to describe the large, sudden changes in angle of the intra-molecular $^1$H--$^1$H vector for water associated with hydrogen-bond breaking and reforming.  The ``fatness" of the tails is determined by the L\'{e}vy coefficient $\alpha$ where $0 \! < \! \alpha \! \leq 2$. The limits $\alpha \! = \! 2$ and $\alpha \! = \! 1$ correspond to Gaussian and Cauchy-Lorentz distributions respectively.

In the present article, the L\'{e}vy coefficient for the ion-$^1$H rotor is approximately 2 corresponding to a Brownian rotational diffuser (see Note 2.1 below).    For a Brownian rotor, 
\[
P(\beta,t) = N(t)  \left[ 1 +2 \sum_{p=1}^\infty e^{-p^2t/\tau} \cos p\beta  \right] \tag{S9} \label{S9:Pbeta}
\]
where the normalisation term $N(t)$ is given by Eq.\,(\ref{S8:LevyPnorm}) with $\alpha \! = \! 2$.
Substituting Eq.\,(\ref{S9:Pbeta}) into Eq.\,(\ref{S4:G3App}) and executing the elementary integrals yields
\[
G(t) = \frac{ 16 \delta N_M N(t)}{a^4} \!\! \sum_{p=2,4...}^\infty \!\! \frac{e^{-p^2 t/\tau} p^2}{(9-p^2) \, (p^2-1)}.   \tag{S10} \label{S10:final}
\]
Finally, combining Eq.\,(\ref{S10:final}) with Eq.\,(\ref{S8:LevyPnorm}) and expanding the exponentials as a series yields the reduced forms for $G(t)$ used in the main text.

\section{Note 2: Molecular Dynamics simulations} \label{SectionMD}

The molecular dynamics (MD) simulation technique requires setting the Cartesian coordinates of all atoms in the system, defining the interatomic potentials between all the combinations of atomic species and letting the system evolve according to Newton's laws of motion.  The MD simulations were performed using the Large-scale Atomic/Molecular Massively Parallel Simulator  (LAMMPS) package \cite{plimpton1995fast} on the High Performance Computing (HPC) facility at the University of Surrey. LAMMPS is an open-source and free MD program from Sandia National Laboratories (USA). 

The MD simulations are used to test the validity of the shell model assumptions through simulations of aqueous NaCl, and to  interpret the results of fits to experimental data from aqueous MnCl$_2$.
Simulations were executed with periodic boundary conditions in a cubic box with side length of about 100\,Å. Software called Moltemplate \cite{jewett2020moltemplate} was used to build the molecule and force-field database system for LAMMPS. Parameters such as mass, charge, coordinates, pair interaction parameters were specified and run in Moltemplate to generate the required database for LAMMPS to read and execute the simulation.

The SPC/E potential \cite{Berendsen.1981,Berendsen.1987} was chosen to describe the water-water interactions. The SPC/E is built from Lennard-Jones interatomic potential functions and Coulombic electrostatic interaction between partial atomic charges.  This potential reproduces the self-diffusion coefficient for water to excellent accuracy \cite{mark2001structure}.

Each set of full simulations was preceded by an equilibration step.  The equilibration procedure was initiated using the NPT ensemble at 300\,K for 100\,ps. This is a statistical mechanical ensemble that keeps the number of particles (N), pressure (P) and temperature (T) constant. The system equilibration was confirmed by the observation of the temperature and pressure fluctuating about constant values. This final configuration was taken as an input to the main simulation which was executed using the NVT ensemble. This is a canonical ensemble which places the system in a heat bath to allow exchange of energy using the Nos\'{e}-Hoover thermostat\cite{nose1984molecular}. The NVT ensemble has constant number of particles (N), volume (V) and temperature (T).

\subsection{Note 2.1: assessing the validity of the Brownian shell model}
\label{results:2.1}
The validity of the Brownian shell model captured by Eq.\,(\ref{S10:final}) is tested through MD simulations of aqueous NaCl solution. The angular probability density function $P(\beta,t)$ derived as Eq.\,(\ref{S6:PbetaLevy}) for the rotational dynamics of the $S\!-\!I$ vector is compared to results for the Na--H vector where H refers to a water proton in the outer sphere of the sodium ion.  

The Lennard-Jones pairwise potential parameters for Na$^+$ and Cl$^-$ are obtained from Joung and Cheatham\cite{joung2008determination} with a charge +1.0 for sodium and combined for two different atoms using the Lorentz-Berthelot combining rules.  The combination with the SPC/E water potential set yields good agreement with experiment for the sodium ion diffusion coefficient\cite{joung2009molecular}.  

The numerical integration was undertaken using a time step of 1\,fs.  Data collection took place at 1\,ps time intervals.  This was sufficient to yield good statistical accuracy for the angular probability density function $P(\beta,t)$ for proton spins in the outer sphere of the sodium ions and a robust estimate of the outer-sphere rotational time constant $\tau_{{\rm o}\beta}$. 

\begin{figure}[tbh!]
	\unitlength1cm
	\begin{center}
		\includegraphics[width=0.7\textwidth, trim={2cm 3cm 2cm 2cm}]{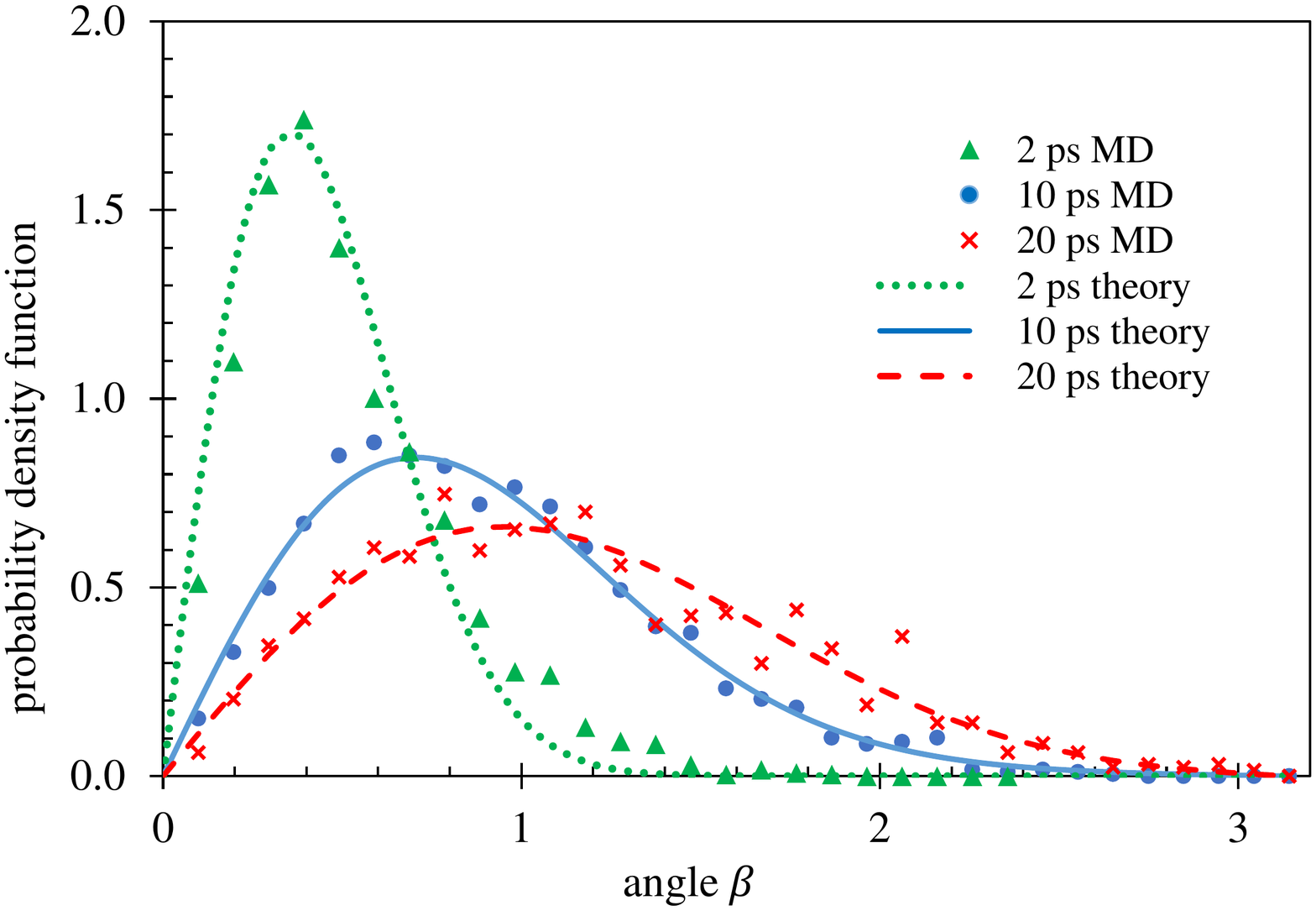}
		\caption{The figure shows the probability density function $ P(\beta,t) \sin \beta$ where $\beta$ is the angle between a Na--H vector at time $t$ and at $t=0$ for outer sphere protons.  The data points are from MD simulations. The theory curves are  obtained from Eq.\,(\ref{S6:PbetaLevy}).  The theory curve at 10\,ps has been fit to the MD data using fit parameters $\alpha$ and $\tau_{{\rm o}\beta}$ (see text).  These fit parameters are used then for the theory curves at 2\,ps and at 20\,ps. }
		\label{Fig_beta}
	\end{center}
\end{figure}

Figure \ref{Fig_beta}  shows the time-dependent probability density function $ P(\beta,t) \sin \beta$ for time intervals $\Delta t$ of 2, 10 and 20\,ps.  The product $ P(\beta,t) \sin \beta$ is normalised to unity when integrated with respect to $\beta$ as Eq.\,(\ref{S7:normalise}). Only Na--H vectors for spins in the outer-sphere (taken as $r\!>\!2\delta \!=\!\,$5.4\,\AA) at both $t\!=\!0$ and at time $t$ are included. The outer-shell L\'{e}vy model of Eq.\,(\ref{S6:PbetaLevy}) is fit to the MD probability density function for $\Delta t \!= \!10$\,ps using the L\'{e}vy parameter $\alpha$ and outer shell rotational time constant $\tau_{{\rm o}\beta}$ as fit parameters. A simple minimisation of the squared difference $Q$ was used to locate the optimum fit.   Fits within 1\% of the optimum $Q$ yield $\alpha = 1.90\pm0.02$ confirming the reasonableness of the Brownian rotational motion ($\alpha=2$) approximation for $S\!-\!I$ spin pairs in aqueous solution.  

The best fit value of  $\tau_{{\rm o}\beta}$ at $\Delta t \!= \!10$\,ps is found to be 30\,ps. The theory curves for $\Delta t \!= \!2$\,ps and 20\,ps in Fig.\,\ref{Fig_beta} are not fits, but use the fit parameters $\alpha \!= \!1.9$ and  $\tau_{{\rm o}\beta}\!=\!30$\,ps obtained from the $\Delta t \!= \!10$\,ps dataset. The match with the equivalent MD simulation data at $\Delta t\! =\! 20$\,ps is impressive establishing the validity of applying these values generally. The small systematic deviations at $\Delta t\! =\! 2$\,ps are expected because, over such a small timescale, sufficient collisions have not taken place to establish Brownian statistics.

\subsection{Note 2.2: aqueous manganese (II) chloride simulations}

The aqueous MnCl$_2$ MD simulations were undertaken using potentials  adapted from the NaCl simulations by changing the mass and charge of the cation.   The system comprised 216 Mn atoms, 432 Cl atoms and 27000 water molecules contained in a box with side lengths approximately 100\AA .

Two MnCl$_2$ simulations of duration 400\,ps and 45\,ns were executed. The short simulation used an integration time step 0.001\,ps and data collection every 1\,ps.  Analysis generated the Mn--H radial distribution function for comparison with the distance fit parameters, and obtained an estimate of the outer-sphere rotational time constant $\tau_{\rm{o}\beta}$ of 32\,ps.  

The long MnCl$_2$ simulation of duration 45\,ns required an integration time step of 0.0009\,ps.  and data collection every 0.9\,ns.  This simulation was an attempt to estimate the {\em inner-sphere} rotational time constant $\tau_{\rm{i}\beta}$ for  Mn--H  where H refers to inner sphere water protons.  All results are discussed in the main text.

\section{Note 3: A collision model for rotational diffusion}

\begin{figure}[H] 
	\unitlength1cm
	\begin{center}
		\includegraphics[width=0.5\textwidth, trim={0cm 1cm 10cm 1cm}]{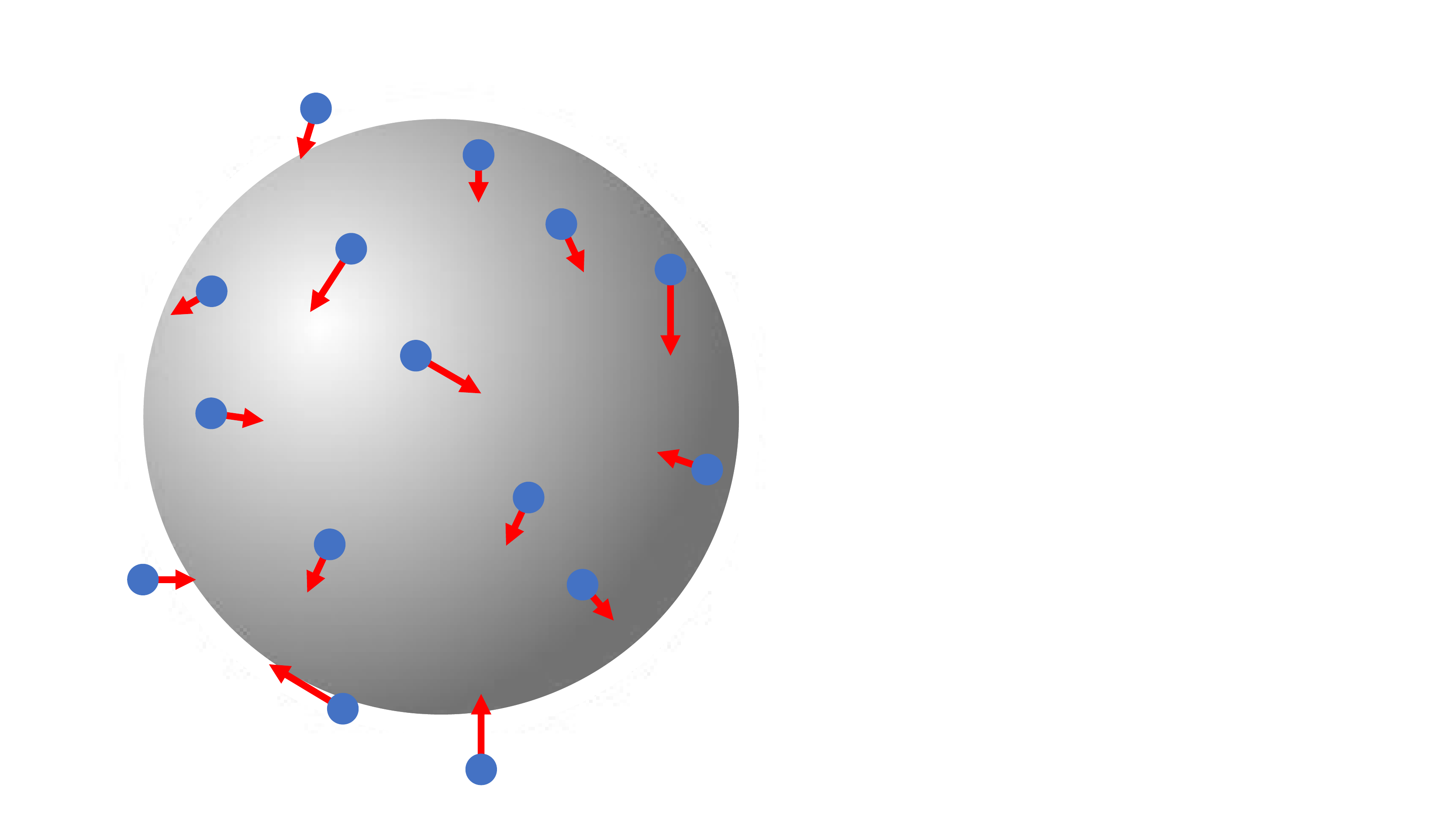}
		\caption{An aqueous complex is modeled as a sphere comprising a central paramagnetic cation surrounded by water and anions. The complex is subject to random collisions by unbound water causing the sphere to execute a rotational random walk.}
		\label{Sphere-complex}
	\end{center}
\end{figure}

The objective is to estimate the angular rotation of an aqueous complex  when subject to repeated  collisions by water. A hydrated ion complex is represented as a sphere of uniform mass density $\rho$ and radius $R$ as illustrated schematically in Fig.\,\ref{Sphere-complex}. 

The sphere is subject to repeated impacts by free water molecules at its surface.  A time interval $\Delta t$ is defined as the characteristic time for a single water molecule collision.  A water molecule in the bulk at room temperature moves the average distance 0.27\,nm (the distance between neighboring  water molecules) in 5\,ps and provides a guide value for  $\Delta t$ although the precise value is not important. The number of water molecules striking the surface of the sphere during time $\Delta t$ is $N$.  Naturally, 
\begin{equation}
N\,=\,4 \pi R^2/A  \tag{S11} \label{eqn:N}
\end{equation}
where $A$ is the collision area at the surface of the sphere associated with a single water molecule.

The change in momentum of the $i^{th}$ water molecule on impact produces a force $\mathbf{F}_i$ at the surface of the sphere.  We are concerned  with rotational motion, rather than translational motion, and so consider only  the force $\mathbf{F}_i$ tangent to the sphere labelled $\mathbf{F}_{i,\beta}$.  $\mathbf{F}_{i,\beta}$ is the force that contributes to rotation about each of the fixed Cartesian $x$, $y$ and $z$ axes centred on the sphere.  The combination of rotations during time interval $\Delta t$ is equivalent to a single rotation about some axis $\mathbf{\hat{r}}$ that runs through the center of the sphere.

Each force $\mathbf{F}_{i,\beta}$ contributes a component  $F_{i,\beta,\hat{r}}$ at a radial distance $R_{i,\hat{r}}$ from the $\mathbf{\hat{r}}$ axis acting to rotate the sphere about the axis.  The net torque associated with the $N$ unbiased collisions at each time interval $\Delta t$ is
\begin{equation}
M_{\hat{r}}  =   \sum_{i=1}^{N} F_{i,\beta, \hat{r}} R_{i,\hat{r}}. \tag{S12} \label{eqn:Mx}
\end{equation}
The radial distance $R_{i,\hat{r}}$ in Eq.\,(\ref{eqn:Mx}) is the distance to $ F_{i,\beta, \hat{r}}$ perpendicular to the $\mathbf{\hat{r}}$  axis.  Its average over $N$ collisions is, 
\[
\langle R_{\hat{r}}  \rangle = \frac{\pi}{2}R \tag{S13}
\]
which is proportional to the sphere radius $R$. 

It is reasonable to assume that the distribution of $N$ forces $F_{i,\beta, \hat{r}}$ is Gaussian with a mean $\langle F_{\beta, \hat{r}} \rangle \,=\,0$ and a standard deviation $\sigma_F$.  On average, no rotation will occur because, on average, the anticlockwise rotations are balanced by clockwise rotations.  The average torque about the $\mathbf{\hat{r}}$ axis as given by Eq.\,(\ref{eqn:Mx}) is zero.  The spread of collision forces is reflected in $\sigma_F$  which is independent of the size of the sphere.   However, statistical fluctuations in {\em mean} force $\langle F_{\beta,\hat{r}} \rangle$ characterised by the standard error in the mean $s_M$ will lead to a rotation about $\mathbf{\hat{r}}$ during each time interval $\Delta t$ given by
\[
s_M \,=\frac{\sigma_F}{\sqrt{N}} \mbox{~~~~so that~~~~}  M_{\hat{r}}   =    \frac{\sigma_F}{\sqrt{N}} \frac{\pi}{2}R .
\tag{S14} \label{eqn:sMsigF}
\]

Each net torque will cause the sphere to rotate about an axis of rotation by some angle $\Delta \beta$. The change in angle will be of a different magnitude and about a different axis of rotation at each time interval $\Delta t$.  The sphere executes a rotational random walk. After a time $t$, comprising a number of collision time intervals $\Delta t$, the sphere is oriented at an angle $\beta$ to its original orientation at $t\,=\,0$.  It is the time evolution of $\beta$ that is required for the shell model described in the main article.

The magnitude of $\beta$ after a time $t$ is proportional to the average change in angle, $\langle \Delta \beta \rangle$, at each time interval  which, in turn, is proportional to the net torque supplied by $s_M$.  Referring to Eq.\,(\ref{eqn:sMsigF}) brings
\begin{equation}
\langle M_{\hat{r}} \rangle =  I \alpha = \left( \frac{8}{15} \rho \pi R^5 \right)  \frac{\langle \Delta \beta \rangle}{\Delta t^2}  \tag{S15} \label{eqn:sMIalpha}
\end{equation}
where $I$ is the moment of inertia of a uniform sphere of density $\rho$, $\alpha$ is the angular acceleration and $\langle \beta \rangle$ is the mean change in angle due to the mean torque $\langle M_{\hat{r}} \rangle$  after a time interval $\Delta t$. Combining Eqs.\,(\ref{eqn:N}), (\ref{eqn:sMsigF}) and (\ref{eqn:sMIalpha}) yields
\begin{equation}
\langle \Delta \beta \rangle =  \frac{15 \sigma_F  A^{1/2} \Delta t^2 }{32 \rho \pi^{1/2} R^5} .  \tag{S16} \label{eqn:beta-ave}
\end{equation}
This result finds that the average change in angle is proportional to $R^{-5}$ because the quantities $A$, $\rho$, $\sigma_F$ and $\Delta t$ are, to a good approximation, independent of the radius of the complex.  

The independence on $R$ of many parameters provides the opportunity to present the ratio of angular changes between the complex and a standard system, taken to be water.  We have
\begin{equation}
\frac{ \beta_c}{\beta_w} = \frac{\langle \Delta \beta \rangle_c}{\langle \Delta \beta \rangle_w} = \frac{R_w^5}{R_c^5} \tag{S17} \label{eqn:betaRatio} .
\end{equation}
where the subscripts $c$ and $w$ refer to the complex and to water respectively.  The final step is to link the rotational correlation time $\tau_\beta$ to  $\langle \Delta \beta \rangle$ and $\beta$.  Equation (\ref{eqn:betaRatio}) will then provide an estimate of $\tau_{\beta,c}$ for the molecular complex in terms of $\tau_{\beta,w}$ and the ratio of characteristic radii. 

First, however, recollect that the diffusion coefficient for the random (Brownian) walk of identical particles moving in one-dimension (1D) along an $x$-axis may be written
\[
D = \frac{ \langle x^2 \rangle}{2 t} = \frac{ \delta^2}{2 \tau}  \tag{S18} \label{eqn:Dx} 
\]
where $\langle x^2 \rangle$ is the mean squared distance moved in a time $ t$, and the diffusion time constant $\tau$ is the time it takes, on average, for particles to move a standard distance $\delta$ in 1D. If $\tau$ is short, the diffusion coefficient is large and the particles are moving more rapidly. Note that $\langle x^2 \rangle \propto \tau^{-1}$ independent of $t$ because there are no limits to the range of diffusion.  This is not so for angular rotation.

The diffusion propagator describing the angular diffusion of a rotor is derived as an adaptation of 1D particle motion in which $x^2$ is replaced by $\beta$ \cite{faux2021a}.   By analogy,
\[
D = \frac{ \langle \beta \rangle}{2 t} = \frac{ \beta_s}{2 \tau_\beta}.  \tag{S19} \label{eqn:Dbeta} 
\]
It appears that $\langle \beta \rangle \propto \tau_\beta^{-1}$, consistent with 1D motion, but this is true only at short times (small angles) where the effect of the maximum limit on $\beta$ of $\pi$ radians has yet to be felt. The consequence is that the ``standard angle'' $\beta_s$ is carefully chosen to be 1 radian \cite{faux2021a} and the rotational correlation time $\tau_\beta$ determined accordingly. 

\begin{figure}[H] 
	\unitlength1cm
	\begin{center}
		\includegraphics[width=0.5\textwidth, trim={2cm 3cm 2cm 3cm}]{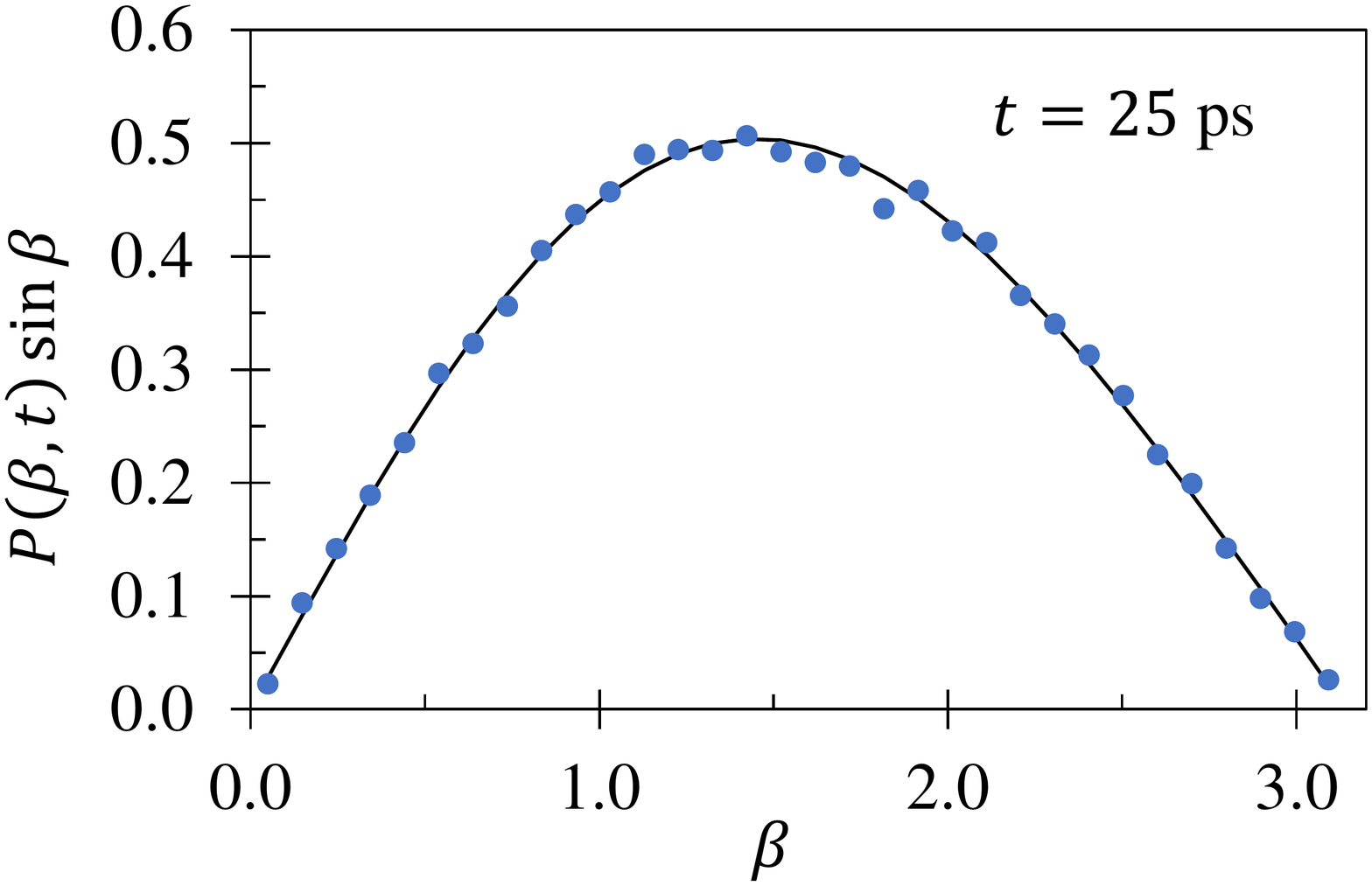}
		\caption{The probability density function describing the distribution of the angular change $\beta$ of the O-H vector in water is presented for a time evolution of 25\,ps.  The molecular dynamics simulation results are displayed as data point and the curve is the best fit using Eq.\,(\ref{S9:Pbeta})}
		\label{PBtvsB}
	\end{center}
\end{figure}

Equation (\ref{eqn:betaRatio}) now becomes
\begin{equation}
\frac{\tau_{\beta,w}}{\tau_{\beta,c}} = \frac{R_w^5}{R_c^5} \tag{S20} \label{eqn:beta-ave-ratio} .
\end{equation}
Equation (\ref{eqn:beta-ave-ratio}) is 
used in the main article to estimate the rotational correlation time $\tau_{\beta,c}$ of the inner sphere of a hydrated ion in terms of the equivalent quantity for water.  An estimate of $\tau_{\beta,c}$  is easily achieved using the data from molecular dynamics simulations previously published \cite{faux2021a} for the rotation of the  O--H intramolecular bond and fitting to Eq.\,\ref{S9:Pbeta}.  A sample fit is presented in Fig.\,\ref{PBtvsB}.


The dynamics of water at picosecond scales shows Brownian motions interspersed by fast H-bond breaking and reforming. The intramolecular H-H bond rotational correlation time was found to be 8\,ps \cite{faux2021a}, and $\tau_{\beta,w}$ for the O-H vector would be expected to be similar.  Figure\,\ref{PBtvsB} presents the simulation data with the fit to Eq.\,(\ref{S9:Pbeta}) which finds $\tau_{\beta,w} \,= \, 9\pm 0.5$\,ps.

\section{Note 4: The inner sphere water fraction}

In the fast-diffusion regime, the inner sphere contribution to the overall measured relaxation rate may be written $xR_{1i}(\omega)$ (see Eq.\,(4) in the main text) where $x$ is the volume fraction of water in the inner sphere.  In this Note, we assess the reasonableness of the value of $x$ obtained from Fit-Mn1 for 1\,mM aqueous Mn(II) chloride. As $x$ is a dimensionless volume fraction,  the inner shell is converted to a volume by assigning a shell thickness $\delta a$  from  $a - \delta a/2$ to $a + \delta a/2$.  The shell thickness calculated from the Fit-Mn1 parameters $a$ and $x$ yields $\delta a =0.066$\,nm.  The result is in good agreement with the full width at half height of the first hydration shell ($\approx$0.05\,nm) obtained from MD simulations and presented in Fig.\,4(b) of the main text.

On the basis of the calculated shell thickness $\delta a = 0.066$\,nm, the first line of Table I presents the equivalent number of inner sphere water molecules assuming the  room temperature bulk water density $\rho_w = 33.3$ molecules/nm$^3$.  The estimate of 2.1 water molecules is reasonable considering the crudeness of the assumptions.  In the second row of Table I, the number of inner sphere water molecules is set to six retaining the assumption that $\delta a = 0.066$\,nm.  The water density required to accommodate six water molecules is approximately $3 \rho_w$ which again is reasonable.

In the third row of Table I, the inner sphere water fraction $x$ obtained from Fit-Mn1 is not used.  Instead, the number of inner sphere water molecules is set to six and the inner sphere water density to  $\rho_w$.  These values are combined with the inner shell radius $a=0.272$\,nm to obtain an estimate for the shell thickness. The result, $\delta a = 0.19$\,nm, compares to a layer of water at a solid surface which has a thickness $\delta a \approx 0.27$\,nm equivalent to the distance between nearest neighbor water molecules in the bulk.

\newpage

\begin{table}[ht]
	\caption{Some fit results and calculations based on the inner sphere water fraction $x$ from Fit-Mn1 are presented.  The underlined quantities are values from Fit-Mn1 in the main article.  The $^\dagger$ labels those quantities calculated with other parameters in the table row set as shown.   The calculation of the inner sphere volume $V_i$ assumes that, at 1\,mM, the volume per ion is $V_{\rm ion}= 1661$\,nm$^3$. $\rho_w$ is the water density at room temperature equal to 33.3 molecules/nm$^3$. }
	
	\begin{tabular}{cccccc}\toprule \\[-3mm]
		~~inner sphere water~~    &  ~~inner sphere volume~~ &inner shell  & ~~~shell thickness~~~& ~~No.\,of inner sphere~~ & water density     \\
		fraction	(from fit)    &  $V_i = xV_{\rm ion}$ &radius	(from fit)  & from $V_i$ and $a$   & water molecules &     \\ \hline \\[-2mm]
		$x$    &  $V_i$ (nm$^3$) &$a$ (nm)   & $\delta a$ (nm)  & $q$  &$\rho_w$ (nm$^{-3}$)    \\ \hline \\[-2mm]
		\underline{3.67$\times$10$^{-5}$} & 0.0610 &\underline{0.272} &  0.066 & 2.1$^\dagger$ & $\rho_W$ \\[1mm]
		\underline{3.67$\times$10$^{-5}$} & 0.0610 &\underline{0.272} &  0.066 & 6& 2.9$\rho_w^\dagger$ \\[1mm]
		& &\underline{0.272} &  0.19$^\dagger$& 6 & $\rho_w$ \\[-3mm]
		\\ \hline \hline
	\end{tabular} 
	\label{Table:samples}
\end{table}

\section{Note 5: Aqueous copper(II) chloride}

Copper(II) chloride has applications in a range of technologies \cite{zhang2021spectroscopic} and consequently has been extensively studied. It is unusual because the interactions between the Cu(II) and water are weak so that the chloride ions may enter the coordination sphere. The stoichiometry of the complex formed is sensitive to both the chloride concentration and the presence of solvents \cite{zhang2021spectroscopic}.  

Work by Bell and co-workers \cite{bell1973solute}, and Magini \cite{magini1981hydration} show that the chloride ions link the copper ions in a chain, such that the number of water molecules surrounding each copper iron is reduced. Results using X-ray diffraction (XRD) at high chloride concentration (2.0-4.35\,M) suggested 2.4-2.8 water molecules per copper ion \cite{bell1973solute,magini1981hydration}.  Later, D'Angelo {\it et al} \cite{dangelo1997structural} summarise in their Table II their own work and that of others \cite{khan1976stability,bjerrum1986weak,ramette1983copper} at chloride molarities of 0.1\,M, 1.0\,M and 3.0\,M. The  number of hydrating water molecules are, on average, 5.6, 4.8 and 4.0 respectively.   The four equatorial water molecules at high concentrations are joined by axial water upon the breakup of the chains as the chloride concentration is reduced.

The FFC NMR experiments reported in the main paper for Cu(II) chloride use chloride concentrations of 3.0--18.4\,mM, much lower than the most dilute system reported by D'Angelo \cite{dangelo1997structural}.  It is reasonable to assume therefore that the dilute aqueous Cu(II) chloride used in the present experiments contain single Cu(II) ions surrounded by six inner sphere water molecules as for the other ions and therefore the use of a shell model is justified.  It is emphasised that the number of inner sphere water molecules is not a parameter used in the fitting process, see Note 4.

A second interesting feature of aqueous Cu(II) chloride is that the six inner sphere water molecules possess an elongated octahedral structure.  Once again, the results from d'Angelo and co-workers using extended X--ray absorption fine structure (EXAFS) in their Table I \cite{dangelo1997structural} is used to elucidate the Cu--$^1$H distances.    The water oxygen cage has an equatorial Cu--O distance of 1.97\,$\AA$ for four oxygen atoms and an axial distance of 2.28\,$\AA$ for the two remaining oxygen atoms. Making the usual correction to secure the Cu--$^1$H distance leads to a range 2.03--2.34\,$\AA$ with the average weighted Cu--$^1$H distance at 2.13\,$\AA$.

\bibliographystyle{apsrev4-1}
\bibliography{References_master_2022a}